\pdfoutput=1
\documentclass[a4paper,11pt]{article}
\usepackage{geometry}
\usepackage{amsmath}
\usepackage{amssymb}
\usepackage{graphicx}
\usepackage{color}
\usepackage[colorlinks=true,citecolor=blue,linkcolor=blue,urlcolor=blue]{hyperref}

\usepackage[square,numbers,comma,sort&compress]{natbib}
\usepackage{verbatim}
\usepackage{textcomp}
\usepackage{nicefrac}
\usepackage{subcaption}
\usepackage{booktabs}
\usepackage{tikz}
\newcommand{\FMDG}[1]{\ensuremath{\vcenter{\hbox{\includegraphics{#1}}}}}

\usepackage{array}
\usepackage{multirow}
\usepackage{slashed}
\usepackage[normalem]{ulem}

\def\varabstract{ }
\def\varkeywords{ }
\def\vararxivnumber{ }
\def\vartitle{ }
\def\varpreprint{ }
\renewcommand{\title}[1]{\gdef\vartitle{#1}}
\renewcommand{\abstract}[1]{\gdef\varabstract{#1}}
\newcommand{\keywords}[1]{\gdef\varkeywords{#1}}
\newcommand{\arxivnumber}[1]{\gdef\vararxivnumber{#1}}
\newcommand{\preprint}[1]{\gdef\varpreprint{#1}}
\newtoks\authtoks
\renewcommand{\author}[2][]{%
	\authtoks=\expandafter{\the\authtoks#2$^{#1}$\ }%
}
\newtoks\affiltoks
\newcommand{\affiliation}[2][]{%
    \affiltoks=\expandafter{\the\affiltoks{\item[$^{#1}$]#2}}%
}
\newtoks\emailtoks\newcounter{emailcounter}%
\newcommand{\emailAdd}[1]{%
\ifnum\theemailcounter>0\emailtoks=\expandafter{\the\emailtoks, \typeemail{#1}}%
\else\emailtoks=\expandafter{\typeemail{#1}}%
\fi
\stepcounter{emailcounter}}
\newcommand{\typeemail}[1]{\href{mailto:#1}{\tt #1}}

\renewcommand\maketitle{
	\newgeometry{margin=2cm}
	\pagestyle{empty}\setcounter{page}{0}
	\if!\varpreprint!\else\begin{flushright}\varpreprint\end{flushright}\fi
	{\huge\flushleft\sffamily\bfseries\vartitle\par}
\vskip6ex
{\large\bfseries\raggedright\sffamily\the\authtoks\par}
\vskip2ex
\begin{list}{}{%
\setlength{\leftmargin}{0.28cm}%
\setlength{\labelsep}{0pt}%
\setlength{\itemsep}{-3pt}%
\setlength{\topsep}{-\parskip}}
\itshape\small%
\the\affiltoks
\end{list}
\vskip2ex
\noindent\hspace{0.28cm}\begin{minipage}[l]{\textwidth}
\begin{flushleft}
\textit{E-mail:} \the\emailtoks
\end{flushleft}
\end{minipage}
\vskip5ex
\noindent{\renewcommand\baselinestretch{.9}\textsc{Abstract:}}\ \varabstract
\vskip5ex 
\if!\varkeywords!\else\noindent{\textsc{Keywords:}}\ \varkeywords \vskip2ex\fi
\newpage
	\newgeometry{margin=2.3cm,top=2.1cm,bottom=2.1cm}
\pagestyle{plain}
\hrule
\bigskip\bigskip

{
	\hypersetup{linkcolor=black}
	\tableofcontents
}
\bigskip\medskip
\hrule
\bigskip\bigskip
\setcounter{footnote}{0}
\restoregeometry
} 

\definecolor{dgreen}{rgb}{0.2,0.7,0.2}       
\definecolor{darkgreen}{HTML}{00AA00}

\newcommand{\braket}[1]{\ensuremath{\left\langle#1\right\rangle}}

\title{\boldmath
Phenomenology of the Generalised Scotogenic Model with Fermionic Dark Matter
}

\author[a]{\textbf{Claudia Hagedorn,}}
\author[b]{\textbf{Juan Herrero-Garc\'ia,}}
\author[c, a]{\textbf{Emiliano Molinaro}}
\author[d]{\textbf{and Michael A.~Schmidt}}

\affiliation[a]{CP$^3$-Origins, University of Southern Denmark, Campusvej 55, 5230 Odense M, Denmark}
\affiliation[b]{ARC Center of Excellence for Particle Physics at the Terascale, University of Adelaide, 5005 Adelaide, South Australia, Australia}
\affiliation[c]{Department of Physics and Astronomy, University of Aarhus, Ny Munkegade 120, DK-8000 Aarhus C, Denmark}
\affiliation[d]{ARC Centre of Excellence for Particle Physics at the Terascale, School of Physics, The University of Sydney, Physics Road, NSW 2006, Australia}

\emailAdd{hagedorn@cp3.sdu.dk}
\emailAdd{juan.herrero-garcia@adelaide.edu.au}
\emailAdd{molinaro@phys.au.dk}
\emailAdd{michael.schmidt@sydney.edu.au}

\abstract{We study a simple extension of the Standard Model that accounts for
neutrino masses and dark matter. The Standard Model is augmented by two
Higgs doublets and one Dirac singlet fermion, all charged under a new dark
global symmetry. It is a generalised version of the Scotogenic Model with Dirac fermion dark matter. Masses for two neutrinos are generated radiatively at one-loop level. We study the case where the singlet
fermion constitutes the dark matter of the Universe. We study in depth the
phenomenology of the model, in particular the complementarity between dark matter direct detection and charged lepton flavour violation observables.
Due to the strong limits from the latter, dark matter annihilations are suppressed and the relic abundance is set by coannihilations with (and annihilations of) the new scalars if the latter and the Dirac fermion are sufficiently degenerate in mass. We discuss how different ratios of charged lepton flavour violating
processes can be used to test the model. 
We also discuss the detection prospects of the
charged scalars at colliders. In some cases these leave ionising tracks and in others have prompt decays, depending on the flavour in the final state and neutrino mass orderings.
}


\arxivnumber{1804.04117}
\preprint{ADP-18-9/T1057\\
CP3-Origins-2018-012-DNRF90}

\makeatother

\begin{document}

\maketitle

\section{Introduction} \label{sec:intro}

Neutrino masses and the missing mass in the Universe are among the most important evidence for physics beyond the Standard Model (SM). Some of the most prominent proposed
explanations for them are radiative neutrino mass models (see Ref.~\cite{Cai:2017jrq} for a recent review) and particle dark matter (DM) (see Ref.~\cite{Bertone:2004pz} for a review), respectively. A simple and elegant candidate of the latter are weakly interacting massive particles (WIMPs). In this work, we study a simple model that has the
interesting feature of explaining simultaneously both neutrino masses and dark matter. In particular, we study a generalised version of the
Scotogenic Model (ScM) with a global $\rm U(1)_{\rm DM}$ symmetry. We denote it the Generalised\footnote{The term \emph{Generalised} should be understood in reference to the original proposal of the ScM. Other symmetry groups are also possible and will be briefly discussed in Sec.~\ref{sec:disc}.} Scotogenic Model (GScM), because the global $\rm U(1)_{\rm DM}$ symmetry contains as a subgroup the discrete $Z_2$ symmetry of the original ScM proposed in Ref.~\cite{Ma:2006km} by E.~Ma. In the last years there have been several studies of the phenomenology of the ScM~\cite{Hessler:2016kwm,Vicente:2014wga,Molinaro:2014lfa,Toma:2013zsa,Racker:2013lua,Schmidt:2012yg,Gelmini:2009xd,Sierra:2008wj,Suematsu:2009ww,Kubo:2006yx}. A systematic study of one-loop neutrino mass models with a viable DM candidate which is stabilised by a $Z_2$ symmetry has been presented in Ref.~\cite{Restrepo:2013aga}. A similar model to the GScM with a gauged $\rm U(1)_{\rm DM}$ symmetry has been introduced in Ref.~\cite{Ma:2013yga}. Several variants of the ScM with a  $\rm U(1)$ symmetry instead of  a $Z_2$ symmetry have been proposed~\cite{Kubo:2006rm,Ma:2008ba} after the original ScM model. 

The GScM involves two scalar doublets and one Dirac fermion, all charged under
the global $\rm U(1)_{\rm DM}$ symmetry. Masses for two neutrinos are generated at the one-loop level, with
a flavour structure different from that involved in processes with charged lepton flavour
violation (CLFV). The model has some definite predictions, as the flavour structure of the
Yukawa couplings is completely determined by the neutrino oscillation
parameters and the Majorana phase. This allows to draw predictions for CLFV processes and decays of the new
scalars, as we discuss in detail.  The constraints from the non-observation of CLFV processes are complementary to the limits from direct detection experiments.

In contrast to the models in Refs.~\cite{Kubo:2006rm,Ma:2008ba} (and some variants in Ref.~\cite{Ma:2013yga}) the $\rm U(1)_{\rm DM}$ symmetry is not broken in the GScM, which leads to several changes in the phenomenology of the model. This makes the study of WIMPs scattering off nuclei in direct detection experiments very interesting, as it is generated via the DM magnetic dipole moment at one loop. The limits from direct detection experiments already imply the need of coannihilations of the Dirac fermion DM and the new scalars in the early Universe to explain the observed DM relic abundance. Scalar DM is disfavoured, because of a generically too large DM-nucleus cross section mediated by $t$-channel $Z$-boson exchange. We focus on the case of fermionic DM, which in this model is a Dirac fermion, unlike the original ScM. 

The paper is structured as follows. In Sec.~\ref{sec:model} we introduce the
GScM and discuss the scalar mass spectrum and neutrino masses. In
Sec.~\ref{sec:pheno} we discuss the most relevant phenomenology of the model,
especially CLFV, the DM abundance as well as
collider searches. In Sec.~\ref{sec:num} we show the results of a numerical
scan of the parameter space of the model. In Sec.~\ref{sec:disc} we discuss variants of the
model with the dark global $U(1)$ symmetry being gauged or replaced by a $Z_2$,  $Z_3$ or $Z_4$ symmetry, and the case where the singlet is substituted by a triplet of the electroweak gauge group. A
comparison to the original ScM is presented in
Sec.~\ref{sec:scot}. Finally, we conclude in Sec.~\ref{sec:conc}. 
Further details of the model are given in the final appendices.
We discuss the stability of the potential in App.~\ref{sec:stab} and neutrino masses and lepton mixing in App.~\ref{app:PMNS}. The parametrisation of the Yukawa couplings in terms of the former is presented in App.~\ref{sec:yukpar}. Loop functions relevant for different processes and input for the computation of the $\mu-e$ conversion ratio are provided in App.~\ref{app:loop}. Expressions for the electroweak precision tests (EWPT) are given in App.~\ref{app:ewpt}.


\section{The Generalised Scotogenic Model} \label{sec:model}

The particle content of the model and its global charges were first outlined in Ref.~\cite{Ma:2013yga}.
It can be viewed as the generalisation of the ScM, since it is based on a global $\rm U(1)_{\rm DM}$ symmetry, while the ScM possesses a $Z_2$ symmetry.
The SM is augmented by two additional scalar doublets and one vector-like Dirac fermion, all charged under the $\rm U(1)_{\rm DM}$ symmetry. The particle content and quantum numbers are given in Tab.~\ref{tab:T3B}. 
  \begin{table}[tb]
\centering
\begin{tabular}{|l|ccc|c|}
\hline
Field & SU(3)$_{\rm C}$ & SU(2)$_{\rm L}$ & U(1)$_{\rm Y}$ & U(1)$_{\rm DM}$\\
\hline\hline
$H$ & 1 & 2 & 1/2 & 0\\ 
$L_\alpha\equiv(\nu_{\alpha L},~\ell_{\alpha L})^T$ & 1 & 2 & -1/2 & 0\\ 
$\alpha_R$ & 1 & 1 & -1 & 0\\ 
\hline
$\Phi\equiv (\phi^+,~\phi_0)^T$ & 1 & 2 & 1/2 & 1\\ 
$\Phi^\prime\equiv (\phi_0^\prime,~\phi^{\prime-})^T$ & 1 & 2 & -1/2 & 1\\ 
$\psi$ & 1 & 1 & 0 & 1\\ 
\hline
\end{tabular}
\caption{Particle content and quantum numbers of the GScM. The upper block corresponds to the SM Higgs doublet and the SM leptons, with flavour index $\alpha=e, \, \mu, \, \tau$.
	The lower part shows the dark sector of the model: two scalar doublets $\Phi$ and $\Phi^\prime$, and one Dirac fermion $\psi$. In the last column we provide the transformation properties under the global $\rm U(1)_{\rm DM}$ symmetry.}\label{tab:T3B}
\end{table}
Without loss of generality we choose the $\rm U(1)_{\rm
DM}$ charge of the new particles as $q=+1$. All new
particles are ${\rm SU(3)}_{\rm C}$ singlets in order to
have a viable DM candidate.\footnote{Alternatively, DM may
be a bound state of coloured octet Dirac
fermions~\cite{DeLuca:2018mzn} (see also
Ref.~\cite{Reig:2018mdk} for a realisation in a radiative
Dirac neutrino mass model). In this case all new particles
are $\rm SU(3)_{\rm C}$ octets.}
In Sec.~\ref{sec:disc} and Ref.~\cite{Ma:2013yga} variants of the model are presented. A comparison to the ScM can be found in Sec.~\ref{sec:scot}.

We denote the SM Higgs doublet by $H$, which is given in unitary gauge after electroweak symmetry breaking by $H \equiv (0,(h+v_H)/\sqrt{2})^T$, with $v_H=246$ GeV the vacuum expectation value (VEV) and $h$ the Higgs boson. Without loss of generality we work in the charged lepton mass basis.
The Lagrangian for the Dirac fermion $\psi$ reads \footnote{It is convenient to use the conjugate for the Yukawa couplings to $\Phi^\prime$, i.e. $\left(y_{\Phi^\prime}^\alpha\right)^*$, so that the expressions for neutrino masses and CLFV are symmetric under simultaneous interchange of  $y^\alpha_{\Phi^\prime} \leftrightarrow y^\alpha_{\Phi}$ and the physical masses $(m_{\eta^\prime_0}, m_{\eta^{\prime+}}) \leftrightarrow (m_{\eta_0}, m_{\eta^+})$.}
\begin{equation}
\mathcal{L}_{\psi}\;=\; i\, \overline \psi\, \slashed \partial\,\psi\,-\, m_\psi \, \overline\psi\, \psi\, -\, \Big( y_{\Phi}^\alpha \, \overline \psi\, \tilde\Phi^\dagger\,L_\alpha\, + \, 
\left(y_{\Phi^\prime}^\alpha\right)^* \, \overline \psi\, \tilde \Phi^{\prime \dagger} \tilde L_\alpha  \,+\,\text{H.c.}\Big)\,,
\label{Lpsi}
\end{equation}
where $\tilde L\equiv i \sigma_2 C\overline L^T$, $C$ the charge conjugation matrix, and $\tilde \Phi \equiv i \sigma_2 \Phi^*$.  
The neutrino Yukawa couplings
$y_{\Phi}^\alpha,\,y_{\Phi^\prime}^\alpha$ are three-component vectors
\begin{equation}
 \mathbf{y_\Phi}\equiv \left(y_\Phi^e,\;y_\Phi^\mu,\;y_\Phi^\tau \right)^T \;\;\;\mathrm{and}\;\;\; \mathbf{y_{\Phi^\prime}}\equiv
\left(y_{\Phi^\prime}^e,\;y_{\Phi^\prime}^\mu,\;y_{\Phi^\prime}^\tau
\right)^T \, .
\end{equation} 
Four phases in the Yukawa vectors $\mathbf{y_\Phi}$ and
$\mathbf{y_{\Phi^\prime}}$ can be removed by phase redefinitions of the lepton doublets $L$ and the fermion $\psi$. In Sec.~\ref{sec:numasses} we discuss neutrino masses and estimate the size and form of neutrino Yukawa couplings for the case of a neutrino mass spectrum with normal ordering (NO) and inverted ordering (IO).

The scalar potential invariant under the $\rm U(1)_{\rm DM}$ symmetry is given by
\begin{eqnarray}
\begin{split}
 \label{eq:potential}
\mathcal{V}  =& \,-\,m_H^2  H^\dagger H \,+\, \lambda_H (H^\dagger H)^2 \,
+ \, m_\Phi^2  \Phi^\dagger\Phi\, +\, \lambda_\Phi (\Phi^\dagger\Phi )^2\,
+\, m_{\Phi^\prime}^2 \Phi^{\prime\dagger} \Phi^\prime \,+ \, \lambda_{\Phi^\prime} (\Phi^{\prime\dagger}\Phi^\prime )^2
\\&
\, + \, \lambda_{H\Phi} (H^\dagger H) (\Phi^\dagger\Phi)  
\,+\, \lambda_{H\Phi^\prime} (H^\dagger H) (\Phi^{\prime\dagger}\Phi^\prime)  
\,+\, \lambda_{\Phi\Phi^\prime} (\Phi^\dagger\Phi) (\Phi^{\prime\dagger}\Phi^\prime)  
\\&
\,+\, \lambda_{H\Phi,2} (H^\dagger \Phi) (\Phi^\dagger H)  
\,+\, \lambda_{H\Phi^\prime,2} (H^\dagger \tilde\Phi^{\prime}) (\tilde\Phi^{\prime \dagger}\,  H)  
\,+\, \lambda_{\Phi\Phi^\prime,2} (\Phi^\dagger\tilde\Phi^{\prime}) (\tilde\Phi^{\prime \dagger}\, \Phi)  \\&
\,+\, \lambda_{H\Phi\Phi^\prime}\left[ (H^\dagger \tilde\Phi^{\prime})(H^\dagger \Phi)\, +\, {\rm H.c.} \right]\,.
\end{split}
\end{eqnarray}
The coupling $\lambda_{H\Phi\Phi^\prime}$ can be chosen real and positive by redefining the scalar doublets $\Phi$ or $\Phi^\prime$. In our numerical analysis we apply the stability conditions outlined in App.~\ref{sec:stab}, which allow for the potential to be bounded from below. 

The lightest neutral particle of the dark sector is stabilised by the global $\rm U(1)_{\rm DM}$ symmetry, which remains unbroken,
 and thus is a potential DM candidate.
 If the DM is identified with the lightest neutral scalar coming from the new scalar doublets $\Phi$ and $\Phi^\prime$, as it carries non-zero 
hypercharge, neutral current interactions mediated the $Z$ boson give
scattering cross sections off nuclei well above current DM direct detection limits and thus disfavour this possibility. This is expected for a scalar doublet with a mass of about 1 TeV, whose relic abundance is set by gauge interactions.\footnote{For the specific case of maximal mixing between the neutral scalars, the contributions from $Z$-boson exchange cancel. The Higgs portal contributions could also be tuned to be small by suppressing the relevant quartic couplings. A study of the case of scalar DM will be presented in a future work.\label{scDM}} The only viable DM candidate is the SM singlet Dirac fermion $\psi$. We study in detail the allowed parameter space of the model. This, indeed, is the most interesting scenario, as
there is a connection between DM phenomenology, neutrino
masses, CLFV and searches at
colliders. The experimental constraints on the model coming from neutrino
masses and CLFV select the scalar mass spectrum and the
possible mechanisms to obtain the correct DM abundance.

\subsection{Scalar mass spectrum} \label{sec:scalar}

We assume in the following that none of the neutral components of $\Phi$ and $\Phi^\prime$ takes a VEV, so that the global $\rm U(1)_{\rm DM}$ symmetry is unbroken.

The physical scalar states of the theory are given by  $(i)$ one real field $h$, which corresponds to the SM Higgs boson, $(ii)$ two complex neutral scalar fields $\eta_0$ and
$\eta_0'$, which are linear combinations of $\phi_0$ and $\phi_0^\prime$ (see Tab.~\ref{tab:T3B}), and $(iii)$ two charged scalars $\eta^+\equiv\phi^+$ and $\eta^{\prime +}\equiv\phi^{\prime +}$ and their charged conjugates. The SM Higgs boson mass is given by
\begin{equation}
	m_h \; = \; \sqrt{2\, \lambda_H}\,v_H\,,
\end{equation}
which we set to $m_h = 125\, \mathrm{GeV}$ in the numerical scan.
The neutral mass eigenstates are defined as
\begin{eqnarray}
	\eta_0 & = & s_\theta \,\phi_0\,+\, c_\theta\, \phi_0^\prime\,,\label{eta0}\\
	\eta_0^\prime & = & -\,c_\theta\, \phi_0 \,+\, s_\theta\, \phi_0^\prime\,,\label{eta0p}
\end{eqnarray}
with $s_\theta\equiv\sin\theta$ and $c_\theta\equiv\cos\theta$. The mixing angle $\theta$ is defined in terms of
\begin{equation}
\label{eq:theta}
\tan 2\theta = \frac{2\,c}{b\,-\,a}\,
\end{equation}
with 
\begin{eqnarray} \label{eq:abc}
\begin{split}
	a &= m_\Phi^2\,+\,\frac 12 \,v_H^2\left(\lambda_{H\Phi}\,+\,\lambda_{H\Phi,2}\right)\,,\\
	b &= m_{\Phi^\prime}^2\,+\,\frac 12 \,v_H^2\left(\lambda_{H\Phi^\prime}\,+\,\lambda_{H\Phi^\prime,2}\right)\,,\\
	c &= -\,\frac 12 \, \lambda_{H\Phi\Phi^\prime}\,v_H^2\,.
\end{split}	
\end{eqnarray}
The corresponding mass eigenvalues are
\begin{eqnarray} 
\label{eq:meta_abc}
\begin{split}
	m_{\eta_0} &= \sqrt{\frac 12 \left(a\,+\,b\,+\,\sqrt{(a\,-\,b)^2\,+\,4\,c^2} \right)}\,,\\
	m_{\eta_0^\prime} &= \sqrt{\frac 12 \left(a\,+\,b\,-\,\sqrt{(a\,-\,b)^2\,+\,4\,c^2} \right)}\,.
\end{split}	
\end{eqnarray}
The minimum of the potential implies the relation $a\, b \,-\, c^2 \,>\, 0$.
Notice that, by definition, the scalar $\eta_0$ is always heavier than $\eta_0^\prime$, i.e.~$m_{\eta_0}\geq m_{\eta_0^\prime}$. 
The two charged scalars of the model do not mix among themselves, so that their masses are simply 
\begin{eqnarray}
\label{eq:metapm}
	m_{\eta^+} & =  \sqrt{m_{\Phi}^2\,+\,\frac 1 2 \lambda_{H\Phi} \,v_H^2} \;\;\; \mbox{and} \;\;\;
	m_{\eta^{\prime +}} & =  \sqrt{m_{\Phi^{\prime}}^2\,+\,\frac 1 2 \lambda_{H\Phi^\prime} \,v_H^2}\,.
\end{eqnarray}
For values of $\lambda_{H\Phi\Phi^\prime}$ small compared to the other quartic couplings $\lambda_i$ motivated by light neutrino masses (see Sec.~\ref{sec:numasses}), the charged scalar masses are related to the neutral ones as
\begin{eqnarray}
	m^2_{\eta_0} & \simeq  m^2_{\eta^+} \, +\,\frac 1 2 \lambda_{H\Phi,2}\,v_H^2 \;\;\; \mbox{and} \;\;\;
	m^2_{\eta_0^\prime} & \simeq  m^2_{\eta^{\prime +}}\,+\,\frac 1 2 \lambda_{H\Phi^\prime,2} \,v_H^2\,.	
\end{eqnarray}
Hence in this case the couplings
$\lambda_{H\Phi,2}$ and $\lambda_{H\Phi^\prime,2}$ determine the relative hierarchy of the neutral scalars with respect to the charged ones. For positive $\lambda_{H\Phi,2}>0$ the neutral scalar $\eta_0$ is heavier than the corresponding charged scalar, $m_{\eta_0}>
m_{\eta^{+}}$, and vice versa. Similarly, $\eta_0^\prime$ is heavier than $\eta^{\prime+}$ for positive $\lambda_{H\Phi^\prime,2}$. Both $\lambda_{H\Phi,2}$ and $\lambda_{H\Phi^\prime,2}$ can be either positive or negative, but there are constraints from the stability of the scalar potential which are discussed in detail in App.~\ref{sec:stab}, that is $\lambda_{H\Phi,2} \geq -\lambda_{H\Phi}$ and $\lambda_{H\Phi^\prime,2} \geq -\lambda_{H\Phi^\prime}$. 
These are sufficient conditions which are imposed in the numerical scan.

\subsection{Neutrino masses} 
\label{sec:numasses}

\begin{figure}[t]
\centering
\FMDG{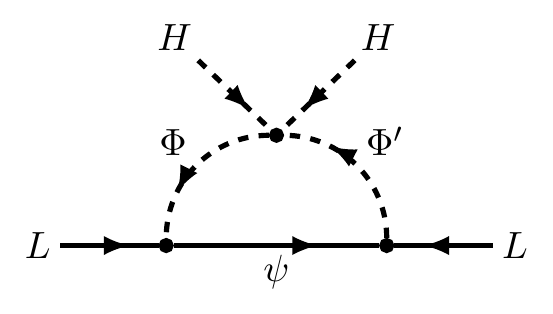}
\caption{Diagram generating neutrino masses at the one-loop level.}
\label{fig:T3B}
\end{figure}

From the Lagrangian for the Dirac fermion $\psi$ in Eq.~\eqref{Lpsi} and the scalar potential in Eq.~\eqref{eq:potential} we see that the total lepton number is violated by the simultaneous presence of the Yukawa couplings $\mathbf{y_{\Phi, \Phi^\prime}}$, the quartic coupling $\lambda_{H\Phi\Phi^\prime}$, and the fermion mass $m_\psi$. Thus Majorana neutrino masses need to be proportional to all of these parameters. They are generated after electroweak symmetry breaking at the one-loop level from the schematic diagram shown in Fig.~\ref{fig:T3B}, which generates the Weinberg operator after integrating out the Dirac fermion $\psi$ and the scalars $\Phi$ and $\Phi^\prime$. In the mass basis, $\eta_0$, $\eta_0^\prime$ and the Dirac fermion $\psi$ run in the loop. The Majorana mass term for the neutrinos is $-1/2\, \overline{\nu_{\rm L}^{c}}\mathcal{M}_{\nu}\nu_{\rm L}+\mathrm{H.c.}$, 
with the neutrino mass matrix given by
\begin{eqnarray} \label{eq:numasses}
	\left( \mathcal{M}_\nu\right)_{\alpha\beta} & = & \,\frac{\sin2\theta\,m_\psi}{32\,\pi^2}\,\Big(y_\Phi^\alpha\,y_{\Phi^\prime}^\beta+ y_{\Phi^\prime}^\alpha\,y_{\Phi}^\beta\Big)\,\,F(m_{\eta_0},m_{\eta_0^\prime},m_\psi)\,,
\end{eqnarray}
where we introduced the loop function
\begin{equation} \label{eq:F}
	F(x,y,z)\;\equiv\;\frac{x^2}{x^2-z^2}\,\ln\frac{x^2}{z^2}- \frac{y^2}{y^2-z^2}\,\ln\frac{y^2}{z^2}\,.
\end{equation}
There is always a suppression induced by the quartic coupling $\lambda_{H\Phi\Phi^\prime}\neq0$, which can be further enhanced by a small splitting of the neutral scalar masses $m^2_{\eta_0^\prime}-m^2_{\eta_0}$, which is approximately given by $|a-b|$ for $\lambda_{H\Phi\Phi^\prime}\ll1$, see Eqs.~\eqref{eq:abc} and \eqref{eq:meta_abc}. 

The resulting neutrino mass matrix is of rank two, provided the Yukawa vectors $\mathbf{y_\Phi}$ and $\mathbf{y_{\Phi^\prime}}$ are not proportional to each other. Hence, the neutrino mass spectrum consists of one massless neutrino and two (non-degenerate) Majorana fermions with masses
\begin{eqnarray}
\label{eq:mnupm}
  m_{\nu}^\pm &=& \frac{\left|\sin 2\theta\right|\,m_\psi}{32\,\pi^2}\, \left(  \left| \mathbf{y_\Phi} \right|\,\left| \mathbf{y_{\Phi^\prime}}\right|\,\pm\, | \mathbf{y_\Phi}^\ast\cdot\mathbf{y_{\Phi^\prime}} | \right)\,F(m_{\eta_0},m_{\eta_0^\prime},m_\psi)\,,
\end{eqnarray}
where $|\mathbf{y}| \equiv \sqrt{\sum_\alpha |y^{\alpha}|^2}$ denotes the norm of $\mathbf{y}$. As $m_{\eta_0}\geq m_{\eta_0^\prime}$, the loop function $F(m_{\eta_0},m_{\eta_0^\prime},m_\psi)\geq0$.
 The flavour structure is determined by the product $y_{\Phi}^\alpha \,y_{\Phi^\prime}^\beta$. 
 
For vanishing solar mass squared difference we can estimate the form of the Yukawa couplings $y^\alpha_\Phi$ and $y^\alpha_{\Phi^\prime}$ 
with the help of the formulae given in App.~\ref{sec:yukpar}. Indeed,
from Eq.~\eqref{eq:YukNO} and taking $m_2=0$, we find both $y^\alpha_\Phi$ and $y^\alpha_{\Phi^\prime}$
to be proportional to the complex conjugate of the third column of the Pontecorvo-Maki-Nakagawa-Sakata (PMNS) mixing matrix, as shown in Eq.~\eqref{UPMNS}. In particular, for neutrino masses with NO
we have
\begin{equation}    \label{eqs:estNO}
y^\mu_\Phi \approx y^\tau_\Phi \;\;\; \mbox{and} \;\;\; y^\mu_{\Phi^\prime} \approx y^\tau_{\Phi^\prime} \, ,
\end{equation}
taking $\theta_{23} \approx \pi/4$ and neglecting $\theta_{13}$. The Yukawa couplings
 $y^e_\Phi$ and $y^e_{\Phi^\prime}$ are expected to be smaller in magnitude than the other ones, since they are proportional to $\theta_{13}$. 
Plugged into the formulae for the two non-vanishing neutrino masses $m_\nu^\pm$
in Eq.~\eqref{eq:mnupm}, we confirm that $m_\nu^- \approx 0$ whereas $m_\nu^+$ does not vanish.

For IO we use Eq.~\eqref{eq:YukIO} with $m_1 \approx m_2$ and find that $y^\alpha_\Phi$ and $y^\alpha_{\Phi^\prime}$ are proportional to the sum and difference of 
the complex conjugate of the first two columns $u_1$ and $u_2$ of the PMNS mixing matrix, respectively, i.e. 
\begin{equation}
\label{eq:yPhiyPhipIO}
\mathbf{y_\Phi} \propto \left(
\begin{array}{c}
c_{12} \, \pm \, i \, e^{-i\, \gamma} \, s_{12}\\
- (s_{12} \, \mp \, i \, e^{-i\, \gamma} \, c_{12})/\sqrt{2}\\
(s_{12} \, \mp \, i \, e^{-i\, \gamma} \, c_{12})/\sqrt{2}
\end{array}
\right) \;\;\; \mbox{and} \;\;\; \mathbf{y_{\Phi^\prime}} \propto \left(
\begin{array}{c}
c_{12} \, \mp \, i \, e^{-i\, \gamma} \, s_{12}\\
- (s_{12} \, \pm \, i \, e^{-i\, \gamma} \, c_{12})/\sqrt{2}\\
(s_{12} \, \pm \, i \, e^{-i\, \gamma} \, c_{12})/\sqrt{2}
\end{array}
\right) 
\end{equation}
for $c_{12}\equiv \cos\theta_{12}$, $s_{12}\equiv \sin\theta_{12}$, $\theta_{13} \approx 0$ and $\theta_{23} \approx \pi/4$. This clearly shows that 
\begin{equation} \label{eqs:estIO}
y^\mu_\Phi \approx -y^\tau_\Phi \;\;\; \mbox{and} \;\;\; y^\mu_{\Phi^\prime} \approx -y^\tau_{\Phi^\prime}
\end{equation}
as well as $y^e_\Phi$ and $y^e_{\Phi^\prime}$ of similar magnitude, but not the same. 
 For the proportionality constant in Eq.~(\ref{eq:yPhiyPhipIO}) being real and positive, as it is assumed in our numerical analysis, we expect the real part of both $y^e_\Phi$ and $y^e_{\Phi^\prime}$ to be positive, 
  since $c_{12} > s_{12}$. Furthermore, the imaginary parts of $y^e_\Phi$ and $y^\mu_\Phi$ ($y^\tau_\Phi$) are proportional to each other with a positive (negative)
  proportionality constant, determined by the ratio $s_{12}/c_{12}$. The same holds for the imaginary parts of the Yukawa couplings $y^e_{\Phi'}$ and $y^\mu_{\Phi'}$ ($y^\tau_{\Phi'}$).
   When plugged into the formula for $m_\nu^\pm$ in Eq.~\eqref{eq:mnupm}, we find $m_\nu^- \approx m_\nu^+$, as expected for neutrino masses with IO. 
   The expectations for the Yukawa couplings $y^\alpha_\Phi$ and $y^\alpha_{\Phi^\prime}$ are confirmed in our numerical analysis to a certain extent,\footnote{For NO the approximation $m_2=0$ is oversimplifying, since we neglect in the estimate for  $y^e_{\Phi^{(\prime)}}$ the contribution proportional to the second column of the 
PMNS mixing matrix, which is relatively suppressed by $(\Delta m_{21}^2/\Delta m_{31}^2)^{1/4} \approx 0.41$ compared to the contribution coming from the third column, which is suppressed by $\theta_{13}\approx 0.15$.} as shown in Figs.~\ref{fig:yukawasNO} and \ref{fig:yukawasIO} in App.~\ref{sec:yukpar}.


\section{Phenomenology} \label{sec:pheno}

\mathversion{bold}
\subsection{$\ell_\alpha\to \ell_\beta\,\gamma$}
\mathversion{normal}

\begin{figure}[hbt]\centering
		\centering
		\FMDG{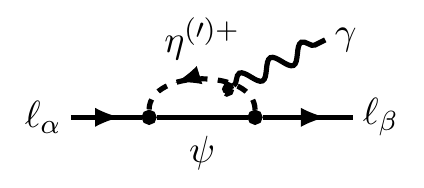}
	\caption{Diagram contributing to $\ell_\alpha\to \ell_\beta\gamma$.}
	\label{fig:llgamma}
\end{figure}

The most general amplitude for the electromagnetic CLFV transition $\ell_\alpha(p)\to \ell_\beta(k)\,\gamma^*(q)$ can be parameterised as 
 \cite{Hisano:1995cp}
\begin{equation}
	\mathcal{A}_\gamma \;=\; e\,\epsilon_\rho^*(q)\,\overline{u}(k)\Big[ q^2\,\gamma^\rho\left( A_1^L\,P_L\,+\,A_1^R\,P_R \right)\,+\,
						 m_\alpha\,i\,\sigma^{\rho\sigma} \left(A_2^L\,P_L\,+\,A_2^R\,P_R\right) q_\sigma  \Big] u(p)\,,\label{ampgamma}
\end{equation}
where $e>0$ is the proton electric charge, $p$ ($k$) is the momentum of the initial (final) charged lepton $\ell_\alpha$ ($\ell_\beta$), and $q = p-k$ is the momentum of the photon and $m_\alpha$ is the mass of the decaying charged lepton $\ell_\alpha$. The form factors in Eq.~(\ref{ampgamma}) are radiatively generated at one-loop level via the diagram shown in Fig.~\ref{fig:llgamma}
 and receive two independent contributions from  the charged scalars running in the loop. 
For the transition $\ell_\alpha^- \to \ell_\beta^{-}\gamma^*$ they are given  by
\begin{equation}\label{gammaform}
	\begin{split}
        A_2^L \;= \; & A_1^R \;=\; 0\,, \\ 
	A_2^R \;=\;  & -\frac{1}{32\,\pi^2}\left[\frac{ y_{\Phi}^{\beta*} \, y_{\Phi}^\alpha}{m_{\eta^+}^2}\,f\left(\frac{m_\psi^2}{m_{\eta^+}^2}\right) \,+\,
	\frac{y_{\Phi^\prime}^{\beta *} \, y_{\Phi^\prime}^{\alpha}}{m_{\eta^{\prime+}}^2}\,f\left(\frac{m_\psi^2}{m_{\eta^{\prime+}}^2}\right) \right], \\
	A_1^L \;= \; & -\frac{1}{48\,\pi^2}\left[\frac{ y_{\Phi}^{\beta*} \, y_{\Phi}^\alpha}{m_{\eta^+}^2}\,g\left(\frac{m_\psi^2}{m_{\eta^+}^2}\right) \,+\,
	\frac{y_{\Phi^\prime}^{\beta *} \, y_{\Phi^\prime}^{\alpha}}{m_{\eta^{\prime+}}^2}\,g\left(\frac{m_\psi^2}{m_{\eta^{\prime+}}^2}\right) \right],
	\end{split}
\end{equation}
where the loop functions $f(x)$ and $g(x)$ are reported in Eq.~\eqref{f-g-loop} in App.~\ref{app:loop}. 
 They are approximately equal to $1/6$ for small $x$.

As is well known, the radiative LFV decays are mediated  by the electromagnetic dipole transitions in Eq.~(\ref{ampgamma}) and are thus described by the form factors $A_2^{L/R}$. The monopole, which is given by the form factors $A_1^{L/R}$, does not contribute to processes with an on-shell photon. Thus, the corresponding branching ratio  (BR) is given by
\begin{equation}
	\text{BR}(\ell_\alpha\to \ell_\beta\,\gamma) \;=\; \frac{48\,\pi^3\,\alpha_{\rm em}}{G_F^2}\,\Big[\left| A_2^L\right|^2\,+\,\left| A_2^R\right|^2  \Big] 
	\times \text{BR}\left(\ell_\alpha\to\ell_\beta\,\nu_\alpha\,\overline{\nu_\beta}\right) \label{BRmueg2}
\end{equation}
with the fine-structure constant $\alpha_{\rm em}=e^2/(4 \pi)$, the Fermi coupling constant $G_F$ and the branching ratios $\text{BR}\left(\ell_\alpha\to\ell_\beta\,\nu_\alpha\,\overline{\nu_\beta}\right)$ are $\text{BR}\left(\mu\to e\,\nu_\mu\,\overline{\nu_e}\right) \approx 1$,
$\text{BR}\left(\tau\to e\,\nu_\tau\,\overline{\nu_e}\right)\approx 0.178$ and $\text{BR}\left(\tau\to \mu\,\nu_\tau\,\overline{\nu_\mu}\right)\approx 0.174$~\cite{Patrignani:2016xqp}. 

Notice that in the branching rations of these CLFV processes, which set the most stringent constraints on the parameters of the model, there is a different dependence on the neutrino Yukawa couplings $y_{\Phi}^\alpha,\,y_{\Phi^\prime}^\alpha$ than in neutrino masses, where the product of both enters, c.f. Eq.~\eqref{eq:numasses}. This is different from the original ScM, where there is only one type of Yukawa interaction, and therefore a very similar combination enters in both neutrino masses and CLFV~\cite{Ma:2006km,Vicente:2014wga}. In Sec.~\ref{sec:scot} we review the structure of the neutrino mass matrix  in the ScM and comment on results for branching ratios of CLFV processes. 

The branching ratios of the different radiative decays $\mu\to e\gamma$, $\tau\to e\gamma$ and $\tau\to \mu\gamma$ are tightly correlated.
Using the estimates for the Yukawa couplings $y_\Phi^\alpha$ and $y_{\Phi^\prime}^\alpha$ for neutrino masses with NO and IO given in Sec.~\ref{sec:numasses}, respectively,
we expect that 
\begin{equation}
\label{eq:ratioradBRNO}
\frac{\text{BR}(\tau\to e \,\gamma)}{\text{BR}(\mu\to e \,\gamma)} \approx 0.2 \;\;\; \mbox{and} \;\;\; \frac{\text{BR}(\tau\to \mu \,\gamma)}{\text{BR}(\mu\to e \,\gamma)} \approx 5 \; , 
\end{equation}
since $y_{\Phi}^{\mu}$ ($y_{\Phi^\prime}^{\mu}$) and $y_{\Phi}^{\tau}$ ($y_{\Phi^\prime}^{\tau}$) are of the same size, 
whereas $y_{\Phi}^e$ and $y_{\Phi^\prime}^e$ are suppressed by $\theta_{13}$ for the case of neutrino masses with NO.
 For IO we instead expect both of the radiative $\tau$-lepton decays to be of similar size and
\begin{equation}
\label{eq:ratioradBRIO}
\frac{\text{BR}(\tau\to e \,\gamma)}{\text{BR}(\mu\to e \,\gamma)} \approx \frac{\text{BR}(\tau\to \mu \,\gamma)}{\text{BR}(\mu\to e \,\gamma)} \approx 0.2 \; ,
\end{equation}
as none of the Yukawa couplings $y_{\Phi}^\alpha$  and $y_{\Phi^\prime}^\alpha$ is suppressed.
Since the experimental bound on $\text{BR}(\mu\to e \,\gamma)$ is several orders of magnitude stronger than the one on radiative $\tau$-lepton decays, once the constraint BR$(\mu\to e\gamma)< 4.2\times 10^{-13}$ at 90\% CL~\cite{TheMEG:2016wtm} has been imposed on the parameter space of the model, the branching ratios of the radiative $\tau$-lepton decays are automatically below the current limits, i.e.
BR$(\tau\to e\gamma)< 3.3\times 10^{-8}$ and
BR$(\tau\to \mu\gamma)< 4.4\times 10^{-8}$ at 90\% CL~\cite{Aubert:2009ag}, 
as well as future experimental sensitivity $\approx \mathcal{O}(3\times 10^{-9})$~\cite{Aushev:2010bq}. 

As the neutrino oscillation parameters are already tightly constrained, it is possible to derive a constraint on the undetermined parameter $\zeta$ which affects the relative size of the Yukawa vectors $\mathbf{y_\Phi}$ and $\mathbf{y_{\Phi^\prime}}$, see Eqs.~(\ref{eq:YukNO}) and (\ref{eq:YukIO}), as a function of the masses $m_{\eta^+}$ and $m_\psi$. For large $|\zeta|$ the branching ratios are dominated by the diagram with $\eta^{+}$ in the loop and for small $|\zeta|$ by the diagram with $\eta^{\prime+}$ and we can always neglect the other contribution. The loop function $f(x)$ in Eq.~(\ref{gammaform}) takes values between $1/12$ and $1/6$ in the relevant parameter range  $0<x<1$. After conservatively approximating $f(x)\approx 1/12$ and the loop function  in the expression for neutrino masses $F$ with one, we find 
\begin{align}\label{eq:zetaEstimate}
\frac{\alpha_{\rm em} |\zeta|^4 |y_{\Phi^\prime}^{\beta*} y_{\Phi^\prime}^{\alpha}|^2} {3072\pi m_{\eta^{\prime+}}^4G_F^2}
\frac{\mathrm{BR}(\ell_\alpha\to \ell_\beta\nu_\alpha\bar\nu_\beta)}{\mathrm{BR}(\ell_\alpha\to  \ell_\beta\gamma)  } 
	\lesssim |\zeta|^4 \lesssim 
	\frac{3072\pi m_{\eta^{+}}^4G_F^2 |\zeta|^4}{\alpha_{\rm em} |y_\Phi^{\beta*} y_\Phi^{\alpha}|^2}
	\frac{\mathrm{BR}(\ell_\alpha\to  \ell_\beta\gamma)} {\mathrm{BR}(\ell_\alpha\to \ell_\beta\nu_\alpha\bar\nu_\beta)} \,,
\end{align}
which, using the Yukawa couplings expressed in terms of the lepton mixing angles given in App.~\ref{sec:yukpar}, translates for $\mu\to e\gamma$ into the following ranges for any neutrino mass ordering
\begin{equation}
	\label{eq:zetaEstimate2}
\begin{split}
	3\times 10^{-4} \left(\frac{100\,\mathrm{GeV}}{m_{\eta^{\prime+}}}\right) \left(\frac{100\,\mathrm{GeV}}{m_\psi\sin2\theta}\right)^{1/2}
	&\lesssim |\zeta| \lesssim 
			4 \times 10^3 \left(\frac{m_{\eta^{+}}}{100\,\mathrm{GeV}}\right) \left(\frac{m_\psi\sin2\theta}{100\,\mathrm{GeV}}\right)^{1/2}\,.
		\end{split}
\end{equation}
In the estimates above we have used the best-fit values for the lepton mixing parameters and neutrino masses and marginalised over the two possible solutions for the Yukawas (see App.~\ref{sec:yukpar}) and the Majorana phase $\gamma$. The lower (upper) bounds on $|\zeta|$ stemming from $\tau\to e\gamma$ and $\tau\to \mu\gamma$ are weaker, of the order of $10^{-5}$ ($10^5$). The stronger limits from Eq.~\eqref{eq:zetaEstimate2} apply unless there are fine-tuned cancellations among the contributions involving $\eta^{+}$ and $\eta^{\prime+}$ to the branching ratio for $\mu\to e\gamma$. 

In the numerical scan we take $\zeta$ as real and positive and vary it in the range of $10^{-3}$ to $10^3$. We obtain a wide range of values, i.e. values for $BR (\mu \to e\gamma)$ as small as $10^{-32}$ and as large as the current experimental bound are obtained, depending on the Yukawa couplings and the quartic coupling $\lambda_{H\Phi\Phi^\prime}$, see Fig.~\ref{fig:various2}. Similar ranges apply for radiative CLFV $\tau$-lepton decays, as shown in Fig.~\ref{fig:various4}. 

\mathversion{bold}
\subsection{$\ell_\alpha\to \ell_\beta\,\overline{\ell}_\gamma\,\ell_\gamma$}
\mathversion{normal}
\begin{figure}[hbt]\centering
	\begin{subfigure}{0.3\linewidth}\centering
		\FMDG{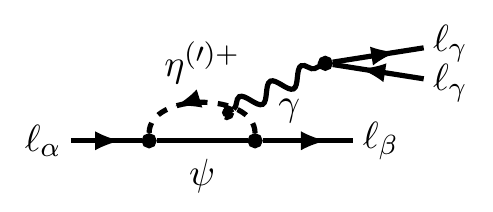}
		\caption{$\gamma$-penguin}
		\label{fig:l2lllgamma}
	\end{subfigure}
	\begin{subfigure}{0.3\linewidth}\centering
		\FMDG{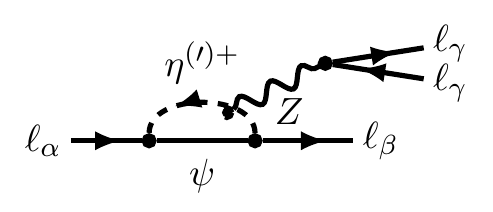}
		\caption{$Z$-penguin}
		\label{fig:l2lllZ}
	\end{subfigure}
	\begin{subfigure}{0.6\linewidth}\centering
		\FMDG{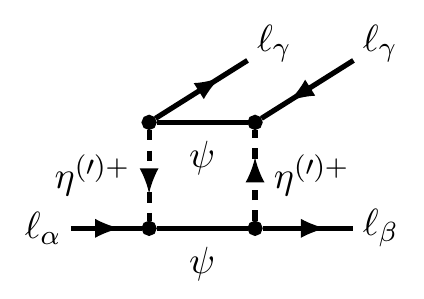}
		\FMDG{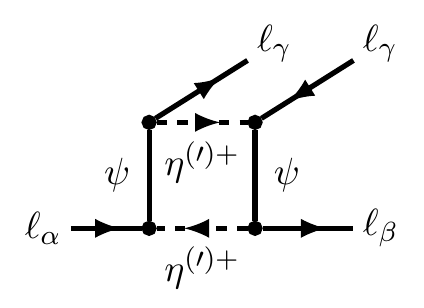}
		\caption{selection of box diagrams}
		\label{fig:l2lllbox}
	\end{subfigure}
		\caption{Diagrams contributing to $\ell_\alpha\to \ell_\beta \bar\ell_\gamma\ell_\gamma$.}
		\label{fig:l2lll}
	\end{figure}

	This type of process receives in general three independent contributions, shown in Fig.~\ref{fig:l2lll}, i.e. from $(i)$ $\gamma$-penguin, $(ii)$ $Z$-penguin and $(iii)$ box-type diagrams. 
We follow the notation of Ref.~\cite{Hisano:1995cp}. The $\gamma$-penguin amplitude for the transition  $\ell_\alpha(p)\to \ell_\beta(k_1)\,\overline{\ell}_\gamma(k_2)\,\ell_{\gamma}(k_3)$ is described by 
\begin{eqnarray}
	\mathcal{A}_{\gamma} &=& \overline{u}(k_1)\Big[ q^2\,\gamma^\rho\left( A_1^L\,P_L\,+\,A_1^R\,P_R \right)\,+\,
						 m_{\alpha}\,i\,\sigma^{\rho\sigma} \left(A_2^L\,P_L\,+\,A_2^R\,P_R\right) q_\sigma  \Big] u(p) \nonumber \\
						 && \times\,\frac{e^2}{q^2}\,\overline{u}(k_3)\,\gamma_\rho\,v(k_2)\,-\,\left( k_1 \leftrightarrow k_3\right),
\end{eqnarray}
where  $q=p-k_1$ and the form factors $A_{1,2}^{L/R}$ are reported in Eq.~\eqref{gammaform}.

The leading order contribution from the $Z$-penguin is proportional to the square of the charged lepton masses and thus negligible compared to the $\gamma$-penguin contribution. There are also box-type diagrams whose contributions are given by
\begin{equation} \label{eq:box}
	\mathcal{A}_{\text{BOX}} \;=\; e^2 B\, \overline{u}(k_1)\,\gamma^\alpha\,P_L\,u(p)\,
	\overline{u}(k_3)\,\gamma_\alpha\,P_L\,v(k_2)\,,
\end{equation}
where for the decay $\ell_\alpha^- \to \ell_\beta^- \ell_\gamma^- \ell_\gamma^+$ the form factor $B$ reads
\begin{equation} 
\begin{split}
	e^2 B\;=\; & \frac{1}{16\,\pi^2}\Bigg[\frac{y_{\Phi}^\alpha y_{\Phi}^{\beta*} y_{\Phi}^{\gamma*}y_{\Phi}^{\gamma}}{m_{\eta^+}^2}\,h_1\left(\frac{m_\psi^2}{m_{\eta^+}^2}\right) 
	+ \frac{y_{\Phi^\prime}^{\alpha} y_{\Phi^\prime}^{\beta *} y_{\Phi^\prime}^{\gamma*}y_{\Phi^\prime}^{\gamma}}{m_{\eta^{\prime+}}^2}\,h_1\left(\frac{m_\psi^2}{m_{\eta^{\prime+}}^2}\right)  \Bigg.\\
	& +\, \Bigg.\frac{\left(y_{\Phi}^{\beta*} y_{\Phi^\prime}^{\gamma*}+y_{\Phi}^{\gamma*} y_{\Phi^\prime}^{\beta *}\right)\left( y_{\Phi^\prime}^{\gamma} y_{\Phi}^{\alpha}+y_{\Phi}^{\gamma} y_{\Phi^\prime}^{\alpha}\right)}{m_\psi^2}\,
	h_2\left(\frac{m_\psi^2}{m_{\eta^+}^2}\,, \frac{m_\psi^2}{m_{\eta^{\prime+}}^2}\right) \Bigg]\,,
\end{split}	
\end{equation}
and
for $\ell_\alpha^- \to \ell_\gamma^- \ell_\gamma^- \ell_\beta^+$ it is given by
\begin{equation} 
\begin{split}
	e^2 B\;=\; & \frac{1}{16\,\pi^2}\Bigg[\frac{y_{\Phi}^\alpha y_{\Phi}^{\beta} (y_{\Phi}^{\gamma*})^2}{m_{\eta^+}^2}\,h_1\left(\frac{m_\psi^2}{m_{\eta^+}^2}\right) 
	+ \frac{y_{\Phi^\prime}^{\alpha} y_{\Phi^\prime}^{\beta} (y_{\Phi^\prime}^{\gamma*})^2}{m_{\eta^{\prime+}}^2}\,h_1\left(\frac{m_\psi^2}{m_{\eta^{\prime+}}^2}\right)  \Bigg.\\
	& +\, 2\Bigg.\frac{\left(y_{\Phi}^{\gamma*} y_{\Phi^\prime}^{\gamma *}\right)\left( y_{\Phi^\prime}^{\beta} y_{\Phi}^{\alpha}+y_{\Phi}^{\beta} y_{\Phi^\prime}^{\alpha}\right)}{m_\psi^2}\,
	h_2\left(\frac{m_\psi^2}{m_{\eta^+}^2}\,, \frac{m_\psi^2}{m_{\eta^{\prime+}}^2}\right) \Bigg]\,,
\end{split}	
\end{equation}
where all external momenta and masses have been neglected. The loop functions $h_1(x)$ and $h_2(x,y)$ are given in Eq.~\eqref{app:looph} in App.~\ref{app:loop}.

One can express the trilepton branching ratios in terms of form factors. Following Ref.~\cite{Abada:2014kba}, the branching ratio of $\ell_\alpha \to \ell_\beta \, \overline{\ell}_\beta\, \ell_\beta$ reads 
\begin{equation}\label{ell3ellBR}
\begin{split}
\text{BR}(\ell_\alpha \to \ell_\beta \, \overline{\ell}_\beta \, \ell_\beta)\;=\;&\frac{6\pi^2\alpha_{\text{em}}^2}{G_F^2} \Bigg[ 
\left|A_1^L\right|^2+\left|A_2^R\right|^2\left(\frac{16}{3}\ln\frac{m_\alpha}{m_\beta}-\frac{22}{3}\right) \\
&\;+\frac16\left|B\right|^2 -4\,\text{Re} \left(A_1^{L*} A_2^{R}-\frac16\left(A_1^L -2A_2^R \right)B^{*}\right)\Bigg]\\
&\,\times \text{BR}\left(\ell_\alpha\to\ell_\beta\,\nu_\alpha\,\overline{\nu_\beta}\right)\,.
\end{split}
\end{equation}
For $\ell_\alpha^- \to \ell_\beta^- \ell_\gamma^- \ell_\gamma^+$, with $\beta \ne \gamma$, the branching ratio is given by
\begin{equation}\label{ell3ellBRb}
\begin{split}
\text{BR}(\ell_\alpha \to \ell_\beta \, \overline{\ell}_\gamma \, \ell_\gamma)\;=\;&\frac{6\pi^2\alpha_{\text{em}}^2}{G_F^2} \Bigg[ 
\frac{2}{3}\left|A_1^L\right|^2+\left|A_2^R\right|^2\left(\frac{16}{3}\ln\frac{m_\alpha}{m_\gamma}-8\right) \\
&\;+\frac{1}{12}\left|B\right|^2 -\frac{8}{3}\,\text{Re} \left(A_1^L A_2^{R*}-\frac18\left(A_1^L -2A_2^R \right)B^{*}\right)\Bigg]\\
&\,\times \text{BR}\left(\ell_\alpha\to\ell_\beta\,\nu_\alpha\,\overline{\nu_\beta}\right)\,.
\end{split}
\end{equation}
For $\ell_\alpha^- \to \ell_\beta^+ \ell_\gamma^- \ell_\gamma^-$, as there are only box-type contributions, we get~\cite{Abada:2014kba}
\begin{equation}\label{ell3ellBRc}
\begin{split}
\text{BR}(\ell_\alpha \to \overline{\ell}_\beta \, \ell_\gamma \, \ell_\gamma)\;=\;&\frac{\pi^2\alpha_{\text{em}}^2}{G_F^2}
\left|B\right|^2 \times \text{BR}\left(\ell_\alpha\to\ell_\beta\,\nu_\alpha\,\overline{\nu_\beta}\right)\,.
\end{split}
\end{equation}
In the \emph{dipole dominance approximation}, we can express $\mu \to 3e$ in terms of $\mu \to e \gamma$ as
\begin{equation}\label{mue3emueg}
\begin{split}
\text{BR}(\mu \to 3e)\;\approx\;&\frac{\alpha_{\text{em}}}{8\pi}\,\left(\frac{16}{3}\ln\frac{m_\mu}{m_e}-\frac{22}{3}\right)\,\times \text{BR}(\mu\to e\,\gamma)\,\\
\;\approx\;& 0.006 \,\times \text{BR}(\mu\to e\,\gamma)\,.
\end{split}
\end{equation}
We have checked in the numerical analysis that the above estimate is fulfilled to great precision, confirming that the box-type contributions are not relevant. As the latter involve two extra Yukawa couplings which are smaller than one, the box-type contributions are suppressed with respect to the dipole $A_2^{L,R}$ and the monopole $A_{1}^{L,R}$ contributions given in Eq.~(\ref{gammaform}). 

The current experimental limit is $\text{BR}(\mu \rightarrow 3e)<1.0 \times 10^{-12}$~\cite{Bellgardt:1987du}, with an expected future sensitivity of $\sim\mathcal{O}(10^{-16})$~\cite{Blondel:2013ia}. The other trilepton decays involve $\tau$ leptons, and the upper limits on their branching ratios are $\mathcal{O}(10^{-8})$~\cite{Hayasaka:2010np}, with future expected sensitivities of $\sim \mathcal{O}(10^{-9})$~\cite{Aushev:2010bq}.

\mathversion{bold}
\subsection{$\mu-e$ conversion in nuclei}
\mathversion{normal}
\begin{figure}[hbt]\centering
	\begin{subfigure}{0.3\linewidth}\centering
		\FMDG{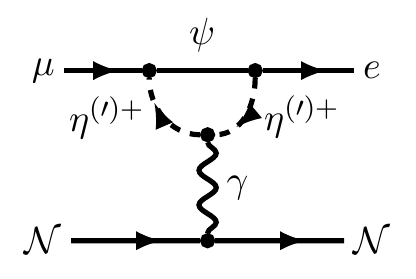}
		\caption{$\gamma$-penguin}
		\label{fig:mu2egamma}
	\end{subfigure}
	\begin{subfigure}{0.3\linewidth}\centering
		\FMDG{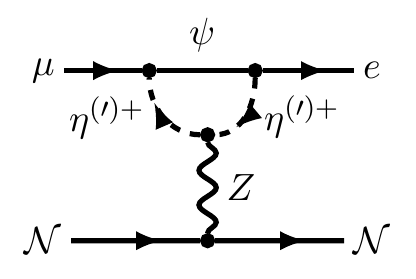}
		\caption{$Z$-penguin}
		\label{fig:mu2eZ}
	\end{subfigure}
		\caption{Diagrams contributing to $\mu-e$ conversion in nuclei.}
		\label{fig:mu2e}
	\end{figure}

The conversion of a muon to an electron in a nucleus also imposes stringent constraints on the parameter space of the model. This process is dominated by coherent conversions in which initial and final states of the nucleus $\mathcal{N}$ are the same. In this case the matrix elements of the axial-vector $\langle \mathcal{N}|\overline{q}\,\gamma_\alpha\,\gamma_5\,q| \mathcal{N}\rangle$, pseudoscalar $\langle \mathcal{N}|\overline{q}\,\gamma_5\,q| \mathcal{N}\rangle$, and tensor quark currents $\langle \mathcal{N}|\overline{q}\,\sigma_{\alpha\beta}\,q| \mathcal{N}\rangle$ vanish identically \cite{Kitano:2002mt}. 
Similar to the leptonic decay of $\tau$ and $\mu$ leptons, the $Z$-penguin contribution to $\mu-e$ conversion in nuclei is proportional to the square of the charged lepton masses and thus negligible compared to the $\gamma$-penguin contribution. See Fig.~\ref{fig:mu2e} for the relevant Feynman diagrams.
Moreover, the contribution involving the SM Higgs boson is suppressed by the small Yukawa couplings of the first generation of quarks. Thus $\mu-e$ conversion is dominated by photon exchange and the relevant terms in the effective Lagrangian contributing to $\mu-e$ conversion can be parameterised as~\cite{Kitano:2002mt}
\begin{eqnarray}
    {\cal L}_{\mu-e\,\, \mathrm{conv}} &=&
- \frac e2
    \left(
    m_\mu A_2^L\, \overline{\ell_e}\, \sigma^{\mu \nu} P_L\, \ell_\mu F_{\mu \nu}
    + m_\mu A_2^R\, \overline{\ell_e}\, \sigma^{\mu \nu} P_R\, \ell_\mu F_{\mu \nu}
    + {\rm H.c.}
    \right) \nonumber \\
    &&
    - 
 \sum_{q = u,d,s}
    \left[ {\rule[-3mm]{0mm}{10mm}\ } \right.
   \left(
    g_{LV(q)}^\gamma\, \overline{\ell_e} \gamma^{\alpha} P_L \ell_\mu
   \right) \overline{q} \gamma_{\alpha} q
   + {\rm H.c.}
    \left. {\rule[-3mm]{0mm}{10mm}\ } \right].\;\;\;\;\;\;\;
    \label{eq:mue}
\end{eqnarray}
The long-range interaction mediating the process is given by the electromagnetic dipole transitions, whose  form factors $A_2^{L/R}$ are introduced in
Eq.~(\ref{gammaform}), taking into account the appropriate flavour indices of the Yukawa couplings. 
The short-range interaction through the $\gamma$-penguin diagrams generate the vector current operator with 
\begin{equation}
\begin{split}
	g_{LV(q)}^\gamma \; = \; & e^2\, Q_q\, A_1^L\,,\\
	\end{split}
\end{equation}
where $Q_q$ is the electric charge of the quark $q$ in units of $e$ and $A_1^L$ is the electromagnetic form factor given in Eq.~(\ref{gammaform}) for the flavour indices $\mu$ and $e$. A right-handed leptonic vector current is not induced at one-loop level because all new particles exclusively couple to the left-handed lepton doublets $L_{\alpha}$.
Accordingly, the $\mu-e$ conversion rate is given in terms of the overlap integrals $D$ and $V^{(p,n)}$ as
\begin{align}\label{eq:mu2eOmega}
	\omega_{\rm conv} \;=\;&
4\,\left| \frac e8 A_2^{R} D+ \tilde{g}_{LV}^{(p)} V^{(p)} + \tilde{g}_{LV}^{(n)} V^{(n)}\right|^2  \ ,
\end{align}
where the effective vector couplings $\tilde{g}_{LV}^{(p,n)}$ for the proton and the neutron are 
\begin{align}
\tilde g_{LV}^{(p)}&\approx 2\, g_{LV(u)}^\gamma + g_{LV(d)}^\gamma = e^2 A_1^L\ ,  & 
\tilde g_{LV}^{(n)}&\approx g_{LV(u)}^\gamma + 2\, g_{LV(d)}^\gamma =0\,.
\end{align}
Notice that the neutron contribution is in our case approximately zero, as we neglected the $Z$-penguin contribution.

We can express $\omega_{\rm conv} $ in terms of $\text{BR}(\mu\to e\,\gamma)$ using that 
\begin{equation}
A_1^{L} \approx \frac23 \,r_{g/f}\,A_2^{R} \, ,
\end{equation}
 where $r_{g/f}$ parametrises the difference due to the different loop functions with $1\lesssim r_{g/f} \lesssim 1.5$ for $x \lesssim1$ ($m_\psi$ is always smaller than
  $m_{\eta^{(\prime)+}}$, since $\psi$ is the DM candidate), see Eqs.~\eqref{gammaform} and~\eqref{f-g-loop}. In this case we can derive the allowed ranges for 
\begin{eqnarray}
\label{eq:mu2emueg2}
	\text{CR}_{\rm conv} &\equiv&  \frac{\omega_{\rm conv}}{\omega_{\rm capt}} \nonumber \\
	& \approx&\; \frac{G_F^2}{192\,\pi^2\,\omega_{\rm capt}}\,\left| D+ \frac{16}{3}\, r_{g/f}\, e\, V^{(p)}\right|^2 \,\times \text{BR}(\mu\to e\,\gamma)  \nonumber  \\
	& \approx& \;   [0.0077,0.011]\,([0.010,0.015])\,\{[0.013,0.019]\}\,\times \text{BR}(\mu\to e\,\gamma)\,,
\end{eqnarray}
for\,\, Al\, (Au)\,\{Ti\}. Here we used the numerical values of the overlap integrals $D$ and $V^{(p,n)}$ and the total capture rate $\omega_{\rm capt}$ reported in Tab.~\ref{tab:overlap} in App.~\ref{app:loop}. We have checked that the numerical results for the $\mu-e$ conversion ratio (CR) are in very good agreement with the estimates in Eq.~(\ref{eq:mu2emueg2}), see Figs.~\ref{fig:various2} and \ref{fig:various2L}.

The currently best experimental limits are set by the SINDRUM II experiment~\cite{Bertl:2006up}:  $\mathrm{CR}_{\rm conv} (\rm Au)< 7\cdot 10^{-13}$ and $\mathrm{CR}_{\rm conv} (\rm Ti) < 4.3 \cdot 10^{-12}$ at 90~\% CL. Future experiments are expected to improve the sensitivity by several orders of magnitude: the COMET experiment at J-PARC~\cite{Cui:2009zz,Kuno:2013mha} may improve down to $\mathcal{O}(10^{-17})$, Mu2e experiment at Fermilab using Al~\cite{Carey:2008zz, Kutschke:2011ux, Donghia:2016lzt} $6 \cdot 10^{-17}$ and at Project X (Al or Ti) $\mathcal{O}(10^{-19})$, and PRISM/PRIME~\cite{Barlow:2011zza, Witte:2012zza} may reach $\mathcal{O}(10^{-18})$. 

\subsection{Lepton dipole moments}
\label{sec:lepton_moments}

Contributions to the electric dipole moments of charged leptons arise in the model
only at the two-loop level.\footnote{See
Ref.~\cite{Abada:2018zra} for the calculation of the
electric dipole moments in the minimal ScM which has two
new Majorana fermions.} However, non-zero contributions to leptonic magnetic dipole
moments are generated, similarly to contributions to radiative CLFV decays, at one-loop level. The relevant Feynman diagram is shown in Fig.~\ref{fig:llgamma} for $\alpha=\beta$. They receive two independent contributions from
the charged scalars $\eta^\pm$ and $\eta^{\prime\pm}$ running in the loop. They are given by (see also Refs.~\cite{Raidal:2008jk,Kopp:2009et})
\begin{equation}
	\Delta a_\ell \equiv \frac{g_\ell-2}{2} \;=\; 2\,m^2_\ell\, {\rm Re} [A_2^R]_\ell\,,
	\label{gminus2}
\end{equation}
where $[A_2^R]_\ell$ is the diagonal part ($\alpha=\beta \equiv \ell$) of the coefficient $A_2^R$ given in Eq.~\eqref{gammaform}.

In the case of the anomalous muon magnetic dipole moment $\Delta a_\mu$,  the
discrepancy between the measured value and the one predicted within the SM is larger than zero (see Ref.~\cite{Lindner:2016bgg} for a recent review), and therefore
the model cannot explain it, as it gives a negative
contribution, see Eq.~\eqref{gammaform} together with the
loop function $f(x)>0$. In any case, the predictions for
$|\Delta a_\mu|$ are always small, $|\Delta
a_\mu|\lesssim 10^{-12}$, as long as limits imposed by the non-observation of CLFV decays are fulfilled.
The magnetic dipole moments of the electron and the $\tau$ lepton are
subject to very weak limits. 

\subsection{Dark matter relic abundance}
\label{sec:dm}

We study the case where the DM candidate is the Dirac
fermion $\psi$. There are several possible production
channels in the early Universe. In particular the value of the Yukawa couplings of the fermionic singlet control the different regimes, see also Ref.~\cite{Khoze:2017ixx}: 
	(i) 
	If the Yukawa couplings are very small ($\ll 1$), direct annihilations of the scalars dominate the relic abundance.
	(ii)
	For intermediate values, fermion--scalar coannihilations dominate and set the relic abundance. 
	(iii)
	For larger values, annihilations of the fermion would in principle dominate. 

\begin{figure}[t!]\centering
	\begin{subfigure}{0.45\linewidth}\centering
		\FMDG{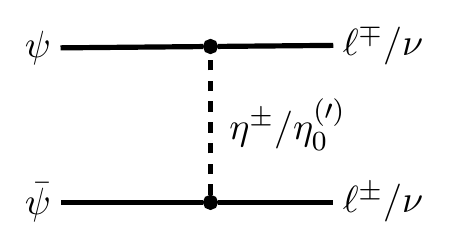}
	\caption{Annihilations of dark matter into leptons.}\label{fig:DMann}
\end{subfigure}
\vspace{3ex}

\begin{subfigure}{\linewidth}\centering
\FMDG{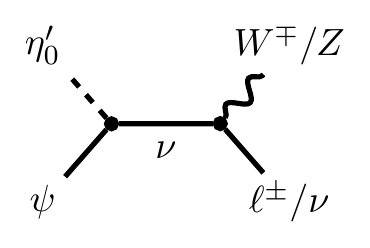}%
\FMDG{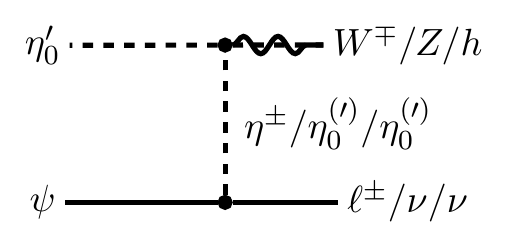}%

\FMDG{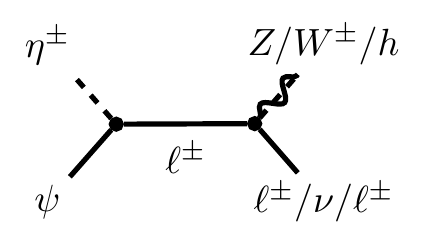}%
\FMDG{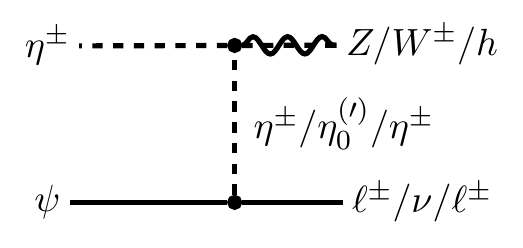}%
\caption{Coannihilations of dark matter with a new scalar.}\label{fig:DMcoan}
\end{subfigure}

\vspace{3ex}
	\begin{subfigure}{\linewidth}\centering
\FMDG{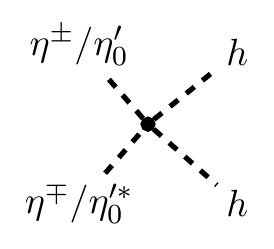}%
\FMDG{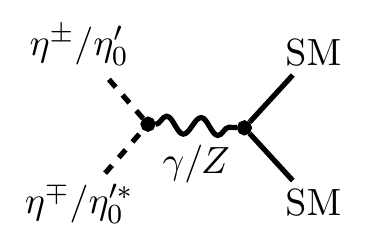}%
\FMDG{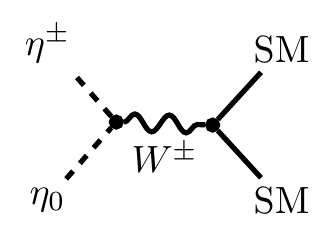}%

\FMDG{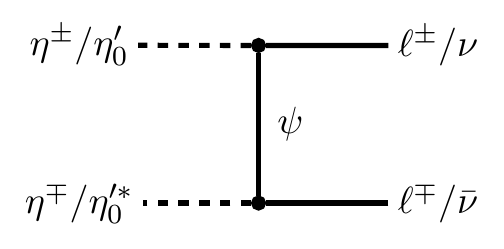}%
\FMDG{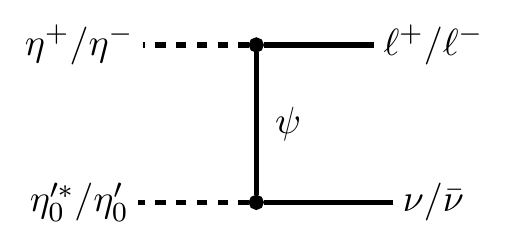}%
\caption{Annihilations of the scalar coannihilation partners.}\label{fig:scalarann}
\end{subfigure}

\caption{Illustration of the relevant annihilation (top panel) and coannihilation (middle panel) channels of the dark matter $\psi$ involving the scalars $\eta^\pm,\,\eta_0^\prime$. We also show diagrams of the annihilation channels of the coannihilating partners $\eta^\pm,\,\eta_0^\prime$ (bottom panel). More diagrams exist involving the other new scalars.}
\label{fig:coannihilation}
\end{figure}

In the next subsection we
argue that DM annihilations to SM leptons $\psi\bar\psi\to
\ell\bar\ell, \nu\bar\nu$ (regime 3) are too small due to constraints from CLFV
processes. We show in Fig.~\ref{fig:DMann} the $t$-channel annihilations mediated by $\eta^\pm,\,\eta_0^\prime$. The correct relic abundance is set by coannihilations (regimes 1 and 2), see the diagrams in Figs.~\ref{fig:DMcoan} and~\ref{fig:scalarann}. The latter are possible if the relative mass splitting between the fermion and the scalars is smaller or equal than $5\%$.

\subsubsection{Dark matter annihilations} \label{sec:annihilations}

The dominant DM annihilation channels are into a pair of charged leptons or a pair of neutrinos, shown in Fig.~\ref{fig:DMann}. In the non-relativistic limit the $s$-wave annihilation cross sections
to neutrinos and charged leptons
are given by
\begin{align} \label{eq:ann}
	\braket{v\sigma(\psi\bar\psi\to \ell_\alpha^-\ell_\beta^+) }& \simeq
	\frac{1}{32\pi m_\psi^2}
	\left|
	y_{\Phi}^\alpha y_{\Phi}^{\beta*}\frac{m_\psi^2}{m_{\eta^{+}}^2+m_\psi^2}
	-y_{\Phi^\prime}^{\alpha} y_{\Phi^\prime}^{\beta*}
	\frac{m_\psi^2}{m_{\eta^{\prime+}}^2+m_\psi^2}
	\right|^2\,,\\\label{eq:annNu}
	\braket{v\sigma(\psi\bar\psi\to \nu_\alpha \nu_\beta) }& \simeq
	\frac{1}{64\pi m_\psi^2(1+\delta_{\alpha\beta})}
       \left|
      	y_{\Phi}^\alpha y_{\Phi}^{\beta*}
	\left(
		\frac{m_\psi^2 s_\theta^2}{m_{\eta_0}^2+m_\psi^2}
	+	\frac{m_\psi^2 c_\theta^2}{m_{\eta_0^\prime}^2+m_\psi^2}
	\right)
	\right.\\\nonumber
	&\hspace{3cm}\left.
	-y_{\Phi^\prime}^{\alpha} y_{\Phi^\prime}^{\beta*}
	\left(
		\frac{m_\psi^2 c_\theta^2}{m_{\eta_0}^2+m_\psi^2}
	+	\frac{m_\psi^2 s_\theta^2}{m_{\eta_0^\prime}^2+m_\psi^2}
	\right)
       \right|^2\,.
\end{align}
The minus sign between the different contributions
originates from the presence of $t$- and $u$-channel diagrams
mediated by $\eta$ and $\eta^\prime$, respectively. As neutrinos are Majorana particles, the corresponding cross section is smaller by a factor of two.
For identical neutrinos in the final state
$\delta_{\alpha \beta}=1$ which leads to another factor of
$1/2$.

Next we conservatively estimate these cross sections in the
limit of large and small $|\zeta|$ using the obtained
limits on $|\zeta|$ in Eq.~\eqref{eq:zetaEstimate2}. In
the limit of equal masses, denoted by $m$, the cross
sections only depend on whether primed or unprimed Yukawa
couplings are present. Larger scalar masses only further
suppress the annihilation cross section. Thus we obtain
the following conservative limit
\begin{align} 
	\sum_{\alpha,\beta}	\braket{v\sigma(\psi\bar\psi\to \ell_\alpha^-\ell_\beta^+, \nu_\alpha \nu_\beta) }& \lesssim
	\sum_{\alpha,\beta}	\frac{	\left|
		y_{\Phi^{(\prime)}}^\alpha y_{\Phi^{(\prime)}}^{\beta*}
	\right|^2}{ 256\pi\,m^2} \frac{3+2\delta_{\alpha\beta}}{1+\delta_{\alpha\beta}}\,.
\end{align}
This allows us to use the limit on $|\zeta|$ from Eq.~\eqref{eq:zetaEstimate2}. A comparison to the typical
freeze-out annihilation cross section, $\braket{v \sigma}_{\rm th} \simeq 2.2\times
10^{-26}\,{\rm cm^3/s}$,
yields that the DM annihilation cross section is always too small in order for the Dirac fermion $\psi$ to account for the observed relic density. 
Using the experimental upper limit on the branching ratio for $\mu\to e\gamma$ on the parameter $|\zeta|$ in Eq.~(\ref{eq:zetaEstimate2}), we obtain
\begin{align}
	\sum_{\alpha,\beta}	\braket{v\sigma(\psi\bar\psi\to \ell_\alpha^-\ell_\beta^+, \nu_\alpha \nu_\beta) }	&\lesssim 2 \times 10^{-4}
	\left(\frac{\braket{v\sigma}_{\rm th}}{2.2\times 10^{-26}\, \mathrm{cm}^3\,\mathrm{s}^{-1}}\right) \left(\frac{m}{100\,\mathrm{GeV}}\right)^4
\end{align}
for any neutrino mass ordering, which is valid unless cancellations occur. 

An interesting way to break the correlation of ${\rm BR}(\ell_\alpha\to
\ell_\beta\,\gamma)$ and DM annihilation cross section is to have the DM relic abundance set
by $\braket{v\sigma(\psi\bar\psi\to\nu\bar\nu)}$. This can be achieved if the
charged scalars $\eta^\pm$ and $\eta^{\prime\pm}$ are much heavier than at least one neutral scalar
($\eta^\prime_0$ in this model), leading therefore to suppressed contributions to
all radiative CLFV processes, and also to $\braket{v\sigma
(\psi\bar\psi\to\ell^-\ell^+)}$. In this scenario the mass of the Dirac fermion DM $\psi$ is typically in the MeV range to obtain the correct DM relic density, with slightly
heavier neutral scalar $\eta_0^{\prime}$. However, gauge
invariance relates the interactions of neutrinos and
charged leptons, and therefore this scenario requires some
tuning of the parameters in order to circumvent experimental
constraints from $Z$-boson decays, the $T$ parameter and CLFV processes. We thus do not consider it any further. Examples of
similar scenarios have been studied in
Refs.~\cite{Boehm:2006mi,Farzan:2009ji, Huang:2014bva,
Arhrib:2015dez}.

\subsubsection{Dark matter coannihilations}
As DM annihilation into charged leptons and neutrinos is strongly constrained by experimental limits on CLFV observables, coannihilation processes may become important. The explanation of the correct DM relic abundance requires a small mass splitting between the DM candidate $\psi$ and the scalars $\eta_0^\prime$ and $\eta^{(\prime)+}$. While this is perfectly plausible in the model, in which the particles naturally are at the TeV scale, in its current version there is no symmetry or dynamic reason to generate similar scalar and fermion masses. Another option is a variant of the model with an fermionic electroweak triplet instead of a singlet, discussed in Sec.~\ref{sec:triplet_model}. This allows to have the relic abundance set by annihilations, without the need of coannihilations. The relative contribution of $(i)$ annihilations of the DM particle with the coannihilation partner into a lepton and a gauge boson or Higgs boson (see Fig.~\ref{fig:DMcoan}), and $(ii)$ annihilations of the coannihilation partner(s) via gauge interactions ($\gamma/Z/W$) into SM particles, direct annihilations to Higgs bosons or DM-mediated $t$-channel annihilations into leptons (shown in Fig.~\ref{fig:scalarann}), depends on the size of the Yukawa couplings and the mass splitting. 
The coannihilation channels dominate the abundance, because the corresponding
cross sections only depend on the square of one of the Yukawa couplings
$y_{\Phi}^\alpha,\,y_{\Phi^\prime}^\alpha$ compared to the annihilation cross section, see
Eq.~\eqref{eq:ann} and \eqref{eq:annNu}, which involves four Yukawa couplings. In our numerical scan we use micrOMEGAs 4.3.5~\cite{Belanger:2013oya} to calculate the DM relic abundance and thus take all relevant (co)annihilation channels into account. See the seminal work~\cite{Griest:1990kh} by
K.~Griest and D.~Seckel for an analytic discussion of coannihilation.

\subsection{Dark matter direct detection}
\label{sec:dirdet}

One of the interesting features of the GScM is that the fermionic DM candidate $\psi$ is a Dirac fermion rather than a Majorana fermion as in the original ScM. A direct detection (DD) signal can not be generated at tree level, but there are sizeable long-range contributions at the one-loop level via photon exchange.\footnote{In the case of Majorana DM these long-range interactions can occur among different fermionic states and give rise to inelastic scattering if the mass splitting among them is sufficiently small, see Ref.~\cite{Schmidt:2012yg}.} It can be parameterised by the magnetic (and electric) dipole interactions, namely
\begin{align}
	\mathcal{L}_\mathrm{DD} & = \mu_\psi \frac{e}{8\pi^2} \bar \psi \sigma_{\mu\nu} \psi F^{\mu\nu} + d_\psi\frac{e}{8\pi^2} \bar \psi \sigma_{\mu\nu}i\gamma_5 \psi F^{\mu\nu} \,.
\end{align}
In this model the electric dipole moment $d_\psi$ vanishes at one-loop level, because $\psi$ only couples to left-handed lepton doublets and not to right-handed charged leptons simultaneously~\cite{Herrero-Garcia:2018koq}. 
The magnetic dipole moment $\mu_\psi$ is given by
\begin{align} \label{eq:magDD}
	 \mu_\psi&=
	 \frac{-1}{4m_\psi} \sum_\alpha \Big(
		 \left|y_{\Phi}^\alpha\right|^2 f_{\rm DD}(m_\psi, m_{\eta^+}, m_{\ell_\alpha})
		 - \left|y_{\Phi^\prime}^\alpha\right|^2 f_{\rm DD}(m_\psi, m_{\eta^{\prime+}}, m_{\ell_\alpha})
		 	 \Big)
\;.
\end{align}
The loop function $f_\mathrm{DD} (x,y,z)$ is defined in Eq.~(\ref{eq:fDD}) in App.~\ref{app:loop}.
We checked our result against the well-known expressions for the magnetic dipole moment for a Yukawa interaction found in Ref.~\cite{Lavoura:2003xp}. Similar results are given in Refs.~\cite{Chang:2010en, Agrawal:2011ze, Dissauer:2012xa, Schmidt:2012yg, Kopp:2014tsa, Ibarra:2016dlb, Herrero-Garcia:2018koq}. 
In Fig.~\ref{fig:various2L} in Sec.~\ref{sec:num} we show how results from the latest Xenon experiments XENON1T~\cite{Aprile:2017iyp}, PandaX-II~\cite{Tan:2016zwf}, and LUX~\cite{Akerib:2015rjg} constrain the parameter space of the model using LikeDM~\cite{Liu:2017kmx}.

\subsection{Electroweak precision tests}
\label{sec:ewpt}
The dominant contribution from new physics to electroweak radiative processes is generally expected to affect the gauge boson self-energies which are parameterised by the oblique parameters $S$, $T$, and $U$~\cite{Peskin:1990zt,Peskin:1991sw}.
The limits on the oblique
parameters are obtained from a global fit to electroweak precision data. The
Gfitter collaboration finds the values: $S=0.05\pm 0.11$, $T= 0.09\pm 0.13$ and
$U= 0.01\pm 0.11$, with correlation coefficients $\rho_{\rm ST}=0.90$,
$\rho_{\rm SU}=-0.59$ and $\rho_{\rm TU}=-0.83$~\cite{Baak:2014ora}. The
strongest constraints on the parameter space of the model are set by the $T$
parameter, which is sensitive to the mass splitting between the neutral and
charged scalar components of the two inert doublets $\Phi$ and $\Phi^\prime$. We use the expressions for
the oblique parameters found in Refs.~\cite{Grimus:2007if,
Haber:2010bw,Herrero-Garcia:2017xdu}. Details are reported in
App.~\ref{app:ewpt}.

\subsection{Production and decay of the new scalars at colliders}

 Searches for neutral and charged scalars at colliders set constraints
 on the scalar mass spectrum of the model. In fact, from the precise
 measurement of the $W$ and $Z$ boson decay widths at LEP-II, the following
 kinematical bounds can be derived:
 $m_{\eta_0^{(\prime)}}+m_{\eta^{(\prime)+}}>m_W$ and
 $2m_{\eta^{(\prime)+}},2m_{\eta_0^{(\prime)}},
 m_{\eta_0}+m_{\eta_0^\prime}>m_Z$ for $m_{W (Z)}$ being the $W$ ($Z$) boson mass.

At the LHC the production of these states proceeds mainly
via neutral and charged current Drell-Yan processes. Other
production channels are via an off-shell $Z/W$ boson.
A sub-leading
contribution is given by Higgs mediated gluon
fusion, provided the relevant couplings in the scalar
potential in Eq.~(\ref{eq:potential}) are
sizeable~\cite{Hessler:2014ssa,Hessler:2016kwm}.

In the case one of the charged scalars is the
next-to-lightest particle in the dark sector, the expected
signature at the LHC consists in the pair production of
$\eta^{(\prime)\pm}$ followed by the prompt decay
$\eta^{(\prime)\pm}\to\psi\ell_{\alpha}^\pm$
($\alpha=e,\mu,\tau$).\footnote{The signature of this
process at the LHC is similar to the one predicted in simplified supersymmetric models with light sleptons and weakly decaying
charginos, which are searched for by the ATLAS
\cite{Aad:2014vma} and CMS \cite{Khachatryan:2014qwa}
collaborations.} The DM particle $\psi$ escapes the detector and is
revealed as missing transverse energy. The decay branching
ratios of $\eta^{(\prime)\pm}$ into the different leptons
only depend on the neutrino Yukawa couplings, namely
\begin{equation}
	\text{BR}(\eta^{(\prime)\pm}\to\psi\,\ell_\alpha^\pm) \;= \; \frac{\Gamma(\eta^{(\prime)\pm} \rightarrow \psi \ell_\alpha^\pm)}{\sum_\beta \Gamma(\eta^{(\prime)\pm} \rightarrow \psi\, \ell_\beta^\pm)}=
	\frac{|y_{\Phi^{(\prime)}}^\alpha|^2}{\sum_\beta |y_{\Phi^{(\prime)}}^{\beta}|^2}\,. \label{BR-MET}
\end{equation}
  Using the estimates for 
 $y^\alpha_\Phi$ and $y^\alpha_{\Phi'}$ reported in
 Sec.~\ref{sec:numasses}
  we expect for neutrino masses with NO that both charged scalars $\eta^{\pm}$ and $\eta^{\prime\pm}$ have very similar branching ratios with the one to $e^\pm$ being 
  suppressed by two powers of  the reactor mixing angle $\theta_{13}$ with respect to those to $\mu^\pm$ and $\tau^\pm$. Since $\theta_{23}\approx \pi/4$, the branching 
  ratios to the two flavours $\mu^\pm$ and $\tau^\pm$ are expected to be very similar for both NO and IO, see Eqs.~\eqref{eqs:estNO} and~\eqref{eqs:estIO}, respectively. Moreover, for neutrino masses with IO the branching ratios of both charged scalars $\eta^{\pm}$ and $\eta^{\prime\pm}$
  to $\mu^\pm$ and $\tau^\pm$ are expected to be very similar, whereas the ones to $e^\pm$ are expected to be different, but of similar size. 
  In particular, we note that for IO ${\rm BR} (\eta^{\pm}\to\psi\,e^\pm) \approx 2 \, {\rm BR} (\eta^{\prime \pm}\to\psi\,\mu^\pm (\tau^\pm))$ and 
  ${\rm BR} (\eta^{\prime \pm}\to\psi\,e^\pm) \approx 2 \, {\rm BR} (\eta^{\pm}\to\psi\,\mu^\pm (\tau^\pm))$. 
  A measurement of at least one of
  the branching ratios may allow to extract information on
  the neutrino mass ordering and the Majorana phase $\gamma$, while there is only a weak dependence on the Dirac phase $\delta$.
 
 In the coannihilation region corresponding to
 $m_{\eta_0^{(\prime)}}>m_{\eta^{(\prime)+}}\gtrsim
 m_\psi$, the decay width of the charged scalar is
 kinematically suppressed.  As we discuss in the numerical analysis in Sec.~\ref{sec:num}, in this case the lightest charged scalar may be
 long-lived,  leaving ionising tracks in the
 detector~\cite{Khoze:2017ixx,Hessler:2016kwm}.

 \mathversion{bold}
 \subsection{Decays of the Higgs and $Z$ bosons}
 \mathversion{normal}
 \label{sec:higgs_gamma}

If the scalars are sufficiently light, the Higgs and $Z$ bosons can decay into them at tree level which is strongly experimentally constrained. In specific
	 cases some of the limits on the neutral scalar
	 masses from $Z$ decays can be evaded by tuning
	 the mixing angle $\theta$, see Eq.~\eqref{eq:theta}. For instance, the
	 $Z$-boson decay rate into the lightest neutral
	 scalar, $\Gamma(Z\to \eta_0^{\prime}
	 \eta^{\prime*}_0)$, is proportional to $\cos^2
	 (2\theta)$ and therefore vanishes in the case of
	 maximal mixing, $\theta=
	 \pi/4$.\footnote{This is also relevant for direct
	 detection, if the DM particle is the scalar $\eta_0^\prime$.}  In this case the mass of the lightest neutral scalar, $m_{\eta_0^\prime}$, can be
 smaller than $m_Z/2$. For the Higgs boson the decay to the lightest neutral scalar can be suppressed for sufficiently small quartic couplings and/or a suitable choice of the mixing angle $\theta$. 

There can also be Higgs and $Z$-boson decays at one-loop level. The charged scalars couple to the Higgs boson and thus modify the decay of the Higgs boson to two photons. The relative change of the Higgs partial decay width to two photons compared to the SM prediction can be parameterised as~\cite{Ellis:1975ap,Shifman:1979eb,Carena:2012xa}
\begin{equation}
	R_{\gamma \gamma} = \frac{{\rm BR}(h \rightarrow \gamma \gamma)_{\rm GScM}}{{\rm BR}(h \rightarrow \gamma \gamma)_{\rm SM}} \simeq \left |1+ \frac{\frac{\lambda_{H\Phi} \, v_H^2}{2\, m_{\eta^+}^2}\,A_0\left( \frac{4m_{\eta^+}^2}{m_h^2}\right)+\frac{\lambda_{H\Phi^\prime}v_H^2}{2\,  m^2_{\eta^{\prime+}}}\,A_0\left(\frac{4m_{\eta^{\prime+}}}{m_h^2}\right)}{A_1\left(\frac{4m_{W}^2}{m_h^2}\right)+\frac{4}{3} A_{1/2}\left(\frac{4m_t^2}{m_h^2}\right)} \right |^2\,, \label{eq:ggratio}
\end{equation}
where $\lambda_{H\Phi}, \lambda_{H\Phi^\prime}$ are the couplings of the
charged scalars $\eta^+,\,\eta^{\prime+}$ to the Higgs boson, see
Eq.~\eqref{eq:potential}.
$A_i(x)$ are loop functions for scalars, fermions and gauge bosons with $(i=0,1/2,1)$ respectively, given
in Eq.~\eqref{loopgamma} in App.~\ref{app:loop}. $m_t$ is the top quark mass.
The ATLAS and CMS experiments have measured the partial width of
$h\rightarrow \gamma\gamma$ and reported it in terms of
the signal strength $\mu_{\gamma\gamma}\equiv
R_{\gamma\gamma} \sigma(pp\to h)/\sigma(pp\to h)_{\rm
SM}$. As the new particles are not coloured and thus the
Higgs production cross section is unchanged the signal
strength is simply given by $\mu_{\gamma\gamma}=R_{\gamma\gamma}$ in this
model. ATLAS observes
$\mu_{\gamma\gamma}=1.14^{+0.27}_{-0.25}$~\cite{Aad:2014eha},
and CMS
$\mu_{\gamma\gamma}=1.11^{+0.25}_{-0.23}$~\cite{Khachatryan:2014ira}
which can be interpreted as a constraint on the 
charged scalars. The combined measurement is $\mu_{\gamma\gamma}=1.14^{+0.19}_{-0.18}$~\cite{Khachatryan:2016vau}.
If the charged scalar masses are light enough,
deviations in the $h\rightarrow \gamma\gamma$ channel are
generically expected, but their size crucially depends on
parameters in the scalar potential, see Eq.~\eqref{eq:potential}
As observed in the numerical scan, this constraint can be fulfilled in the GScM.

In principle, there can be new invisible decay channels of
the Higgs and $Z$ bosons to the DM particle $\psi$ as well as of the Higgs to neutrinos.
Generically, the new scalars are constrained to be heavier than $\sim 100$ GeV due to a combination of collider searches,
the limits from the invisible decay width of the $Z$ boson, and EWPT.
Consequently, also the DM particle $\psi$ cannot be light in the case of
coannihilations, see Sec.~\ref{sec:dm}, and thus the Higgs and the
$Z$ bosons cannot decay into $\psi$, which would otherwise
occur at one-loop level, see
Ref.~\cite{Herrero-Garcia:2018koq}.

Other possible processes are CLFV (and lepton flavour universality violating) Higgs and $Z$-boson decays, like
$h\rightarrow \tau\mu$ and $Z\rightarrow \tau \mu$.
These, however, are very suppressed by a loop factor
and due to experimental constraints arising from other CLFV processes (like $\tau \rightarrow \mu
\gamma$). They are therefore well beyond the expected
sensitivity of future experiments~\cite{Herrero-Garcia:2016uab}.


\section{Numerical analysis}
\label{sec:num}

The Yukawa couplings of the model are determined by neutrino oscillation data, the Majorana phase $\gamma$ and the parameter $\zeta$ (which can be taken positive without loss of generality), as explained in App.~\ref{sec:yukpar}. 
We scan over the 3$\sigma$ range of the neutrino oscillation parameters using the results from NuFIT 3.1 (November 2017)~\cite{Esteban:2016qun,nufitweb}, reproduced for convenience in Tab.~\ref{tab:neutrinos}, as well as over the rest of the parameters of the model and $\zeta$ as outlined in Tab.~\ref{tab:parameters}. The points indicate the currently allowed parameter space. The varying density of points is mostly due to the efficiency of the scan and does not have a meaningful statistical interpretation.
\begin{table} [t!]
\begin{center}
  \begin{tabular}{ | c | c | c | }
    \hline
    Observable & NO & IO\\ \hline \hline
    $\sin^2\theta_{12}$ & $[0.272, 0.347]$ & $[0.272, 0.347]$\\ \hline
    $\sin^2\theta_{23}$ & $[0.401, 0.628]$ & $[0.419, 0.628]$\\ \hline
    $\sin^2\theta_{13}$ & $[0.01971, 0.02434]$ & $[0.01990, 0.02437]$\\ \hline
    $\Delta m^2_{21}/10^{-5}\,[{\rm eV}^2]$ & $[6.80, 8.02]$& $[6.80, 8.02]$\\ \hline
    $\Delta m^2_{3\ell}/10^{-3}\,[{\rm eV}^2]$ & $[2.408, 2.621]$& $[-2.589, -2.379]$\\ \hline
    $\delta$~$[{}^\circ]$ &$[0,30] \cup [128,360]$ & $[0,7]\cup[182,360] $\\ \hline
    $\gamma$~$[{}^\circ]$ &$[0,180]$ & $[0,180]$\\
  \hline
  \end{tabular}
  \caption{The $3 \sigma$ ranges for the lepton mixing parameters and mass-squared differences from NuFIT 3.1 (November 2017)~\cite{Esteban:2016qun,nufitweb}. Here $\Delta m^2_{3\ell}=\Delta m^2_{31}>0$ for normal ordering (NO) and $\Delta m^2_{3\ell}=\Delta m^2_{32}<0$ for inverted ordering (IO). We scan over these using flat priors.} \label{tab:neutrinos} 
\end{center}
\end{table}

\begin{table} [h!]
\begin{center}
  \begin{tabular}{ | c | c | c | }
    \hline
     Parameter & Range  \\ \hline \hline
    $\lambda_i$ & $\pm$$[10^{-3}, 4\pi]$  \\  \hline
     $\lambda_{H\Phi\Phi^\prime}$ & $[10^{-8}, 4\pi]$ \\  \hline
    $m_{\Phi^{(\prime)}}$~[GeV] & $[100,10^{5}]$  \\   \hline
    $m_{\psi}$~[GeV] & $[10,10^{5}]$   \\  \hline
     $\zeta$& $[10^{-3},10^{3}]$  \\   \hline
  \end{tabular}
  \caption{Priors on the 12 free real parameters used in the scan. $\lambda_i$ includes the following 8 quartic couplings of the potential: $\lambda_{\Phi^{(\prime)}},\,\lambda_{H\Phi^{(\prime)}},\,\lambda_{H\Phi^{(\prime)},2}, \, \lambda_{\Phi\Phi^\prime},\,\lambda_{\Phi\Phi^\prime,2}$. The parameter $\zeta$ is defined in App.~\ref{sec:yukpar}. We scan over these parameters using logarithmic priors.} \label{tab:parameters} 
\end{center}
\end{table}

We impose several constraints directly in the scan: 
$(i)$ Direct searches for singly-charged scalars from LEP II imply $m_{\eta^{({\prime})+}}\gtrsim 100$ GeV, with some dependence on the search channel; 
$(ii)$ constraints from the Higgs or $Z$-boson decay widths and
$(iii)$ $3\sigma$ constraints from EWPT, see Sec.~~\ref{sec:ewpt}. These constraints
restrict the mass splittings of the scalars, specially the one from the $T$ parameter; thus we also impose a lower bound of 100 GeV for all the new scalars;
$(iv)$ we apply the stability
conditions on the scalar potential given in App.~\ref{sec:stab};
$(v)$ we use the
experimental limits on the branching ratios from radiative $\ell_\alpha \rightarrow \ell_\beta \gamma$;
$(vi)$ we assume that the Dirac fermion $\psi$ constitutes all of the DM in the
Universe and thus require its relic abundance to lie within the $3\sigma$
range of the latest results from Planck~\cite{Ade:2015xua}, $\Omega_{\rm DM} h^2 = 0.1198\pm
0.0026$. All the observables for which we impose constraints in the numerical scan are
provided in Tab.~\ref{tab:observables}.

\begin{table} [h]
\begin{center}
  \begin{tabular}{ | c | c |c | c |  }
   \hline
   Observable & Upper bound & Observable & Measurement \\ \hline \hline
    $\mathrm{Br}(\mu^+\to e^+ \gamma)$& $2.55\cdot 10^{-13}$~\cite{TheMEG:2016wtm} &  $S$ & $0.05\pm 0.11$~\cite{Baak:2014ora} \\ \hline
    $\mathrm{Br}(\tau^-\to \mu^- \gamma)$& $4.4\cdot 10^{-8}$~\cite{Patrignani:2016xqp} &$T$ & $0.09\pm 0.13$~\cite{Baak:2014ora}\\ \hline
   $\mathrm{Br}(\tau^-\to e^- \gamma)$ & $3.3\cdot 10^{-8}$~\cite{Patrignani:2016xqp} &$U$ & $0.01\pm 0.11$~\cite{Baak:2014ora} \\ \hline
    $\sum m_i$~$[\mathrm{eV}]$ & $ 0.23$~\cite{Ade:2015xua} & $\Omega_{\rm DM}\,h^2$ & $0.1198 \pm 0.0026$~\cite{Ade:2015xua} \\ \hline
  \end{tabular}
  \caption{The current experimental upper bounds (two left columns) at 90\% CL and the measurements with their errors (two right columns) used in the parameter scan. They are required to lie within the $3\sigma$ range for the measurements. The correlation coefficients for the oblique parameters are $\rho_{ST}=0.90$, $\rho_{SU}=-0.59$ and $\rho_{TU}=-0.83$.~\cite{Baak:2014ora}} \label{tab:observables} 
\end{center}
\end{table}

In the following subsections we show the results of a numerical scan with about $10^4$ random points, using the input parameters in Tabs.~\ref{tab:neutrinos} and~\ref{tab:parameters}. 
Most of the results are shown for NO. Those for IO, unless explicitly shown, are basically identical. 
\begin{figure}[tb]
	\centering
		\includegraphics[width=0.5\textwidth]{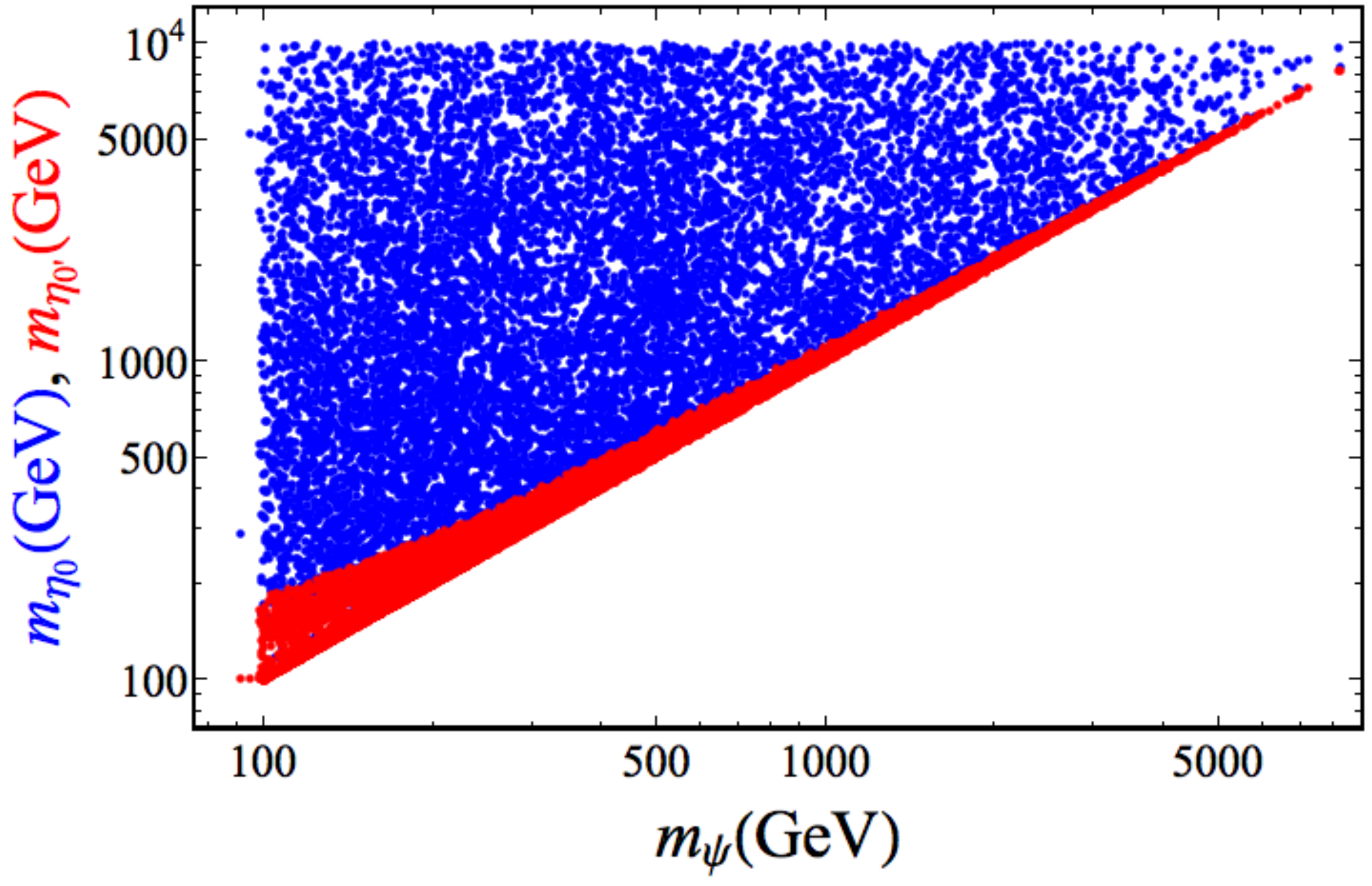}~~\includegraphics[width=0.5\textwidth]{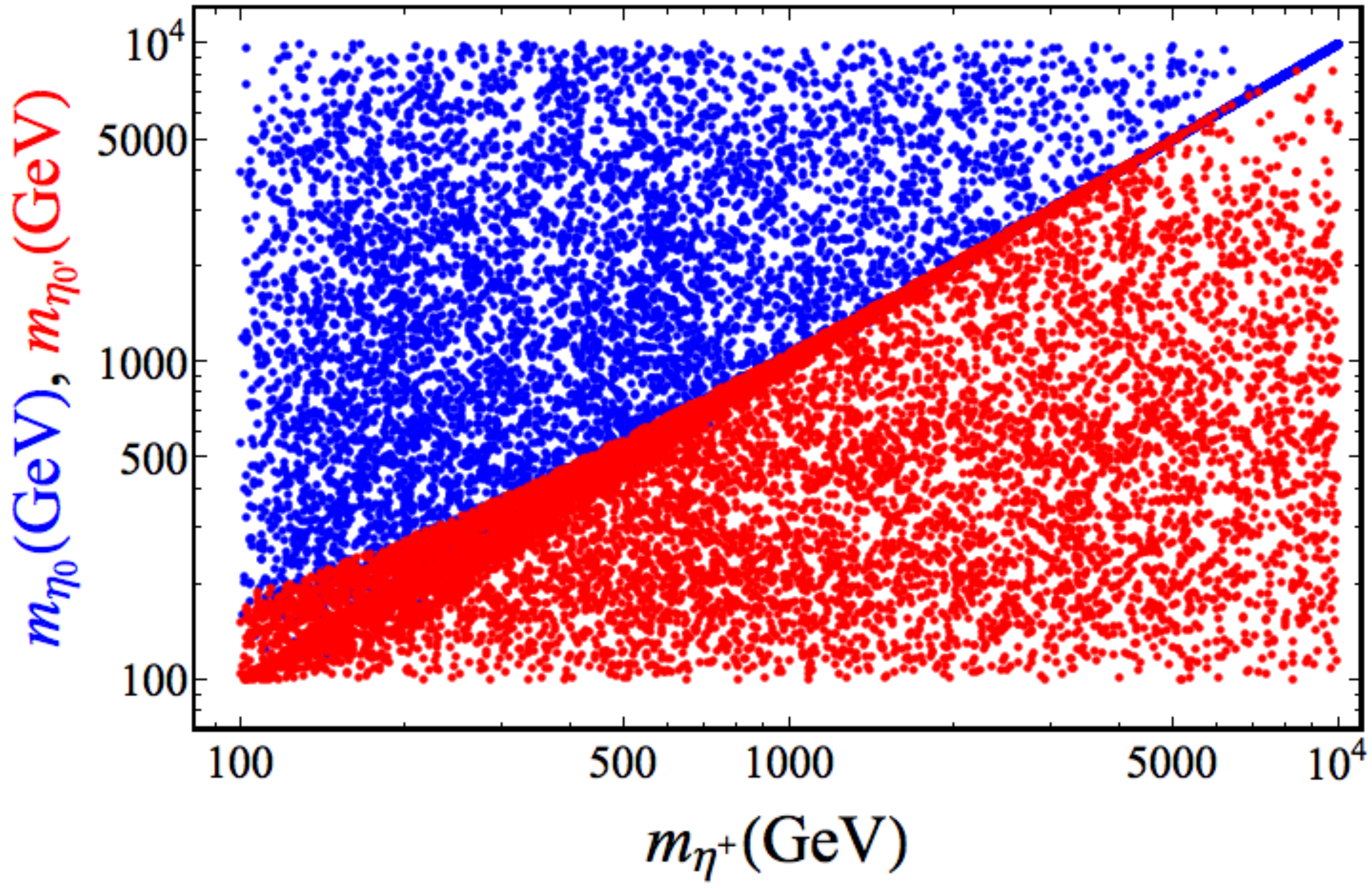}
\caption{The lighter and heavier neutral scalar masses, $m_{\eta^\prime_0}$ and $m_{\eta_0}$, in red and blue, respectively, versus the dark matter mass $m_\psi$ in the left panel, and versus the charged scalar mass $m_{\eta^+}$ in the right panel.}
\label{fig:various1}
\end{figure}

\subsection{Masses and lifetimes of the new scalars}

\begin{figure}[tb]
	\centering
	\includegraphics[width=0.65\textwidth]{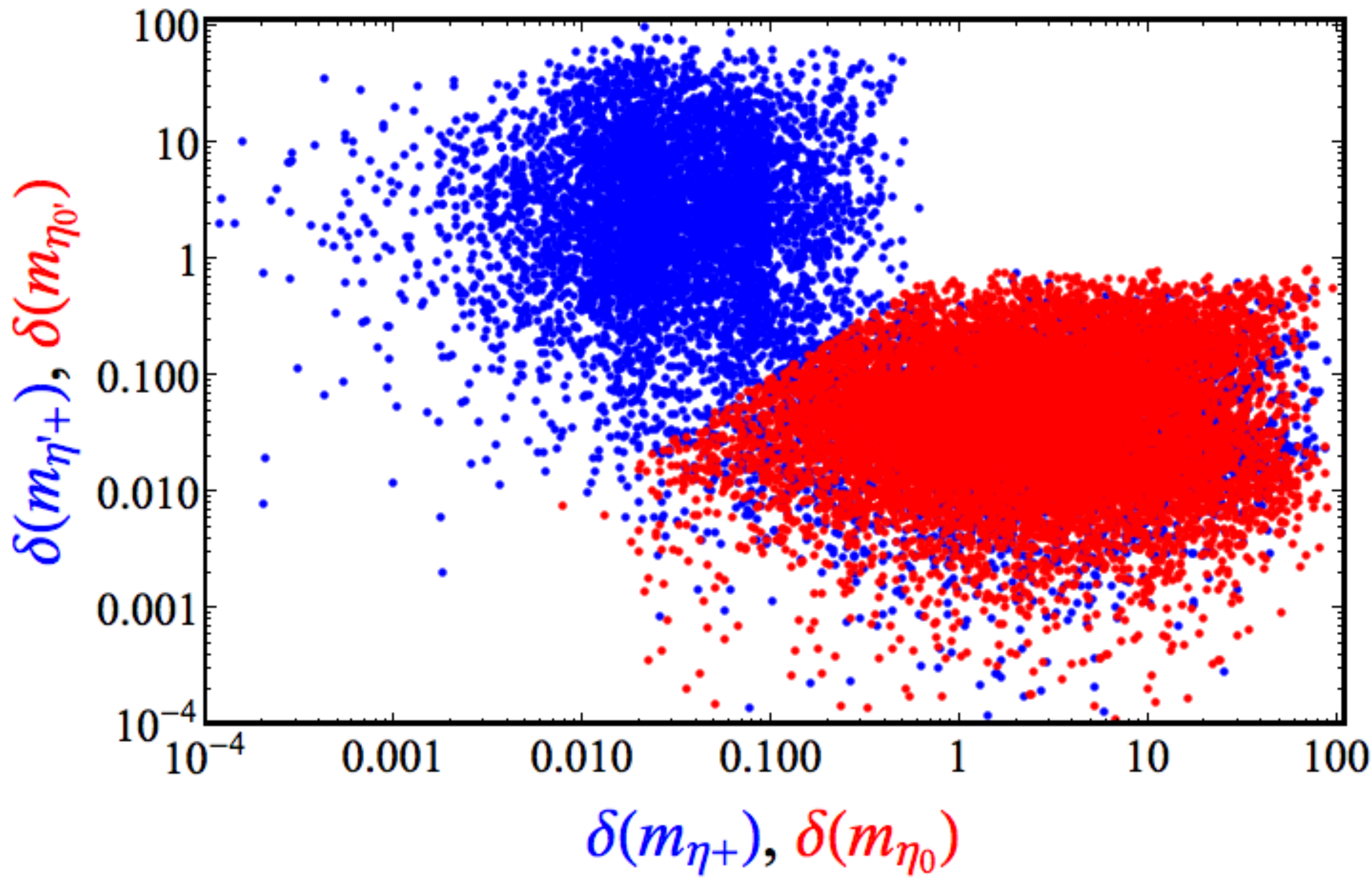}
\caption{Normalised mass splitting $\delta(m_x)=(m_x-m_\psi)/m_\psi$ of the charged scalar $\eta^{\prime+}$ versus $\eta^{+}$, in blue, and of the neutral scalar $\eta_0^\prime$ versus $\eta_0$, in red. Notice that the red region of points is superimposed over part of the blue one.}
\label{fig:various1b}
\end{figure}
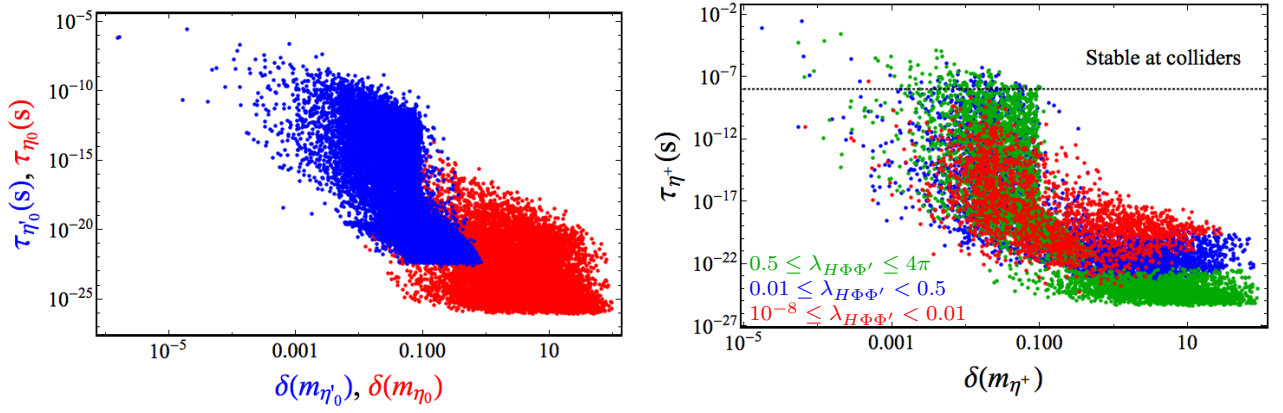
\begin{figure}[bt]
	\centering
	\includegraphics[width=0.5\textwidth]{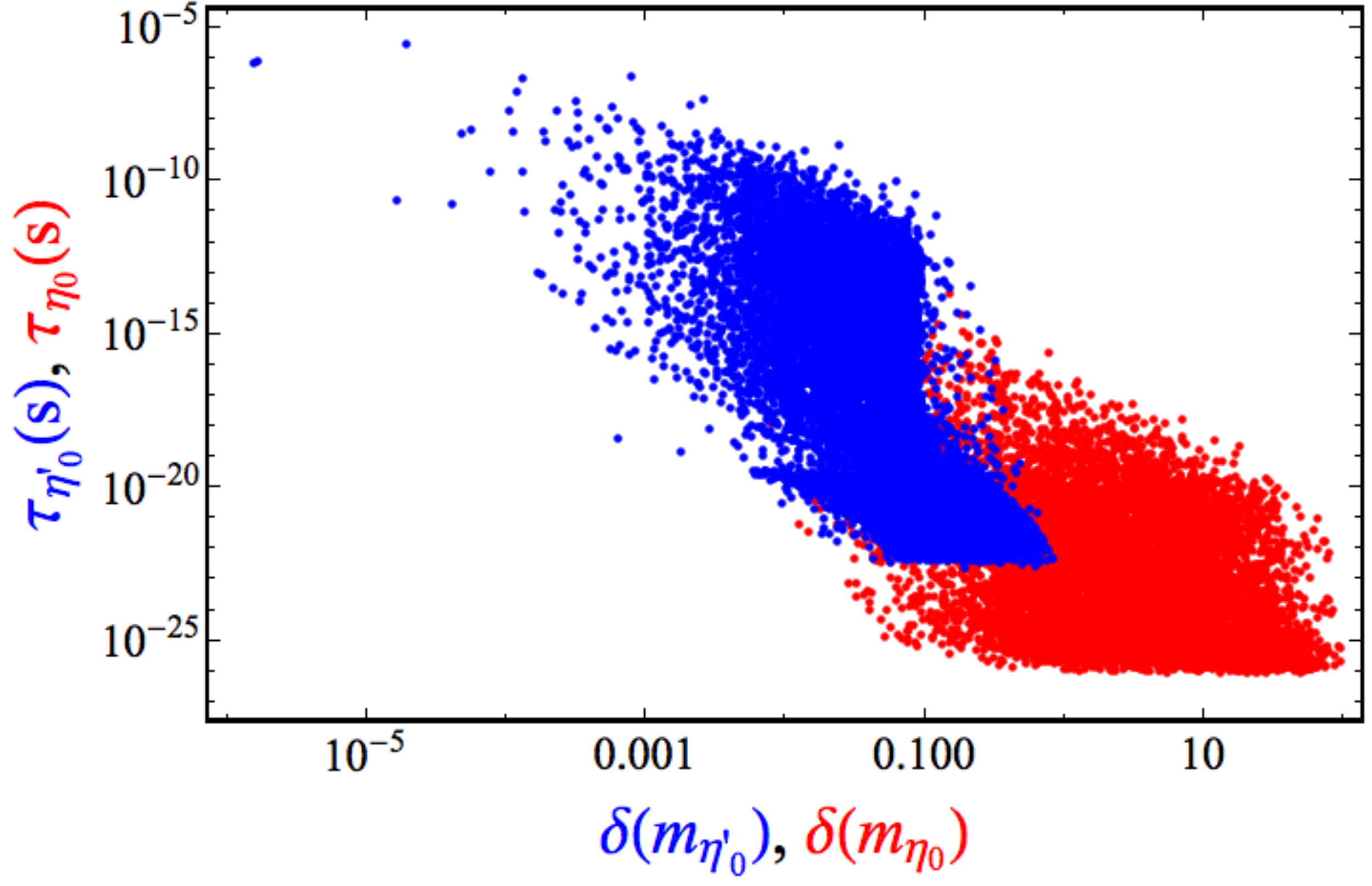}~~%
\begin{tikzpicture}
		\node (pic) {\includegraphics[width=0.5\textwidth]{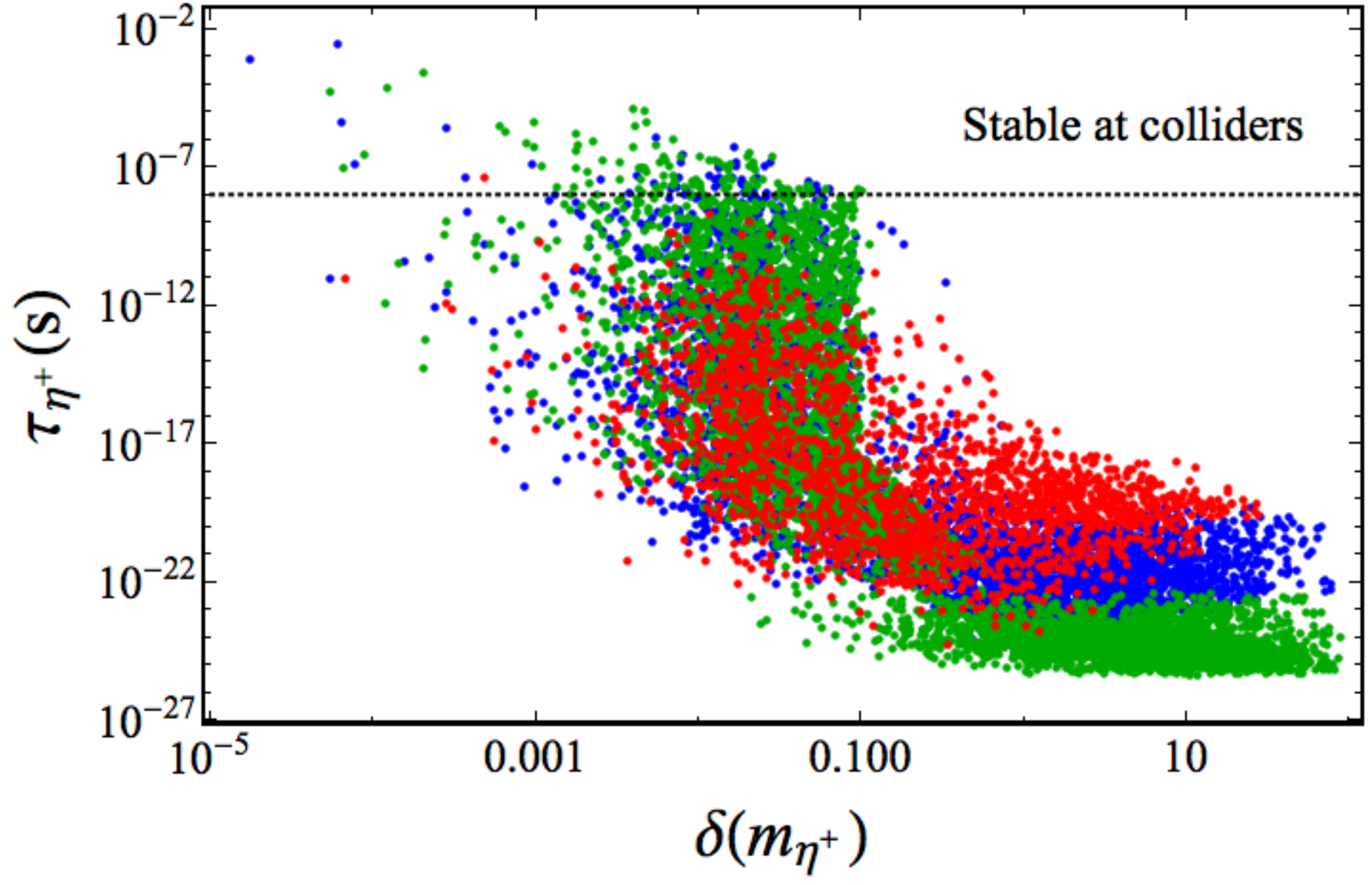}};
		\node[darkgreen,above right,shift={(8ex,10ex)}] at (pic.south west) {\scriptsize $0.5\leq\lambda_{H\Phi\Phi^\prime}\leq 4\pi$};
		\node[blue,above right,shift={(8ex,8ex)}] at (pic.south west) {\scriptsize $0.01\leq\lambda_{H\Phi\Phi^\prime}< 0.5$};
		\node[red,above right,shift={(8ex,6ex)}] at (pic.south west) {\scriptsize $10^{-8}\leq\lambda_{H\Phi\Phi^\prime}< 0.01$};
	\end{tikzpicture}
\caption{\textit{Left panel:} Lifetime of the neutral scalars $\eta_0^\prime$ (blue) and $\eta_0$ (red) versus the normalised mass splitting $\delta(m_x)=(m_x-m_\psi)/m_\psi$. \textit{Right panel:} Lifetime of the charged scalar $\eta^{+}$ versus $\delta(m_{\eta^+})$ for
	different ranges of
$\lambda_{H\Phi\Phi^\prime}$.
We indicate with a horizontal dotted black line the minimum lifetime needed for the charged
scalar to be long-lived at collider scales. A similar plot is obtained for
the charged scalar $\eta^{\prime+}$.}
\label{fig:various2a}
\end{figure}
\begin{figure}[bht]
	\hspace{-0.3in}
	\begin{tikzpicture}
		\node (pic) {\includegraphics[width=0.5\textwidth]{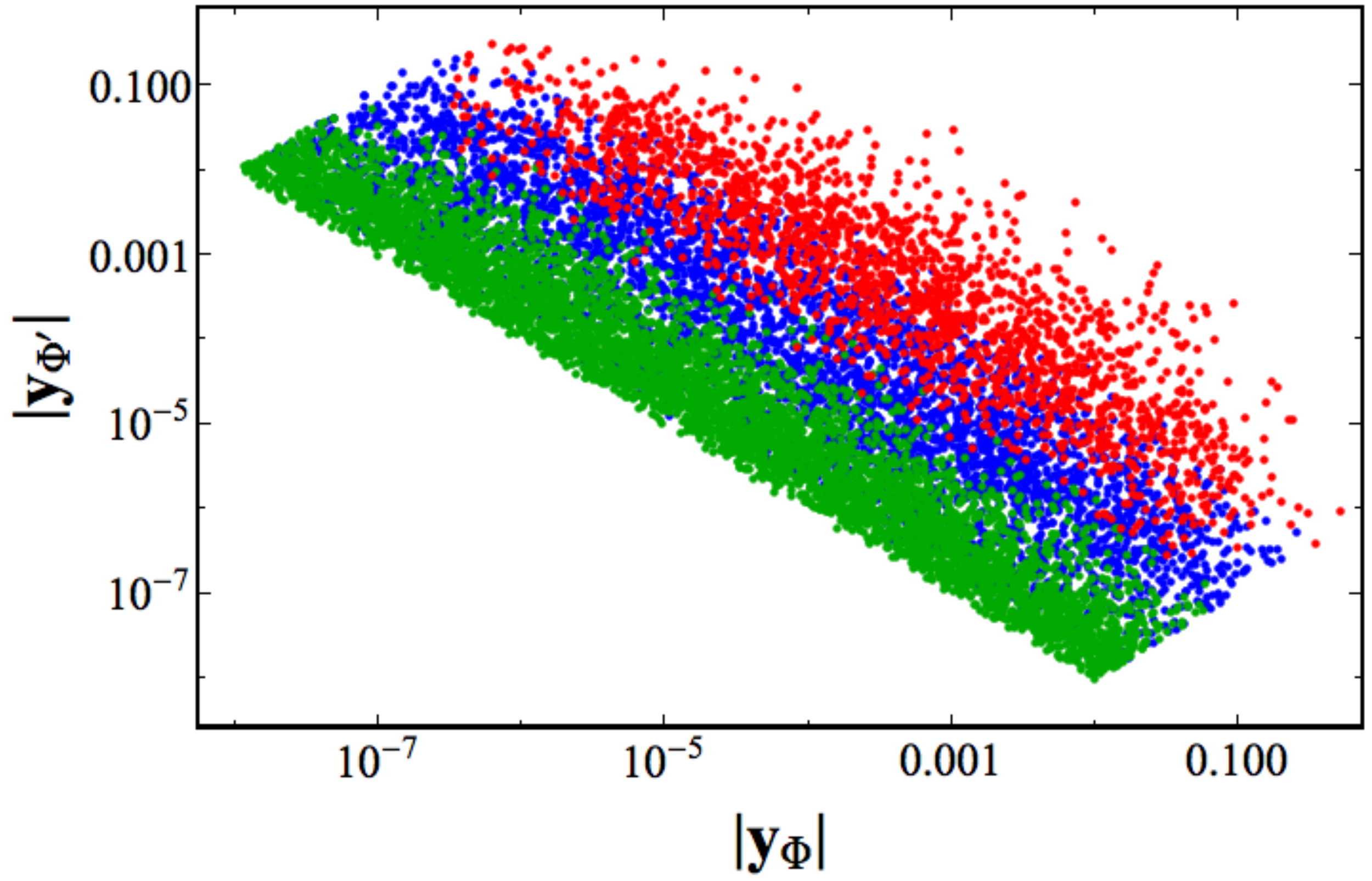}};
		\node[darkgreen,above right,shift={(8ex,10ex)}] at (pic.south west) {\scriptsize $0.5\leq\lambda_{H\Phi\Phi^\prime}\leq 4\pi$};
		\node[blue,above right,shift={(8ex,8ex)}] at (pic.south west) {\scriptsize $0.01\leq\lambda_{H\Phi\Phi^\prime}< 0.5$};
		\node[red,above right,shift={(8ex,6ex)}] at (pic.south west) {\scriptsize $10^{-8}\leq\lambda_{H\Phi\Phi^\prime}< 0.01$};
	\end{tikzpicture}%
	~~\begin{tikzpicture}
		\node (pic) {\includegraphics[width=0.5\textwidth]{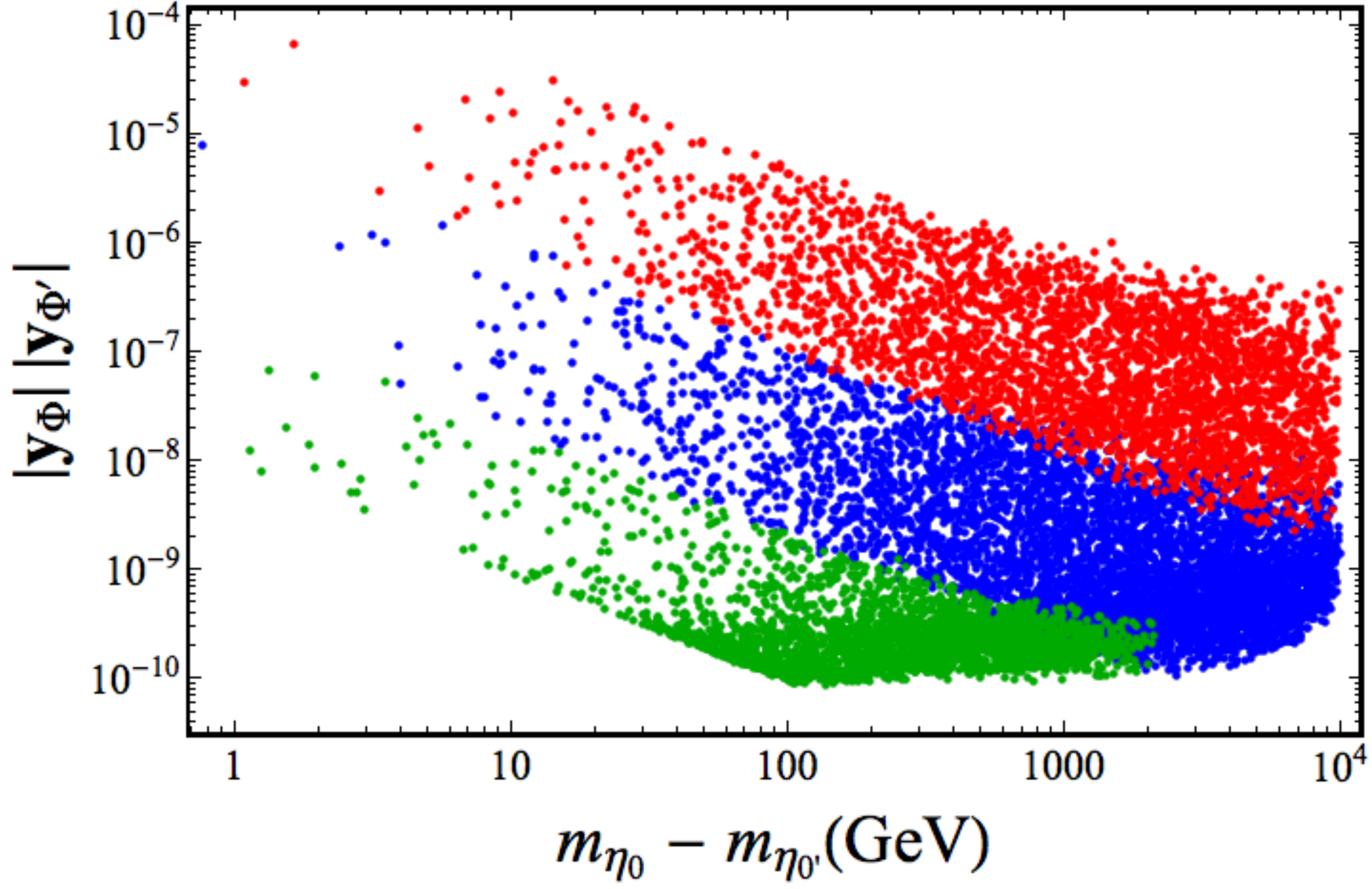}};
		\node[red,below left,shift={(-1ex,-1ex)}] at (pic.north east) {\scriptsize $10^{-5}\leq |\sin2\theta|\leq 0.001$};
		\node[blue,below left,shift={(-1ex,-3ex)}] at (pic.north east) {\scriptsize $0.001\leq |\sin2\theta|< 0.05$};
		\node[darkgreen,below left,shift={(-1ex,-5ex)}] at (pic.north east) {\scriptsize $0.05\leq |\sin2\theta|< 1$};
	\end{tikzpicture}
\caption{
\textit{Left panel:}
$|\mathbf{y_{\Phi^\prime}}|$ versus $ |\mathbf{y_\Phi}|$ for different ranges of
$\lambda_{H\Phi\Phi^\prime}$.
\textit{Right panel:}  $|\mathbf{y_\Phi}| |\mathbf{y_{\Phi^\prime}}|$ versus the mass splitting of the
	new neutral scalars $m_{\eta_0}-m_{\eta^\prime_0}$ for different ranges of
$|\sin2\theta|$.
}
\label{fig:various0}
\end{figure}

We show in Fig.~\ref{fig:various1} (left panel) the lightest and the heaviest neutral
scalar masses, $m_{\eta^\prime_0}$ and $m_{\eta_0}$,   versus the DM mass
$m_\psi$, in red and blue, respectively. The lightest neutral scalar mass is
close to the DM mass. This is driven by the fact that
coannihilations need to be efficient enough in order to obtain the correct relic
abundance. In Fig.~\ref{fig:various1} (right panel), we show the neutral scalar masses versus the
charged scalar mass $m_{\eta^+}$. We observe that $\eta_0$ is always the heaviest
state. Roughly in 50\% (30\%) of the points the mass of the lightest neutral (one of the charged) scalar(s) is very degenerate with the mass of the DM particle (with a normalised mass splitting smaller than $5\%$) and contributes to
coannihilations. In addition, there are significant regions of the parameter space of the model (18\% of the points) where
both masses of lightest neutral and one of the charged scalars ($\eta^\pm$ or
$\eta^{\prime\pm}$) are nearly degenerate with the DM mass. Only in around $1\%$ of the points the masses of the three new scalars $\eta^\prime_0,\,\eta^\pm,\,\eta^{\prime\pm}$ are very degenerate with the DM mass. There is no difference between the cases with neutrinos with NO and IO.

In Fig.~\ref{fig:various1b} we show the normalised mass splitting
$\delta(m_x)=(m_x-m_\psi)/m_\psi$ of the charged scalars, $\eta^{\prime+}$
versus $\eta^{+}$, in blue, and of the neutral scalars, $\eta_0^\prime$ versus
$\eta_0$, in red. Notice that the red region of points is superimposed over part of the blue one. The normalised mass splitting  $\delta(m_x)=(m_x-m_\psi)/m_\psi$ needs to be below $\sim50\%$,
and typically $\sim 5\%$, for at least one of the scalars
$\eta_0^\prime$, $\eta^{\pm}$ and/or $\eta^{\prime\pm}$ in order for coannihilations to be efficient. It is typically much larger for the heaviest neutral scalar $\eta_0$, as can be seen in Fig.~\ref{fig:various1b}.

In Fig.~\ref{fig:various2a} (left panel) we plot the lifetime of the neutral
scalars, $\eta_0^\prime$ (blue) and $\eta_0$ (red), versus the normalised mass
splitting $\delta(m_x)=(m_x-m_\psi)/m_\psi$. 
One can observe how the lifetime of $\eta_0^\prime$ can be much larger than that of $\eta_0$. Indeed, when the splitting $\delta(m_{\eta^\prime_0})$ with the DM mass is small, the only decays of $\eta_0^\prime$ are into charged leptons, and even those can be impossible for very small mass splittings and/or suppressed for small Yukawa couplings. In Fig.~\ref{fig:various2a} (right panel) we plot the lifetime of the charged scalar $\eta^{+}$ versus $\delta(m_{\eta^+})$, for different ranges of $\lambda_{H\Phi\Phi^\prime}$:
$10^{-8}\lesssim\lambda_{H\Phi\Phi^\prime}\lesssim 0.01$ in red,
$0.01\lesssim\lambda_{H\Phi\Phi^\prime}\lesssim 0.5$ in blue, and
$0.5\lesssim\lambda_{H\Phi\Phi^\prime}\lesssim 4\pi$ in green. 
The plot for the
charged scalar $\eta^{\prime+}$ is analogous to that of $\eta^{+}$. We observe
two effects: firstly, for large mass splittings, $\delta(m_{\eta^+})\gtrsim 0.1$ which corresponds to $m_{\eta^+}-m_{\psi}
\gtrsim 80$ GeV, the main decay channel is $\eta^{+} \rightarrow W^+
\eta_0^\prime$, and the larger the quartic coupling $\lambda_{H\Phi\Phi^\prime}$, the larger the
neutral scalars mixing $\cos\theta$, see Eq.~\eqref{eq:theta}, and the larger this decay; secondly, the larger the normalised mass splitting with the DM mass, the
smaller the lifetime. Indeed, the charged scalar can be long-lived at collider scales, meaning $\tau_{\eta^+}\gtrsim 10^{-8}$ s, as shown with a horizontal dotted black line for
mass splittings $m_{\eta^+} - m_{\eta^{\prime}_0}$ smaller than $\sim 80$ GeV, when the decay channel $\eta^{+}
\rightarrow W^+ \eta_0^\prime$ is closed. In that region, the decays $\eta^{+}
\rightarrow \ell_\alpha^+ \psi$, that are mediated by the Yukawa couplings
$y^\alpha_\Phi$, dominate. Therefore, the larger the quartic coupling
$\lambda_{H\Phi\Phi^\prime}$, the smaller the Yukawa couplings, and the larger
the lifetime, see blue and green points in Fig.~\ref{fig:various2a} (right panel).

\subsection{Neutrino Yukawa couplings}
In the left panel of Fig.~\ref{fig:various0} we show $|\mathbf{y_{\Phi^\prime}}|$
versus $ |\mathbf{y_\Phi}|$ for fixed intervals of $\lambda_{H\Phi\Phi^\prime}$:
$10^{-8}\lesssim\lambda_{H\Phi\Phi^\prime}\lesssim 0.01$ in red,
$0.01\lesssim\lambda_{H\Phi\Phi^\prime}\lesssim 0.5$ in blue, and
$0.5\lesssim\lambda_{H\Phi\Phi^\prime}\lesssim 4\pi$ in green. The Yukawa couplings are inversely proportional to each other as expected from neutrino masses, see Eq.~\eqref{eq:numasses}. Also, the larger the quartic coupling
$\lambda_{H\Phi\Phi^\prime}$, the smaller the Yukawa couplings.  For $\lambda_{H\Phi\Phi^\prime} \lesssim
0.5$ the product of the Yukawa couplings is constrained to $10^{-9}\lesssim |\mathbf{y_\Phi}|
|\mathbf{y_{\Phi^\prime}}| \lesssim 10^{-7}$ as shown in the plot. This is a direct consequence of the appearance of these couplings in the expression for the neutrino masses,
see Eqs.~(\ref{eq:numasses}) and (\ref{eq:mnupm}).

We show in Fig.~\ref{fig:various0} (right panel) the product of the absolute
values of the neutrino Yukawa couplings $|\mathbf{y_\Phi}| |\mathbf{y_{\Phi^\prime}}|$ versus the mass
splitting of the neutral scalars $m_{\eta_0}-m_{\eta^\prime_0}$ for different
ranges of $|\sin2\theta|$: $10^{-5}\lesssim |\sin2\theta| \lesssim 0.001$
in red, $0.001\lesssim |\sin2\theta| \lesssim 0.05$ in blue, and
$0.05\lesssim |\sin2\theta| \lesssim 1$ in green. We observe that the larger the mixing $|\sin2\theta|$ among the neutral scalars, the smaller the Yukawa couplings. This
is expected as $|\mathbf{y_\Phi}| |\mathbf{y_{\Phi^\prime}}| |\sin2\theta|$ is proportional to
the scale of neutrino masses, see Eqs.~\eqref{eq:numasses} and \eqref{eq:mnupm}. In addition, we see that neutrino masses are also proportional to
the mass splitting of the neutral scalars, and for a given range of
$|\sin2\theta|$, the larger the product of the neutrino Yukawa couplings, the smaller
this mass splitting has to be.

\subsection{Charged lepton flavour violating processes}

\begin{figure}[tb]
	\centering
	\begin{tikzpicture}
		\node (pic) {\includegraphics[width=0.7\textwidth]{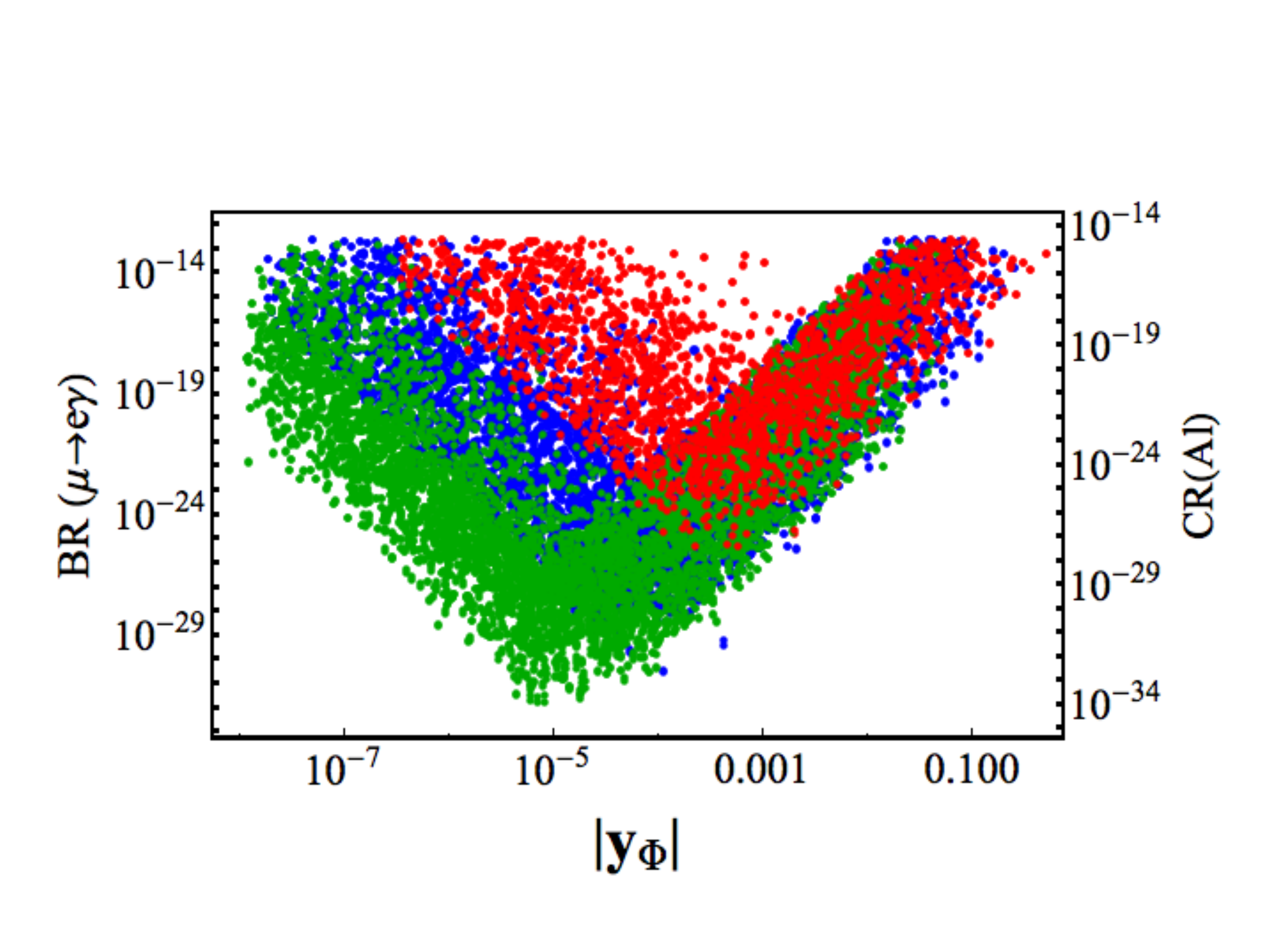}};
		\node[darkgreen,above left,shift={(-12.5ex,16ex)}] at (pic.south east) {\scriptsize $0.5\leq\lambda_{H\Phi\Phi^\prime}\leq 4\pi$};
		\node[blue,above left,shift={(-12.5ex,14ex)}] at (pic.south east) {\scriptsize $0.01\leq\lambda_{H\Phi\Phi^\prime}< 0.5$};
		\node[red,above left,shift={(-12.5ex,12ex)}] at (pic.south east) {\scriptsize $10^{-8}\leq\lambda_{H\Phi\Phi^\prime}< 0.01$};
	\end{tikzpicture}
\caption{Branching ratio of $\mu \rightarrow e \gamma$ (left axis) and the $\mu-e$ conversion ratio in Al (right axis) 
versus $|\mathbf{y_\Phi}|$ for different ranges of $\lambda_{H\Phi\Phi^\prime}$.}
\label{fig:various2}
\end{figure}
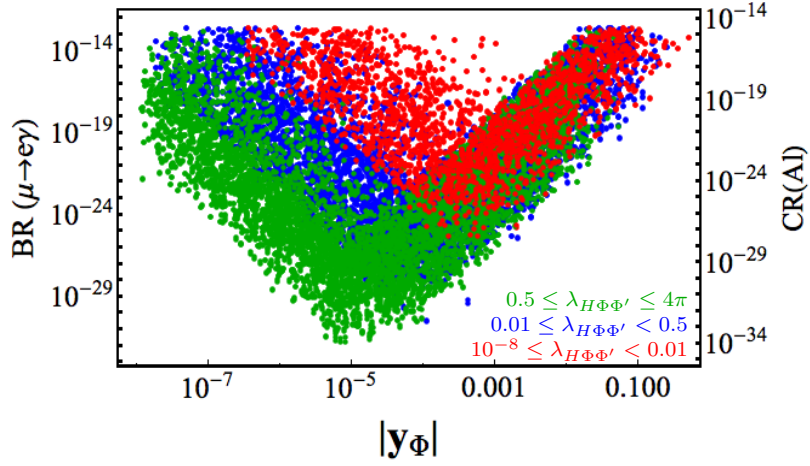
\begin{figure}[t!]
	\centering
	\includegraphics[width=0.5\textwidth]{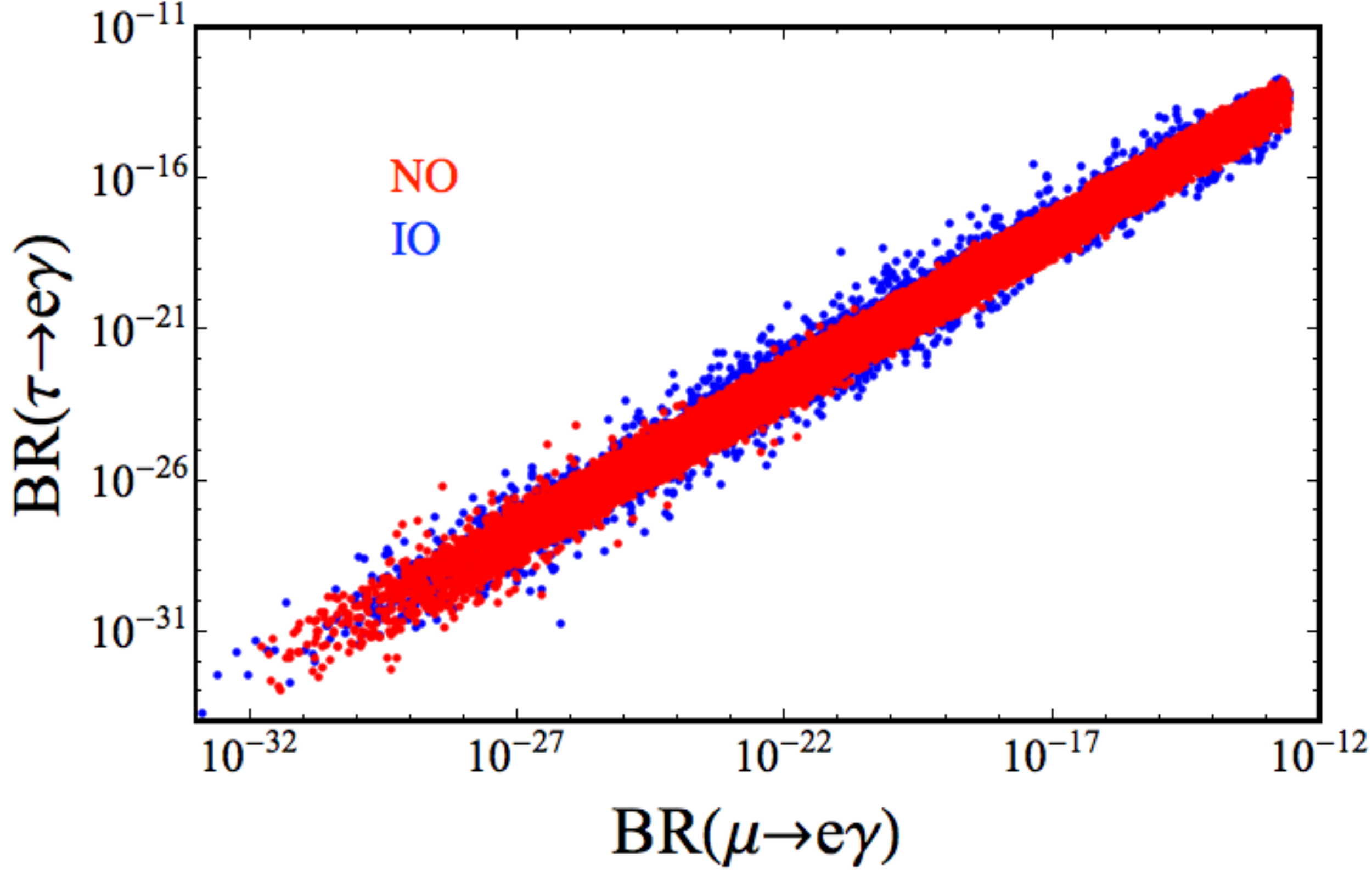}~~\includegraphics[width=0.5\textwidth]{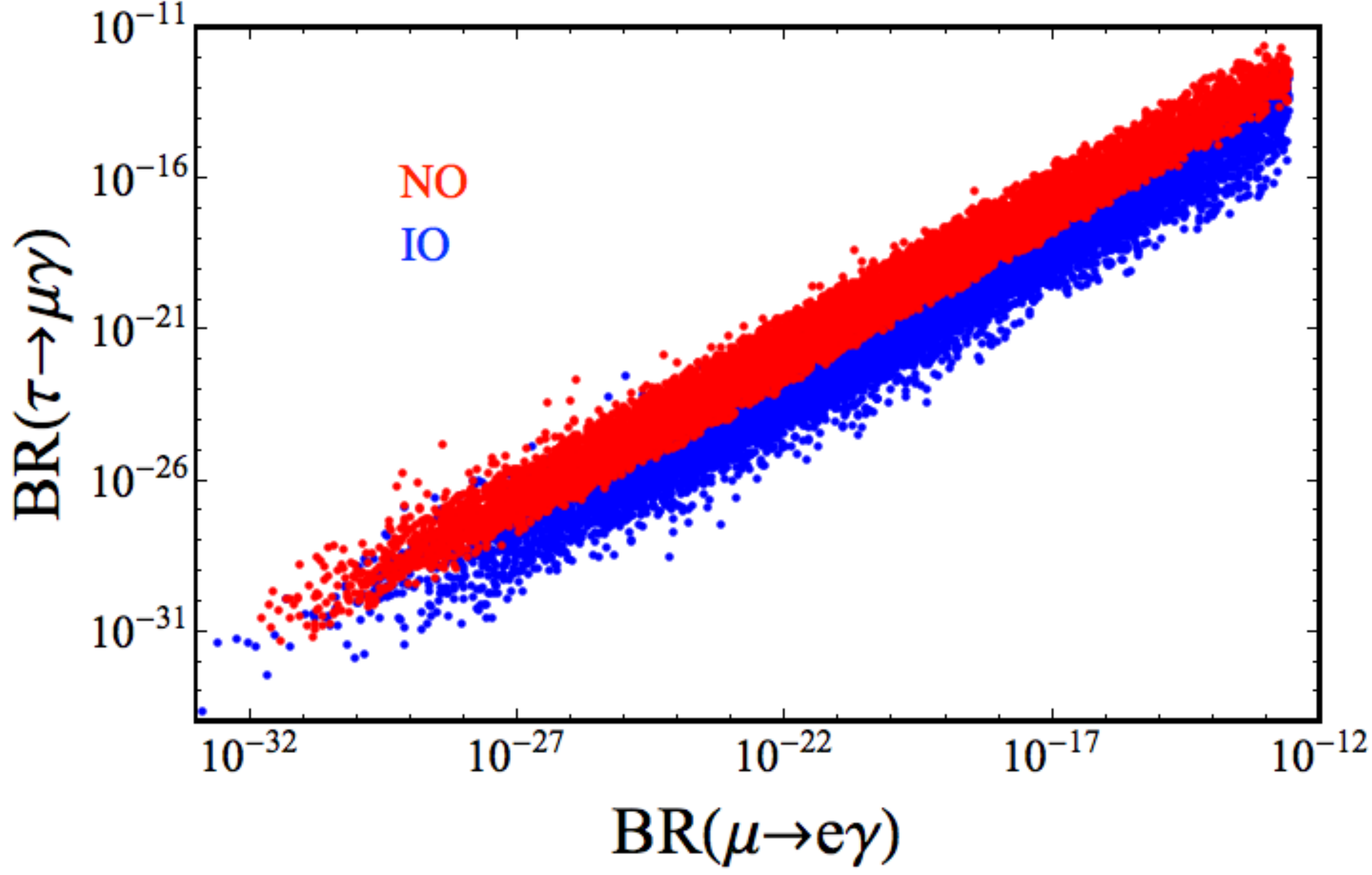}
\caption{Branching ratios of $\tau \rightarrow e \gamma$ (\textit{left panel}) and $\tau \rightarrow \mu \gamma$ (\textit{right panel}) versus that of $\mu \rightarrow e \gamma$ for neutrino masses with normal ordering (in red) and inverted ordering (in blue).}
\label{fig:various4}
\end{figure}

In Fig.~\ref{fig:various2} we plot the branching ratio of $\mu \rightarrow e
\gamma$ versus $|\mathbf{y_\Phi}|$ for the
same ranges of the quartic coupling $\lambda_{H\Phi\Phi^\prime}$ used in Fig.~\ref{fig:various2a}
(right panel). The different sets of points form
V-shaped regions whose minimum value for $\mathrm{BR}(\mu\to e\gamma)$ is larger the smaller 
$\lambda_{H\Phi\Phi^\prime}$. For $|\mathbf{y_\Phi}| \gtrsim 10^{-4}$,
the branching ratio scales as $|\mathbf{y_\Phi}|^4$, independently of $\lambda_{H\Phi\Phi^\prime}$.
In this region the contribution due to the scalar $\eta^{\prime+}$ is suppressed because $|\mathbf{y_{\Phi^\prime}}|\lesssim|\mathbf{y_\Phi}|$. If, however, $|\mathbf{y_\Phi}| \lesssim 10^{-4}$ 
the scalar $\eta^{\prime+}$ dominates the branching ratio. The dependence on
$\lambda_{H\Phi\Phi^\prime}$ again sets the scale of
$|\mathbf{y_{\Phi^\prime}}|$ and thus the branching ratio of $\mu
\rightarrow e \gamma$, i.e., the larger the quartic coupling $\lambda_{H\Phi\Phi^\prime}$ the smaller $\text{BR}(\mu \rightarrow e \gamma)$. The minimum value of the branching ratio of these CLFV processes
occurs for $|\mathbf{y_\Phi}| \sim |\mathbf{y_{\Phi^\prime}}| \sim 10^{-4.5}$,
when both charged scalar contributions are of similar order, such that the overall result is suppressed.

In Fig.~\ref{fig:various4} we plot  ${\rm BR}(\tau \rightarrow e \gamma)$ (left panel) and ${\rm BR}(\tau \rightarrow \mu \gamma)$ (right panel) versus 
${\rm BR}(\mu \rightarrow e \gamma)$, for neutrino masses with NO (in red) and IO (in blue). The
central values of these ratios agree with our analytical estimates given in
Eqs.~\eqref{eq:ratioradBRNO} and \eqref{eq:ratioradBRIO}, although the entire range of these ratios is about two orders of magnitude. 
In particular, we see that $\text{BR}(\tau\to e \,\gamma)$ is
suppressed compared to $\text{BR}(\tau\to \mu \,\gamma)$ for neutrino masses
with NO, while they are very similar for IO. The largest branching ratio is achieved for
$\tau \rightarrow \mu \gamma$ for NO, which can be larger than the one for IO. Therefore a measurement could in principle discriminate between the neutrino mass orderings. 

\subsubsection{Interplay with dark matter direct detection}

\begin{figure}[tb]
	\centering
	\includegraphics[width=0.7\textwidth]{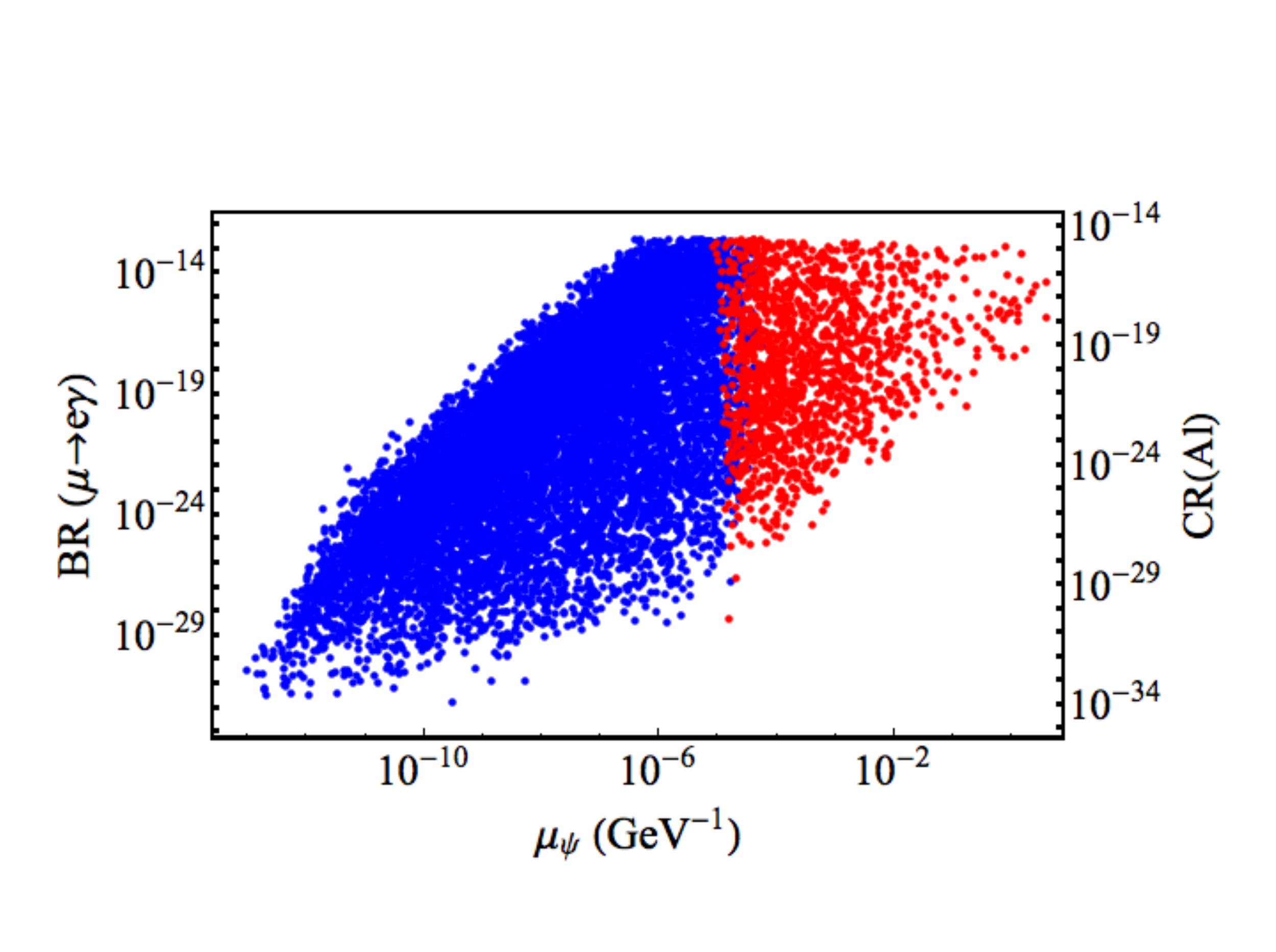}
\caption{Branching ratio of $\mu \rightarrow e \gamma$ (left axis) and the $\mu- e$ conversion ratio in Al (right axis) versus the dark matter magnetic dipole moment $\mu_\psi$ relevant for DM direct detection. We plot in red (blue) the points that are  excluded (allowed) by DM direct detection limits. 
}
\label{fig:various2L}
\end{figure}
In Fig.~\ref{fig:various2L} we plot the branching ratio of $\mu \rightarrow e \gamma$ (left axis) and the $\mu - e$ conversion ratio in Al (right axis) versus the DM magnetic dipole moment $\mu_\psi$, see Eq.~\eqref{eq:magDD}, which is relevant for DM direct detection. The size of the magnetic dipole moment $\mu_\psi$ is correlated with the branching ratios of CLFV processes, because the structure of the loop diagrams is similar, with a charged scalar in the loop and the same Yukawa couplings. The points in red (blue), corresponding to larger (smaller) values of $\mu_\psi$, are excluded (allowed) by the combined constraint from Xenon-based DM direct detection experiments, which are implemented in LikeDM~\cite{Liu:2017kmx}. We can see the interesting complementarity between DM direct detection and CLFV processes in constraining the parameter space of the model. This interplay is further discussed in the generic context for a fermionic SM singlet DM particle in Ref.~\cite{Herrero-Garcia:2018koq}.

\mathversion{bold}
\section{Variants of the model}
\mathversion{normal}
\label{sec:disc}

\mathversion{bold}
\subsection{$\rm U(1)_{DM}$ as gauge symmetry}
\mathversion{normal}

The global $\rm U(1)_{DM}$ symmetry is anomaly-free, because the fermion $\psi$ is vector-like and thus $\rm U(1)_{DM}$ can be straightforwardly gauged. In fact, a similar model with a gauge symmetry has been discussed in Ref.~\cite{Ma:2013yga}. 
Three scenarios can be envisaged: $(i)$ $\rm U(1)_{DM}$ is unbroken and the corresponding dark photon $\gamma_{\rm DM}$ is massless, $(ii)$  $\rm U(1)_{DM}$ can be realised non-linearly and the dark photon $\gamma_{\rm DM}$ obtains its mass from the St\"uckelberg mechanism, or $(iii)$ $\rm U(1)_{DM}$ is spontaneously broken  to a residual $Z_N$ symmetry, which stabilises  the DM candidate.  In this case an additional scalar field $\rho$, charged under $\rm U(1)_{\mathrm{DM}}$, has to take a non-vanishing VEV.
In case $(i)$ $\gamma_{\rm DM}$ contributes to extra radiation and lead to large
self-interactions, see Ref.~\cite{Agrawal:2016quu} for a discussion. If in case $(ii)$ and $(iii)$ the mass of the dark photon $\gamma_{\rm DM}$ is smaller than
that of the DM candidate, the DM relic abundance is set by annihilations into dark photons. Connections of DM phenomenology to neutrino and flavour physics are then lost so that this case is not interesting to us. In addition, in case $(iii)$ the new scalar field $\rho$ mixes with the SM Higgs doublet $H$. Such mixing is experimentally constrained by invisible Higgs decays, if these are kinematically accessible, and by DM direct detection limits, especially for sub-GeV scalar mediators, thus requiring $\rho$ to be either much heavier than the electroweak scale or the mixing to be small. That in turn is in conflict with the fact that the mediator needs to decay before big bang nucleosynthesis, basically ruling out this possibility~\cite{Ren:2018gyx}.

Another effect of gauging $\rm U(1)_{DM}$ is the kinetic
mixing with $\rm U(1)_Y$, that is the term $\epsilon\,
B^{\mu\nu}_{\rm DM} B_{\mu\nu}$ with the field strength
tensors $B^{\mu\nu}$ ($B^{\mu\nu}_{\rm DM}$) of
$\rm U(1)_Y$ ($\rm U(1)_{DM}$). Even if this is tuned to
vanish at a certain scale, it arises at one-loop level,
since  $\Phi$ and $\Phi^\prime$ are charged
under both $\rm U(1)_{\rm Y}$ and $\rm U(1)_{\rm DM}$.
This effect can be estimated as follows: Assuming
$\Lambda_{\rm UV}>m_\Phi, m_{\Phi^\prime}>v_H$ and a vanishing kinetic mixing at a certain high scale $\Lambda_{\rm UV}$, $\epsilon (\Lambda_{\rm UV})=0$,
 renormalisation group running can induce a sizeable kinetic mixing at lower scales.
 Above the mass scales of  $\Phi$ and $\Phi^\prime$, the opposite values of their hypercharge  lead to an exact cancellation of $\epsilon$, but as soon as one of the scalar fields decouples, the kinetic mixing  is induced via renormalisation group running, giving the approximate lower bound 
\begin{equation}
	|\epsilon| \gtrsim \frac{\sqrt{\alpha_Y \alpha_{\rm DM}}}{4\pi} \left|\ln\left(\frac{m_\Phi}{m_{\Phi^\prime}}\right)\right|\,,
\end{equation}
with the dark gauge coupling $\alpha_{\rm DM}=g_{\rm DM}^2/4\pi$. The annihilation into  dark photons is given by $\braket{v\sigma}_{\rm ann} \simeq \pi \alpha_{\rm DM}^2/m_\psi^2$, which implies that in order to reproduce the DM relic abundance $\alpha_{\rm DM} \simeq 10^{-4}\,(m_\psi/{\rm GeV})$. Using the experimental value of $\alpha_Y(m_Z)$ \cite{Patrignani:2016xqp}, and taking the logarithm to be $\mathcal{O}(1)$, we can estimate a lower bound on the kinetic mixing: $|\epsilon|\gtrsim 10^{-4}\,(m_\psi/{\rm GeV})$. As in the case of scalar mediators, there are very strong upper limits from DM direct detection, especially if the dark photon is lighter than a few GeV, with only much smaller mixings still allowed, see the recent analysis by the PandaX-II collaboration~\cite{Ren:2018gyx}. In this case, the mediators should decay before big bang nucleosynthesis sets in. Therefore, a certain amount of fine-tuning is required in this case.

\mathversion{bold}
\subsection{Replacing $\rm U(1)_{DM}$ with $Z_N$}
\mathversion{normal}

Instead of a continuous symmetry we can consider a $Z_N$ symmetry, being a subgroup of $\rm U(1)_{DM}$, i.e. we 
regard the charge of $\Phi$, $\Phi^\prime$ and $\psi$ as given modulo $N$.

For $N > 4$ the model effectively possesses a global $\rm U(1)$ symmetry, as long as we only consider renormalizable terms in the Lagrangian.
Conversely, for $N=2,\, 3,$ and $4$,  additional terms arise at the renormalizable level.

For $N=2$, i.e. the smallest possible symmetry, we can identify $\psi \leftrightarrow \psi^c$ and $\Phi^\prime \leftrightarrow \tilde \Phi$. 
The model thus contains one Majorana fermion $\psi$ and one scalar doublet $\Phi$ 
which are the only particles odd under the $Z_2$
symmetry. The Lagrangian for the Majorana fermion $\Psi=\psi+\psi^c$ reads
\begin{equation}
	\mathcal{L}_{\Psi}\;=\; \frac12 \left(\overline \Psi\, i\,\slashed \partial\,\Psi\,-\, m_\psi \, \overline{\Psi}\, \Psi\right)\, -\, \Big( y^\alpha \, \overline \Psi\, {\tilde \Phi}^\dagger\,L_\alpha\,+\,\text{H.c.}\Big) \,. 
\label{LpsiSc}
\end{equation}
 The scalar potential  for $\Phi$ and the SM Higgs doublet $H$ becomes
\begin{align}
\label{potentialSc}
\mathcal{V}_{Z_2}  =& \,-\,m_H^2  H^\dagger H \,+\, \lambda_H (H^\dagger H)^2 \,
+ \, m_\Phi^2  \Phi^\dagger\Phi\, +\, \lambda_\Phi (\Phi^\dagger\Phi )^2\,
\\\nonumber&
\, + \, \lambda_{H\Phi} (H^\dagger H) (\Phi^\dagger\Phi)  
\,+\, \lambda_{H\Phi,2} (H^\dagger \Phi) (\Phi^\dagger H)   
\,+\, \frac{\lambda_{H\Phi,3}}{2}\left[ (H^\dagger \Phi )^2 \, +\, {\rm H.c.} \right]
\,. 
\end{align}
The model with a $Z_2$ symmetry, a Majorana fermion and one additional scalar doublet has the same symmetries
and types of particles as the original ScM, which has been extensively discussed in the literature~\cite{Ma:2006km}.
We comment on similarities and differences in phenomenology between the latter and the GScM with a global $\rm U(1)$ symmetry 
in Sec.~\ref{sec:scot}.

For $N=3$ and $N=4$ the Lagrangian ${\cal L}_\psi$ remains the same as in Eq.~\eqref{Lpsi}, but additional quartic terms  
 appear in the scalar potential, see also Ref.~\cite{Keus:2013hya}. For $N=3$
there are two new quartic couplings
\begin{equation}
\mathcal{V}_{Z_3}=\lambda_1\, (\Phi \Phi^\prime) (H\Phi^\prime) + \lambda_2 \,(\Phi \Phi^\prime) (\tilde H\Phi) + {\rm H.c.}
\end{equation}
In the case in which one of the new neutral scalars is the lightest particle with non-trivial $Z_3$ charge, these terms give rise to DM semi-annihilations~\cite{DEramo:2010keq,Belanger:2012zr,Ma:2007gq,Aoki:2014cja}. 
In principle, these new couplings are directly testable at colliders.
Furthermore, for $N=4$  the following term can be added to the scalar potential
\begin{equation}
\mathcal{V}_{Z_4}=\lambda_3\, (\Phi \Phi^\prime)^2 + {\rm H.c.}
\end{equation}
If one of the new neutral scalars is the lightest particle with non-trivial $Z_4$ charge, this term gives rise to DM self-interactions.


\subsection{The Generalised Scotogenic Triplet Model} \label{sec:triplet_model}

We can construct an interesting variant of the GScM by replacing the fermion singlet with a fermion triplet (we denote it the GScTM). The Lagrangian for the triplet Dirac fermion $\vec\Sigma=(\Sigma_1,\Sigma_2,\Sigma_3)$ is
\begin{equation}
\mathcal{L}_{\vec\Sigma}\;=\; i\, \overline{ \vec\Sigma}\, \slashed D \vec\Sigma\,-\, m_\Sigma \, \overline{ \vec\Sigma} \vec\Sigma\, -\, \Big [ y_{\Phi}^\alpha \,\tilde\Phi^\dagger\,\left( \overline{\vec\Sigma \cdot \vec \sigma}\right)\, L_\alpha\, + \, 
\left(y_{\Phi^\prime}^\alpha\right)^* \,\tilde \Phi^{\prime \dagger}\, \left( \overline{\vec\Sigma \cdot \vec \sigma}\right)\, \tilde L_\alpha  \,+\,\text{H.c.}\Big]\,,
\end{equation}
where $\vec \sigma=(\sigma_1,\sigma_2,\sigma_3)$ and the covariant derivative for the electroweak triplet fermion, $D_\mu=\partial_\mu+i g \tau_3^a W_\mu^a$, with the three generators $\tau_3^a$ of the triplet representation. The three physical Dirac fermion fields are as usual $\Sigma_0=\Sigma_3$ and $\Sigma_\pm=\frac{1}{\sqrt{2}}(\Sigma_1 \mp \Sigma_2)$, which are degenerate in mass at tree level. Radiative corrections lift the charged components by 166 MeV~\cite{Ma:2008cu,Cirelli:2009uv}. Notice that due to the dark symmetry the components of the fermion triplet do not mix with SM leptons. 

As we saw, in the case of GScM with singlet fermionic DM, the relic abundance cannot be explained by annihilations due to the strong limits from CLFV. In the case of the GScTM with triplet fermionic DM (we take its mass to be smaller than the scalars mass), the phenomenology is very different. The relic abundance, driven by gauge interactions, is decoupled from the neutrino and LFV phenomenology. This implies that the coannihilations of the GScM with the scalars, which involve some degree of fine tuning, are not needed. The dominant annihilation channels of the triplets are through gauge interactions, like $\Sigma_0 \Sigma_0 \rightarrow W^+ W^-$, mediated by the charged scalar $\Sigma_+$. Due to the small splitting between the neutral and the charged components, there are also important contributions from coannihilation channels like $\Sigma_0 \Sigma_+ \rightarrow Z/\gamma/H +W^+$ mediated by the charged $\Sigma_+$ (t-channel) or by a $W^+$ (s-channel), and $\Sigma_0 \Sigma_+ \rightarrow f \bar f$, mediated also by a $W^+$ (s-channel), where f are SM fermions. In this case reproducing the relic abundance fixes the mass of the fermion triplet to be equal to $2.7$ TeV~\cite{Cirelli:2007xd}.  

Regarding direct detection, the Z does not couple to the neutral fermion, so there is no tree level scattering. Moreover, the splitting with the charged fermion (166 MeV) being larger than the typical recoil momentum in direct detection experiments makes it impossible to have inelastic scattering mediated by a W. There are extra one-loop penguin diagrams in addition to those present for a fermion singlet, with the photon/H/Z attached to the $W^+$ in the loop, and with the photon/Z attached to the $\Sigma_+$ in the loop, as well as box diagrams with W in the loop, see Refs.~\cite{Hisano:2004pv,Cirelli:2005uq,Cirelli:2009uv}.

The presence of charged fermion components generate also extra contributions to CLFV, as well as new collider signatures, similar to the wino in SUSY. However the large triplet mass makes its production at the LHC very suppressed, being necessary a future collider to probe directly the model. The new charged fermions or scalars can be pair produced at colliders via the Drell-Yan process with a photon or $Z$ boson. Another important production channel is $\overline u d \rightarrow W^- \rightarrow \Sigma_- \Sigma_0$. The interesting feature is that the lifetime of the $\Sigma_+$ is fixed, such that it generates charged tracks at colliders of length equal to $5.5$ cm. The charged fermions $\Sigma_+ $ will decay into the DM (MET) $\Sigma_0$ plus a very soft W, which in turn decays into pions and leptons with the branching ratios~\cite{Cirelli:2009uv}:
${\rm BR} (\Sigma_+ \rightarrow \Sigma_0 \pi^+ ) =0.977$, 
${\rm BR} (\Sigma_+ \rightarrow \Sigma_0 e^+ \nu_e) =0.0205$, 
${\rm BR} (\Sigma_+ \rightarrow \Sigma_0 \mu^+\nu_\mu) =0.0025$.
One can also produce the scalars via $\overline u d \rightarrow W^- \rightarrow \eta^+ \eta^0$, which decay into $\Sigma_0+\ell^+$ or $\Sigma_++\nu_\ell$, with $\ell=e,\mu,\tau$. These last decays involve the neutrino Yukawas, and therefore there are definite predictions for the ratios of lepton flavours. Other collider studies of the fermion triplet in the context of seesaw type III have been performed in Refs.~\cite{Franceschini:2008pz,Eboli:2011ia}. 


\section{Comparison with the Scotogenic Model}
\label{sec:scot}

As already mentioned in Sec.~\ref{sec:disc}, if the global $\rm U(1)_{DM}$ symmetry is replaced by a $Z_2$ symmetry, the GScM 
coincides in symmetries and types of particles with the original ScM~\cite{Ma:2006km}. In the following, we highlight similarities and
differences between the latter and the GScM discussed here. 

In order to generate at least two neutrino masses, at least two Majorana fermions $\psi_{1,2}$ (with masses $m_{\psi_{1,2}}$) and one additional inert scalar doublet field are needed in the ScM,
whereas in the GScM one Dirac fermion and two new scalar doublet fields are needed. 
We have thus one more charged and one more neutral complex scalar filed  in the GScM compared to the ScM.
In the latter model the scalar mass spectrum, derived from the potential in Eq.~\eqref{potentialSc}, reads
\begin{eqnarray}
	m_{\phi_0^R} & = & \sqrt{m_\Phi^2\,+\,\frac 1 2 \left(\lambda_{H\Phi}\,+\,\lambda_{H\Phi,2}\,+\,\lambda_{H\Phi,3} \right) v_H^2} \,,\\
	m_{\phi_0^I} &=&\sqrt{m_\Phi^2\,+\,\frac 1 2 \left(\lambda_{H\Phi}\,+\,\lambda_{H\Phi,2}\,-\,\lambda_{H\Phi,3} \right) v_H^2}\,,\\
	m_{\phi^+} &=& \sqrt{m_{\Phi}^2\,+\,\frac 1 2 \lambda_{H\Phi}  \,v_H^2} 
\end{eqnarray}
with $\phi_0^R$ and $\phi_0^I$ being the real and imaginary components of the neutral component $\phi_0$ of the additional scalar doublet, $\phi_0\equiv (\phi_0^R\,+\, i\, \phi_0^I)/ \sqrt{2}$, and $\phi^+$ the charged component of the scalar doublet.
The  scalars $\phi_0^R$ and $\phi_0^I$ acquire a mass splitting proportional to the quartic coupling $\lambda_{H\Phi,3}$.
In contrast, in the GScM the mass spectrum, given in Eq.~\eqref{eq:meta_abc}, clearly shows that real and imaginary parts of the neutral scalars have the same mass and form complex neutral scalars, denoted $\eta_0$ and $\eta_0^\prime$. 
 
 In the ScM neutrino masses are generated by diagrams with the  neutral scalars $\phi_0^R$ and $\phi_0^I$ running in the loop. The neutrino mass matrix is given by
\begin{eqnarray} 
\label{nuSc}
	\left( \mathcal{M}_\nu\right)_{\alpha\beta} & = & \,\sum_{k}\,\frac{y_{\alpha k}\,y_{\beta k}\,m_{\psi_k}}{32\,\pi^2}\,\,F(m_{\phi_0^R},m_{\phi_0^I},m_{\psi_k})\,,
\end{eqnarray}
where the loop function is defined in Eq.~\eqref{eq:F}. 
We can see how the \emph{difference} in mass between the two complex neutral scalars $\eta_0$ and $\eta_0^\prime$ in the GScM, that appears in neutrino masses, see Eq.~\eqref{eq:numasses}, 
is traded for the difference in mass between the neutral scalars $\phi_0^R$ and $\phi_0^I$ in the ScM.
As is well-known, in the ScM lepton number is broken by the simultaneous presence of the Yukawa couplings, the masses of the Majorana fermions $\psi_{k}$
and the quartic coupling $\lambda_{H\Phi,3}$ of the  potential in Eq.~\eqref{potentialSc}. 
Thus neutrino masses crucially depend on all three of them. While the dependence on the first two ones is
 obvious from Eq.~\eqref{nuSc}, the one on $\lambda_{H\Phi,3}$ is best revealed in the limit $m_{\phi_0^R}^{2}-m_{\phi_0^I}^{2}=\lambda_{H\Phi,3}v_H^{2}\ll m_{0}^{2}\equiv(m_{\phi_{0}^R}^{2}+m_{\phi_{0}^I}^{2})/2$ where the expression for the neutrino mass matrix takes the form
\begin{equation}
\left( \mathcal{M}_\nu\right)_{\alpha\beta} =\frac{\lambda_{H\Phi,3}v_H^{2}}{32\,\pi^{2}}\sum_{k}\frac{y_{\alpha k}\,y_{\beta k}\, m_{\psi_k}}{m_{0}^{2}-m_{\psi_k}^{2}}\left[1-\cfrac{m_{\psi_k}^{2}}{m_{0}^{2}-m_{\psi_k}^{2}}\ln\cfrac{m_{0}^{2}}{m_{\psi_k}^{2}}\right] \, .
\end{equation} 
This is similar to what happens in the GScM, where the simultaneous presence of both Yukawa couplings $y^\alpha_\Phi$ and $y^\alpha_{\Phi^\prime}$, the fermion mass $m_\psi$ 
and the quartic coupling $\lambda_{H\Phi\Phi^\prime}$ is required in order to break lepton number. Consequently, neutrino masses are proportional
to all these quantities, as can be read off from Eq.~\eqref{eq:numasses} together with Eqs.~(\ref{eq:theta}) and (\ref{eq:meta_abc}). 

In the original ScM CLFV processes have been studied in detail in Refs.~\cite{Ma:2006km,Vicente:2014wga} (see also Ref.~\cite{Lindner:2016bgg}). 
It turns out that the individual penguin diagram contributions of a charged scalar and a fermion to CLFV processes in the original ScM are the same as the ones in the GScM, see Sec.~\ref{sec:pheno}.
However, the number of charged scalars and fermions
differs in the two models. We therefore obtain a different number of contributions to CLFV processes in the two models. Moreover, there are new box diagrams for trilepton decays in the GScM.

The DM phenomenology is different in the original ScM and in the GScM. For
scalar DM the main channels for DM direct detection in the
former are the tree-level mediated processes by the $Z$ 
and the Higgs boson~\cite{Hambye:2009pw}. For $\lambda_{H\Phi,3}\neq 0$ the scalar
($\phi_0^R$) and pseudoscalar ($\phi_0^I$) have different
masses and DM scattering off nuclei is an inelastic
process, with the $Z$-boson exchange typically dominating.
This imposes a lower bound on $\lambda_{H\Phi,3}$ in order
to kinematically forbid such scattering. 
In the GScM scalar DM, with $\eta_0^\prime$ being the DM particle, also naturally has a  large DM direct detection
cross section mediated by the $Z$ boson, unless the interaction with the $Z$ boson is suppressed, like for maximal mixing $\theta = \pi/4$. Moreover, there is an elastic contribution via  Higgs-boson exchange.

For fermionic DM in the ScM DM-nucleon scattering occurs at
one-loop level~\cite{Schmidt:2012yg}  via penguin diagrams, which happens similarly in the GScM, see Sec.~\ref{sec:dirdet}. If the
mass splitting between the Majorana fermions $\psi_k$ is sufficiently
small, there is a transitional magnetic dipole moment interaction with
charged leptons running in the loop. This leads to inelastic
DM-nucleon scattering. As discussed in Sec.~\ref{sec:dirdet}, in the GScM the
dominant DM-nucleon scattering  occurs via a magnetic dipole moment
interaction with charged leptons running in the loop. 

\section{Summary and conclusions}
\label{sec:conc}

We have studied a model in which masses for neutrinos are generated at one-loop level with Dirac fermion DM running in the loop. The model can be viewed as a generalised version of the ScM  (GScM) with a global U(1)$_{\rm DM}$ symmetry. Both neutrino mass orderings (NO and IO) can be accommodated. The flavour structure of the neutrino Yukawa couplings is determined by the neutrino oscillation parameters and the Majorana phase $\gamma$. The model is has some definite predictions. 
The flavour structure relevant  for neutrino masses differs from the one appearing in the expressions for the branching ratios of CLFV processes,  in contrast to the original ScM. We have obtained interesting correlations among the ratios of different CLFV processes, which may allow to test the GScM and to discriminate between the two neutrino mass orderings. 

In this work we have focused on fermionic DM, given the fact that scalar DM would require some  fine-tuning. The main DM annihilation channels are into charged leptons and neutrinos. As they depend on the same Yukawa couplings relevant for CLFV processes, the corresponding cross sections are too small in order to explain the observed DM relic density and thus coannihilations are important. In roughly  half of the parameter space of the model, the next-to-lightest particle is the lightest
neutral scalar ($\eta_0^\prime$), and in the other half it is the lightest
charged scalar (which can be either $\eta^{\pm}$ or $\eta^{\prime\pm}$).

Experimental limits on the branching ratios of  CLFV processes and on DM direct detection give complementary
information on the parameter space of the model. Future experiments, searching for 
$\mu - e$ conversion in nuclei and $\mu \rightarrow 3e$,
will probe the remaining
allowed parameter space of the model best, but also DM direct detection
experiments will further test a complementary region of
the available parameter space of the model. Another interesting signature
of the model  is the 
 production of new (neutral and charged) scalars at colliders and the decay of 
the charged scalars  to a charged lepton and DM. For neutrino masses with  NO the dominant channels are into muons and $\tau$ leptons, while for neutrino masses with IO the decay into electrons is of
similar magnitude. 

In comparison to the original ScM, the GScM has more
degrees of freedom in the scalar sector (two additional doublets versus one in the ScM),
and possesses one vector-like Dirac fermion, unlike the ScM which contains
at least two Majorana fermions. The flavour structure in the GScM is more restricted by the neutrino oscillation parameters than in the original ScM with three Majorana fermions. Nonetheless, they both are simple explanations for
neutrino masses and DM with a rich and testable phenomenology.

\vspace{1cm}

\section*{Acknowledgements} 
We are grateful for useful discussions with Takashi Toma and Wei-Chih Huang. We thank Yue-Ling Sming Tsai for providing a preliminary version of LikeDM and for answering many questions. We thank Diego Restrepo for pointing out the possibility of bound state DM in our scenario. JHG thanks Concha Gonz\'alez-Garc\'ia and Roberto Franceschini for useful discussions on the triplet variant of the model. This  work  was  supported  by  the  DFG  Cluster  of  Excellence  `Origin  and  Structure of the Universe' SEED project ``Neutrino mass generation mechanisms in (grand) unified flavour models and phenomenological  imprints''. This work has been supported in part by the Australian Research Council. J.H.-G acknowledges the support from the University of Adelaide and the Australian Research Council through the ARC Centre of Excellence for Particle Physics at the Terascale (CoEPP) (CE110001104). The CP3-Origins centre is partially funded by the Danish National Research Foundation, grant number DNRF90 (C.H.).

\appendix
\section{Stability of the potential}
\label{sec:stab}

We follow closely the vacuum stability conditions derived from co-positivity criteria in Refs.~\cite{Kannike:2012pe,Kannike:2016fmd}.
For large field values only the quartic part ${\cal V}_4$ of the potential is relevant. Using the following parametrisation 
\begin{eqnarray}  \label{eq:potentialbilinears}
&&  H^\dagger H = h_{1}^{2}, \qquad\qquad\qquad \Phi^\dagger \Phi = h_{2}^{2}, \qquad\qquad\qquad \Phi^{\prime\dagger} \Phi^\prime = h_{3}^{2}, \\ \nonumber
&& {H^\dagger} \Phi = h_{1} h_{2} \rho_{12} e^{i \phi_{12}},  \;\;\;\;\;\; {H^\dagger} \tilde\Phi^\prime = h_{1} h_{3} \rho_{13} e^{i \phi_{13}},  \;\;\;\;\;\;  {\Phi^\dagger} \tilde\Phi^\prime = h_{2} h_{3} \rho_{23} e^{i \phi_{23}} \; ,
\end{eqnarray}
with $0 \leq \rho_{ij} \leq 1$ and $0 \leq \phi_{ij} \leq 2 \, \pi$, ${\cal V}_4$ reads
\begin{align} \label{eq:potentialV4}
{\cal V}_4  =& \, \lambda_H \, h_1^4 \, +\, \lambda_\Phi \, h_2^4 \, + \, \lambda_{\Phi^\prime} \, h_3^4
\, + \, \left( \lambda_{H\Phi} \, + \, \lambda_{H\Phi,2} \, \rho_{12}^2 \right) \, h_1^2 \, h_2^2
\\\nonumber&
\,+\, \left( \lambda_{H\Phi^\prime} \, + \,  \lambda_{H\Phi^\prime,2} \, \rho_{13}^2 \right) \, h_1^2 \, h_3^2
\,+\, \left( \lambda_{\Phi\Phi^\prime}  \, + \, \lambda_{\Phi\Phi^\prime,2} \, \rho_{23}^2 \right) \, h_2^2 \, h_3^2
\\\nonumber&
\,+\, 2 \, \lambda_{H\Phi\Phi^\prime} \, h_1^2 \, h_2 \, h_3 \, \rho_{12} \, \rho_{13} \, \cos (\phi_{12}+\phi_{13}) \,.
\end{align}
If we neglect $\lambda_{H\Phi\Phi^\prime}$, we can write ${\cal V}_4$ as a bilinear form
\begin{align}
	{\cal V}_4 &= \left( \begin{array}{ccc} h_1^2 & h_2^2 & h_3^2 \end{array} \right)\, Q 
\, \left( \begin{array}{c} h_1^2 \\ h_2^2 \\ h_3^2 \end{array} \right)
\end{align}
with
\begin{align}
Q &\equiv\left(
\begin{array}{ccc}
\lambda_H & \frac 12 \, (\lambda_{H \Phi} + \lambda_{H \Phi,2} \, \rho_{12}^2) &  \frac 12 \, (\lambda_{H \Phi^\prime} + \lambda_{H \Phi^\prime,2} \, \rho_{13}^2) \\
 \frac 12 \, (\lambda_{H \Phi} + \lambda_{H \Phi,2} \, \rho_{12}^2) & \lambda_\Phi &  \frac 12 \, (\lambda_{\Phi \Phi^\prime} + \lambda_{\Phi \Phi^\prime,2} \, \rho_{23}^2) \\
  \frac 12 \, (\lambda_{H \Phi^\prime} + \lambda_{H \Phi^\prime,2} \, \rho_{13}^2) & \frac 12 \, (\lambda_{\Phi \Phi^\prime} + \lambda_{\Phi \Phi^\prime,2} \, \rho_{23}^2) & \lambda_{\Phi^\prime}
\end{array}
\right)
\end{align}
and determine the necessary conditions for the stability of the potential. We find 
\begin{equation}
\lambda_H \geq 0 \;, \;\; \lambda_\Phi \geq 0 \;, \;\; \lambda_{\Phi^\prime} \geq 0 
\end{equation}
and 
\begin{eqnarray}
c_1 = \frac 12 \, (\lambda_{H \Phi}+\lambda_{H \Phi,2} \, \rho_{12}^2) + \sqrt{\lambda_H \, \lambda_\Phi} \geq 0 \; ,
\\ \nonumber
c_2 = \frac 12 \, (\lambda_{H \Phi^\prime}+\lambda_{H \Phi^\prime,2} \, \rho_{13}^2) + \sqrt{\lambda_H \, \lambda_{\Phi^\prime}} \geq 0 \; ,
\\ \nonumber
c_3 = \frac 12 \, (\lambda_{\Phi\Phi^\prime}+\lambda_{\Phi \Phi^\prime,2} \, \rho_{23}^2) + \sqrt{\lambda_\Phi \, \lambda_{\Phi^\prime}} \geq 0 \; ,
\end{eqnarray}
together with
\begin{eqnarray}
&& \sqrt{\lambda_H \, \lambda_\Phi \, \lambda_{\Phi^\prime}} + \frac 12 \, (\lambda_{H \Phi}+\lambda_{H \Phi,2} \, \rho_{12}^2) \sqrt{\lambda_{\Phi^\prime}}
+ \frac 12 \, (\lambda_{H \Phi^\prime}+\lambda_{H \Phi^\prime,2} \, \rho_{13}^2) \sqrt{\lambda_\Phi} 
\\ \nonumber
&& \phantom{xxxxxxxxxxxxxxxxxxxxxxx}+ \frac 12 \, (\lambda_{\Phi \Phi^\prime}+\lambda_{\Phi \Phi^\prime,2} \, \rho_{23}^2) \sqrt{\lambda_H}
+ \sqrt{2 \, c_1 \, c_2 \, c_3} \geq 0 \; .
\end{eqnarray}
Depending on the sign of $\lambda_{H \Phi, 2}$, $\lambda_{H \Phi^\prime, 2}$ and $\lambda_{\Phi\Phi^\prime,2}$ the necessary conditions are
given for $\rho_{ij}=0$ or $\rho_{ij}=1$, i.e. 
for  $\lambda_{H \Phi, 2} > 0$ the necessary conditions are obtained for $\rho_{12}=0$, while for $\lambda_{H \Phi, 2} < 0$ these are given for $\rho_{12}=1$.  
The same is true for  $\lambda_{H \Phi^\prime, 2}$ and $\rho_{13}$ as well as for $\lambda_{\Phi \Phi^\prime, 2}$ and $\rho_{23}$.

For non-zero $\lambda_{H\Phi\Phi^\prime}$ we derive sufficient (but not necessary) conditions for the stability of the potential from considering co-positivity. We re-write ${\cal V}_4$ as 
\begin{eqnarray}
{\cal V}_4 &=& h_1^4 \, \left[ \lambda_H + 
\left( \begin{array}{cc} \tilde{h}_2^2 & \tilde{h}_3^2 \end{array} \right)\, 
\left(
\begin{array}{cc}
 \lambda_\Phi &  \frac 12 \, (\lambda_{\Phi \Phi^\prime} + \lambda_{\Phi \Phi^\prime,2} \, \rho_{23}^2) \\
 \frac 12 \, (\lambda_{\Phi \Phi^\prime} + \lambda_{\Phi \Phi^\prime,2} \, \rho_{23}^2) & \lambda_{\Phi^\prime}
\end{array}
\right)
\, \left( \begin{array}{c} \tilde{h}_2^2 \\ \tilde{h}_3^2 \end{array} \right)
\right.
\\ \nonumber
&&\left.
+ \, \left( \begin{array}{cc} \tilde{h}_2 & \tilde{h}_3 \end{array} \right) \,  \left( 
\begin{array}{cc}
 \lambda_{H \Phi} + \lambda_{H \Phi,2} \, \rho_{12}^2 & \lambda_{H \Phi \Phi^\prime} \, \rho_{12} \, \rho_{13} \, \cos (\phi_{12}+\phi_{13}) \\
 \lambda_{H \Phi \Phi^\prime} \, \rho_{12} \, \rho_{13} \, \cos (\phi_{12}+\phi_{13}) &  \lambda_{H \Phi^\prime} + \lambda_{H \Phi^\prime,2} \, \rho_{13}^2
\end{array}
\right) \, \left( \begin{array}{c} \tilde{h}_2 \\ \tilde{h}_3 \end{array} \right)
\right]
\end{eqnarray}
with $\tilde{h}_{2,3}$ being $h_{2,3}$ rescaled by $h_1$. We require then co-positivity of the three different terms of ${\cal V}_4$. This leads to 
\begin{equation}
\lambda_H \geq 0
\end{equation}
and to
\begin{equation}
 \lambda_\Phi \geq 0 \;\; , \;\; \lambda_{\Phi^\prime} \geq 0 \;\;\; \mbox{and} \;\;\;
 \frac 12 (\lambda_{\Phi \Phi^\prime} + \lambda_{\Phi \Phi^\prime,2} \, \rho_{23}^2) + \sqrt{\lambda_\Phi \, \lambda_{\Phi^\prime}} \geq 0\,,
\end{equation}
with the last condition depending on the sign of $ \lambda_{\Phi \Phi^\prime,2}$, i.e.~$\rho_{23}=0 \, (1)$ for positive (negative) $ \lambda_{\Phi \Phi^\prime,2}$. 
In order to use co-positivity of the last term we first minimise ${\cal V}_4$ with respect to $\rho_{12}$, $\rho_{13}$ and $\cos (\phi_{12}+\phi_{13})$. 
The term with $\lambda_{H \Phi \Phi^\prime}$ alone is minimised for $\cos (\phi_{12}+\phi_{13})=-1$. From the extreme values of $\rho_{12}$ and $\rho_{13}$ that minimise ${\cal V}_4$
we derive then the sufficient conditions
\begin{eqnarray}
&&\lambda_{H \Phi} \geq 0 \;\; , \;\; \lambda_{H \Phi^\prime} \geq 0 \;\; , \;\; \lambda_{H \Phi} + \lambda_{H \Phi,2}  \geq 0 \;\; , \;\; \lambda_{H \Phi^\prime} + \lambda_{H \Phi^\prime,2}  \geq 0 \;\; ,
\\ \nonumber
&&\mbox{and} \;\;\; \sqrt{(\lambda_{H \Phi} + \lambda_{H \Phi,2})(\lambda_{H \Phi^\prime} + \lambda_{H \Phi^\prime,2})} - \lambda_{H \Phi \Phi^\prime} \geq 0 \; .
\end{eqnarray}
Only the last condition involves $ \lambda_{H \Phi \Phi^\prime}$ and bounds the latter from above. 

${\cal V}_4$ is also minimised for non-extremal values
of $\rho_{12}$ and $\rho_{13}$. This, however, does not imply conditions different from those already shown above, but 
only leads to an equality involving $\lambda_{H \Phi \Phi^\prime}$, $\lambda_{H \Phi, 2}$ and $ \lambda_{H \Phi^\prime, 2}$
which needs to be fulfilled in addition.
We have checked that the presented conditions can also be applied to the special directions in which one or two of $h_{1,2,3}$ vanish.


\section{Neutrino masses and lepton mixing}
\label{app:PMNS}
We can diagonalize the neutrino mass matrix in Eq.~(\ref{eq:numasses}) as 
\begin{equation}
\mathcal{M}_{\nu}= U^* D_\nu U^\dagger \,, 
\label{eq:numass}
\end{equation}
where $D_\nu$ is a $3 \times 3$ diagonal matrix with positive semi-definite eigenvalues (in our model with $m_1=0$ for NO, and $m_3=0$ for IO). $U$ is the PMNS mixing matrix, which relates the neutrino mass eigenstates $\nu_i$
($i=1,2,3$) with masses $m_i$ to the neutrino flavour eigenstates $\nu_\alpha$ ($\alpha=e,\mu,\tau$):
\begin{equation}
\nu_\alpha = \sum_{i=1}^3 \,U^*_{i\alpha} \, \nu_i\,.
\end{equation} 
The standard parametrisation for $U$ for one massless neutrino is
\begin{align}
U=\left(\begin{array}{ccc}
c_{13}c_{12} & c_{13}s_{12} & s_{13}e^{-i\delta}\\
-c_{23}s_{12}-s_{23}s_{13}c_{12}e^{i\delta} & c_{23}c_{12}-s_{23}s_{13}s_{12}e^{i\delta} & s_{23}c_{13}\\
s_{23}s_{12}-c_{23}s_{13}c_{12}e^{i\delta} & -s_{23}c_{12}-c_{23}s_{13}s_{12}e^{i\delta} & c_{23}c_{13}\end{array}\right)\left(\begin{array}{ccc}
1 & 0 & 0\\
0 & e^{i\gamma} & 0\\
0 & 0 & 1\end{array}\right)\,,
\label{UPMNS}
\end{align}
where $c_{ij}\equiv\cos\theta_{ij}$ and $s_{ij}\equiv\sin\theta_{ij}$ ($\theta_{12}$, $\theta_{13}$, and $\theta_{23}$ being the three lepton mixing angles). $\gamma$ is the Majorana and $\delta$ the Dirac phase. Since the lightest neutrino is massless in the GScM, there is only one physical Majorana phase.

\section{Parametrisation of the neutrino Yukawa couplings}
\label{sec:yukpar}
\begin{figure}[p]
	\centering
	\includegraphics[width=0.45\textwidth]{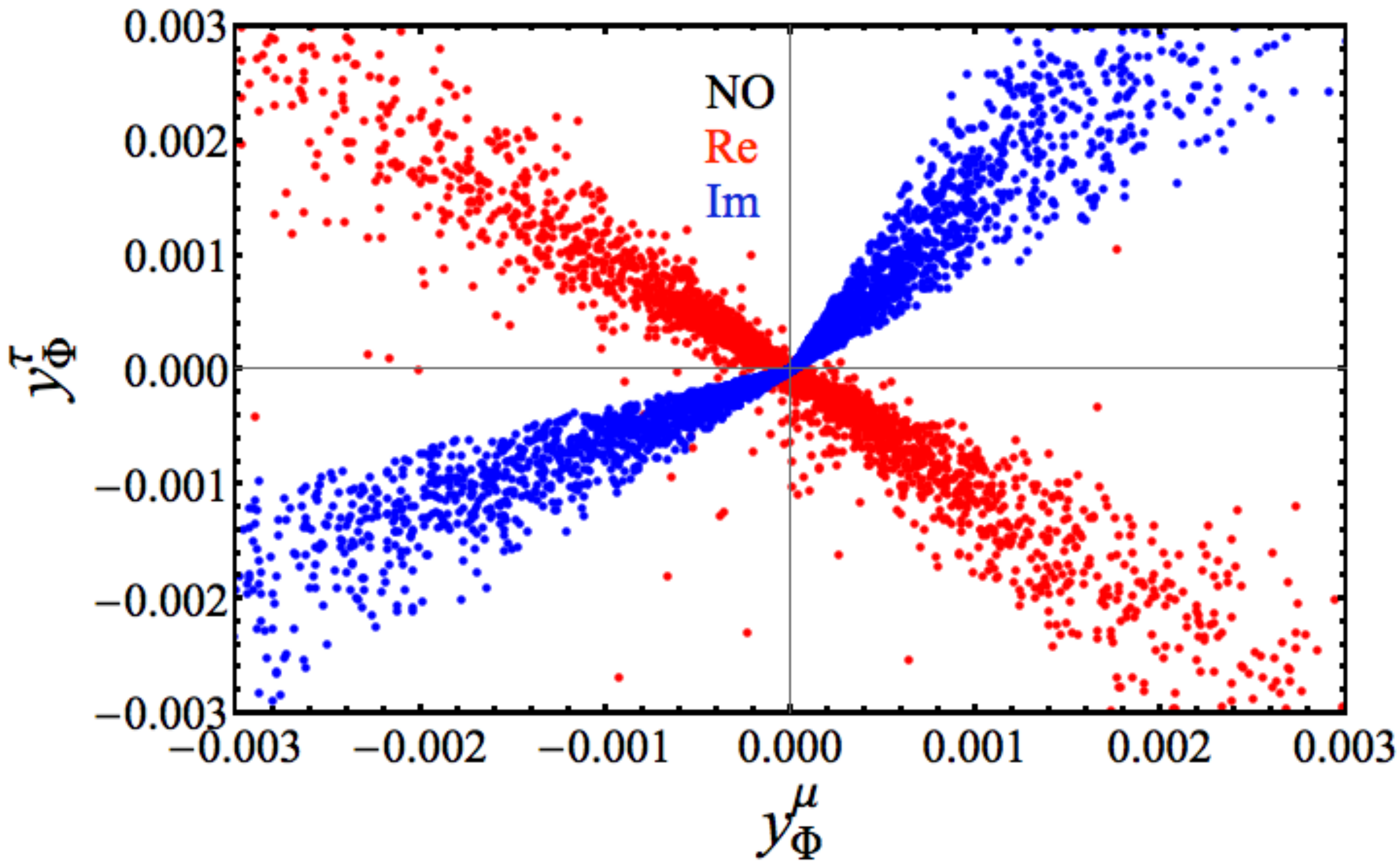}~~\includegraphics[width=0.45\textwidth]{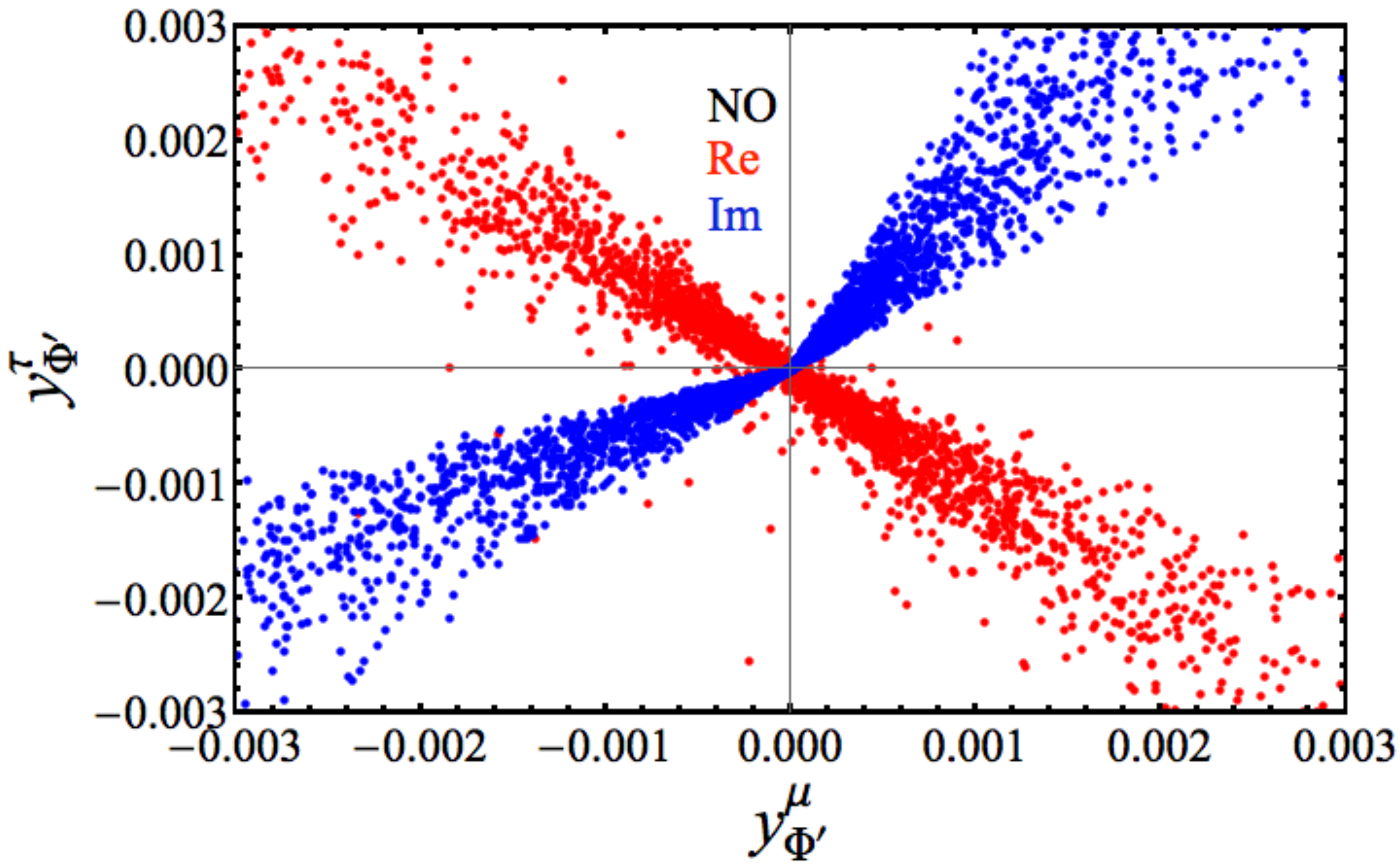}\\
	\includegraphics[width=0.45\textwidth]{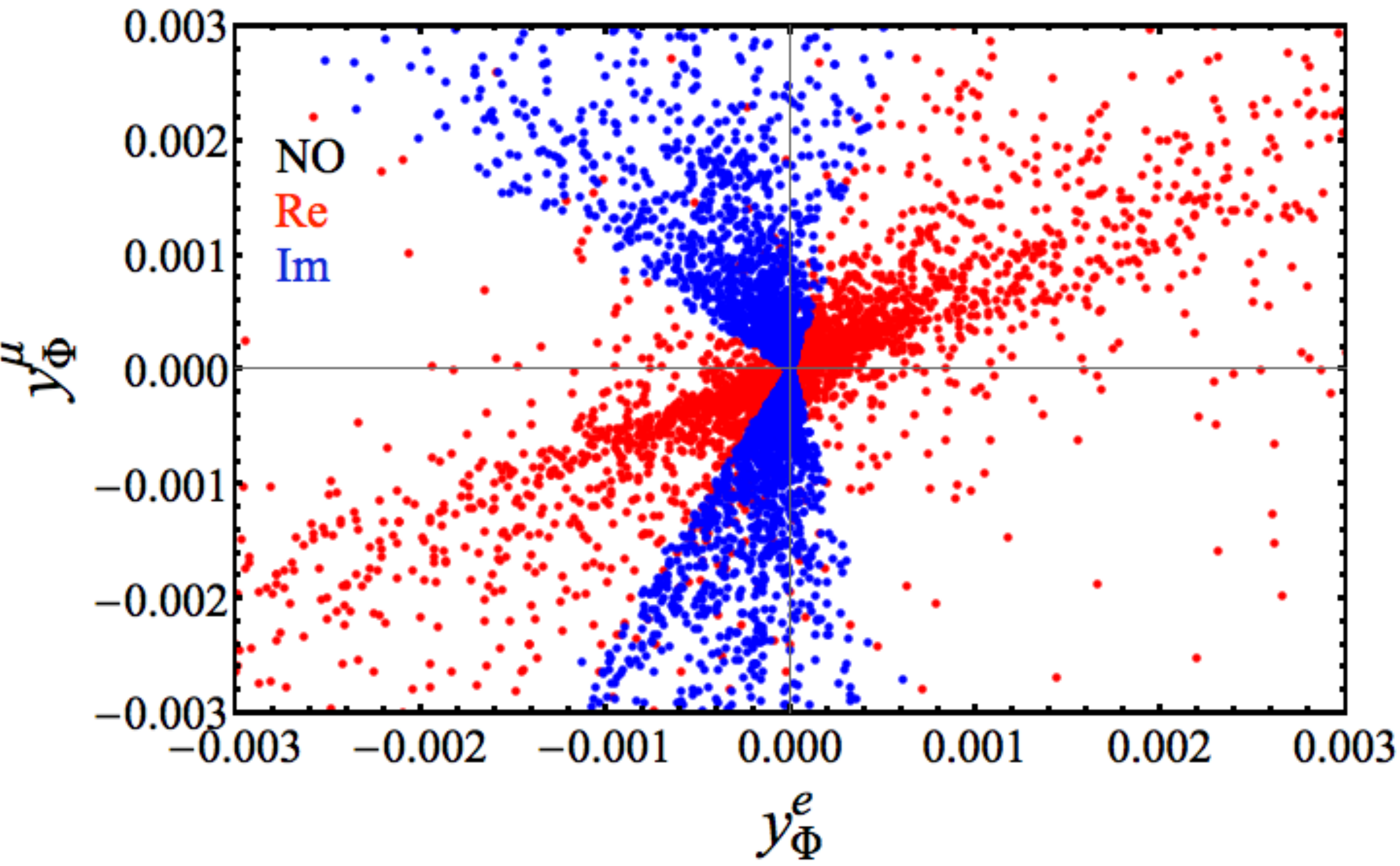}~~\includegraphics[width=0.45\textwidth]{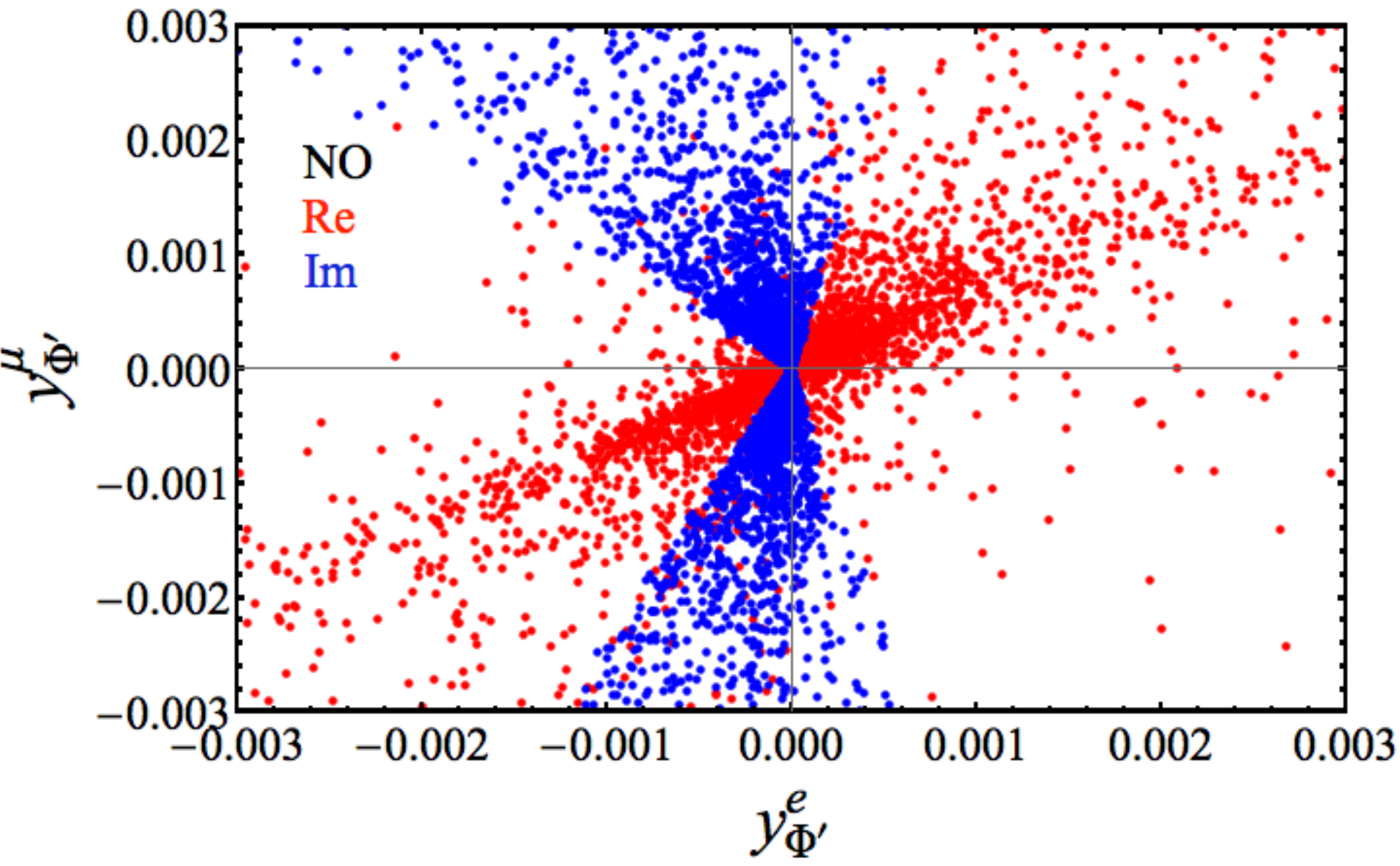}\\
	\caption{Neutrino Yukawa couplings $y_\Phi^\alpha$ (\textit{left panels}) and $y_{\Phi^\prime}^\alpha$ (\textit{right panels}) for neutrino masses with normal ordering, separated according to real (in red) and imaginary parts (in blue). The upper panel shows the flavour $\tau$ versus $\mu$, while the lower one shows $\mu$ versus $e$.}
\label{fig:yukawasNO}
\end{figure}

\begin{figure}[p]
	\centering
	\includegraphics[width=0.45\textwidth]{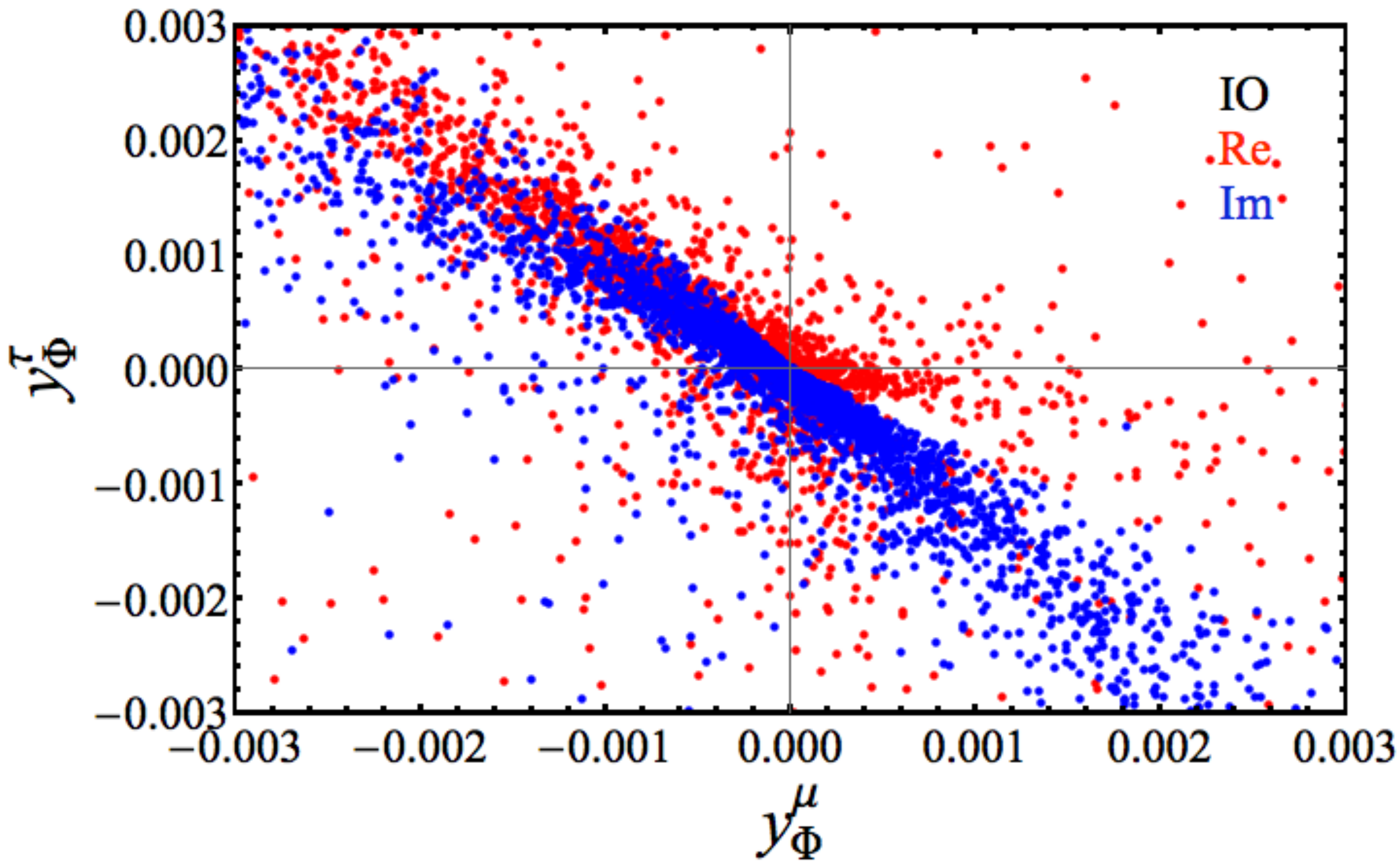}~~\includegraphics[width=0.45\textwidth]{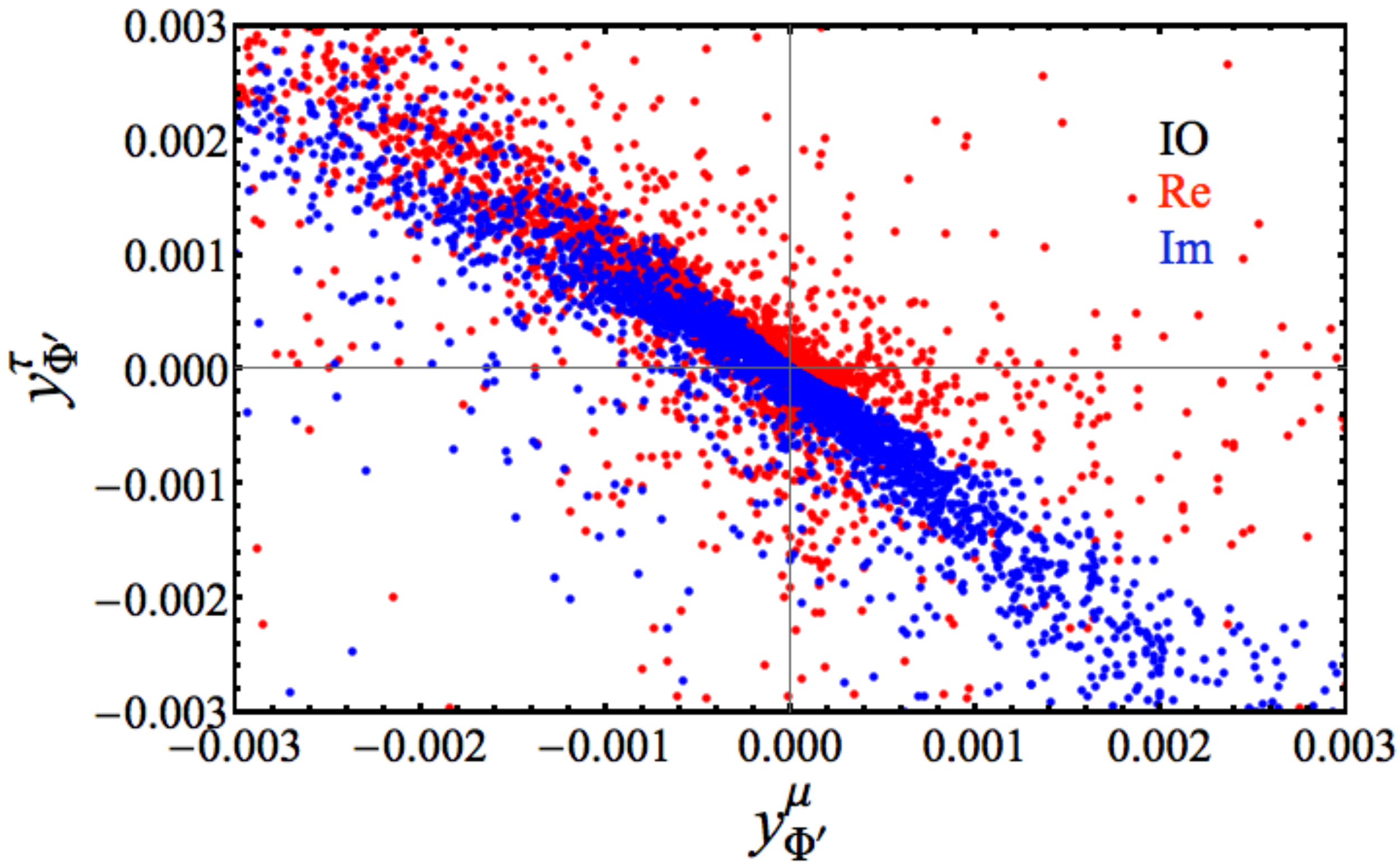}\\
	\includegraphics[width=0.45\textwidth]{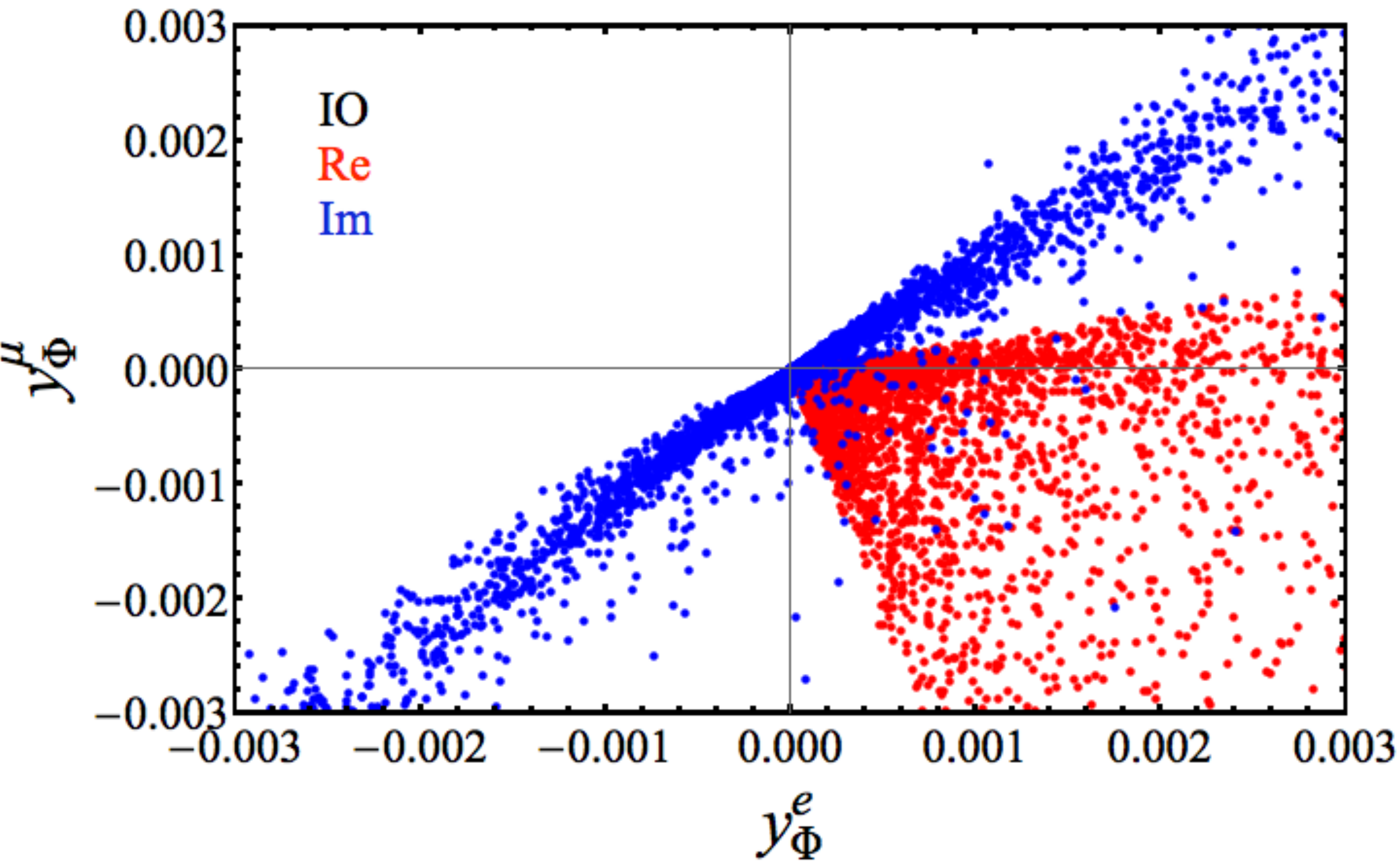}~~\includegraphics[width=0.45\textwidth]{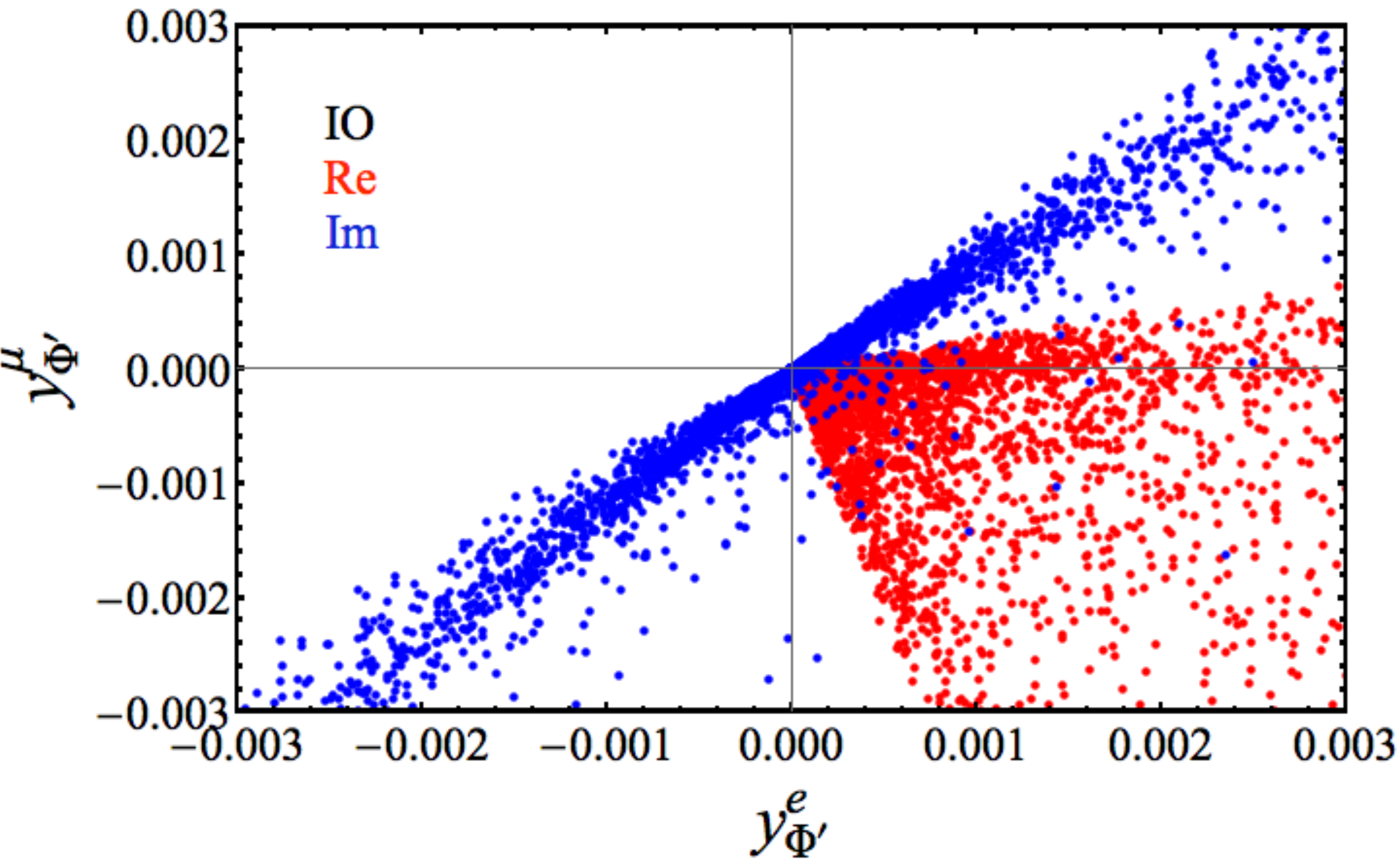}\\
		\caption{The same as in Fig.~\ref{fig:yukawasNO} for neutrino masses with inverted ordering. 
		}\label{fig:yukawasIO}
\end{figure}

We want to express the neutrino Yukawa couplings in terms of neutrino masses and lepton mixing parameters. We follow the discussion in Ref.~\cite{Cai:2014kra}.
On the one hand, the rank-two neutrino mass matrix can be expressed in terms of the two
non-vanishing mass eigenvalues and two columns of the PMNS mixing matrix, $u_i$, in the flavour
basis
\begin{equation}
	{\cal M}_\nu = \sum_{i} m_i u_i^* u_i^\dagger = \sum_{i} v_iv_i^T\;.
\end{equation}
The vectors $v_{i}\equiv \sqrt{m_i} u_i^*$ are linearly independent and span a two-dimensional vector space.
On the other hand, the calculation of the neutrino mass matrix results in the
following form
\begin{equation}
	{\cal M}_\nu = x_1 x_2^T + x_2 x_1^T
\end{equation}
with 
\begin{eqnarray}
	x_1= \,\sqrt{\frac{\sin2\theta \,m_\psi}{32\,\pi^2}\,F(m_{\eta_0},m_{\eta_0^\prime},m_\psi)}\,y_\Phi \; , \; x_2= \,\sqrt{\frac{\sin2\theta \,m_\psi}{32\,\pi^2}\,F(m_{\eta_0},m_{\eta_0^\prime},m_\psi)}\,y_{\Phi^\prime} \,,
\end{eqnarray}
see Eqs.~\eqref{eq:numasses} and \eqref{eq:F}. If $\sin2\theta < 0$ the square roots yield a complex number.
 The two linearly independent vectors $x_i$ can be written in terms of the vectors $v_i$
\begin{equation}
x_i = a_{ij} v_j\,,
\end{equation}
where $(a_{ij})$ forms an invertible $2\times 2$ matrix, i.e. $\det a
=a_{11}a_{22} -a_{12} a_{21}\neq 0$. Using the two different
parametrisations of the neutrino mass matrix, we can find possible solutions of $a_{ij}$
\begin{align}
	{\cal M}_\nu &=  x_1x_2^T + x_2 x_1^T \\
	&= \sum_{i,j} a_{1i}
	a_{2j} v_i v_j^T +\sum_{i,j} a_{2i} a_{1j} v_i v_j^T = \sum_{i,j}
	(a_{1i} a_{2j} + a_{2i} a_{1j}) v_i v_j^T\\
	&=\sum_{i} v_i v_i^T\,.
\end{align}
As the vectors $v_i$ form a basis of the two-dimensional vector space, we find
\begin{align}
	a_{11}a_{21}+a_{21}a_{11} &=1&\Rightarrow \quad a_{11}a_{21}&=\frac12\,,\\
	a_{12}a_{22}+a_{22}a_{12}&=1&\Rightarrow \quad a_{12}a_{22} &=\frac12\,,\\
	a_{11}a_{22}+a_{21}a_{12}&=0&\Rightarrow \quad a_{11}^2 a_{22}^2 &= a_{12}^2 a_{21}^2=-\frac14\,.
\end{align}
In particular the matrix elements are non-zero, $a_{ij}\neq 0$. There are two sets of solutions. Writing the complex matrix element
$a_{11}=r e^{i\alpha}$ in terms of two real parameters $r,\,\alpha$ we obtain 
\begin{align}
	a_{11}& = r e^{i\alpha}\, \equiv \frac{\zeta}{\sqrt{2}} , & 
	a_{21}& = \frac{1}{2r} e^{-i\alpha}\,,&
a_{22} & = \frac{1}{2r} e^{-i \left(\alpha\pm\frac{\pi}{2}\right)}\,,& 
	a_{12} & =  r e^{i\left(\alpha\pm\frac{\pi}{2}\right)}
\end{align}
and the condition of a non-vanishing determinant 
\begin{equation}
	0\neq\det a = a_{11}a_{22}-a_{12}a_{21} = 2 \, a_{11}a_{22} = e^{\mp i
	\frac{\pi}{2}}
\end{equation}
is trivially satisfied for the two solutions.
 Thus the two vectors can be uniquely written as
\begin{align}
	x_1 &= \frac{\zeta}{\sqrt{2}} \left( v_1 \pm i v_2\right),\, & 
	x_2 &= \frac{1}{\sqrt{2} \zeta} \left(v_1 \mp i v_2\right)\, .
\end{align}
For NO we obtain
\begin{align}
\label{eq:YukNO}
	x_1^{\mathrm{(NO)}}= \frac{\zeta}{\sqrt{2}} \left(\sqrt{m_2} u_2^* \pm i \sqrt{m_3} u_3^*\right),\,\qquad\, 
	x_2^{\mathrm{(NO)}}= \frac{1}{\sqrt{2} \zeta} \left(\sqrt{m_2} u_2^* \mp i \sqrt{m_3} u_3^*\right),\,\end{align}
while for IO, we have that
\begin{align}
\label{eq:YukIO}
	x_1^{\mathrm{(IO)}}= \frac{\zeta}{\sqrt{2}} \left(\sqrt{m_1} u_1^* \pm i \sqrt{m_2} u_2^*\right),\,\qquad\,	
	x_2^{\mathrm{(IO)}}= \frac{1}{\sqrt{2} \zeta} \left(\sqrt{m_1} u_1^* \mp i \sqrt{m_2} u_2^*\right)  \,.
\end{align}
Without loss of generality we choose $\zeta$ to be real and positive. Any phase of $\zeta$ can be absorbed via phase redefinitions of the lepton doublets $L_\alpha$ and the  Dirac fermion $\psi$. In Figs.~\ref{fig:yukawasNO} and \ref{fig:yukawasIO} we show the results for $y_\Phi^\alpha$ (left) and $y_{\Phi^\prime}^\alpha$ (right) for neutrino masses with NO and IO, respectively,
separated according to real (in red) and imaginary (in blue) parts, as obtained in the numerical scans. We show different flavours: the upper panel shows the flavour $\tau$ versus $\mu$, while the lower one shows $\mu$ versus $e$. We clearly see that $y_\Phi^\alpha$ and $y_{\Phi^\prime}^\alpha$ of different flavour $\alpha$ are correlated. These correlations can be understood analytically to a certain extent, see Sec.~\ref{sec:numasses}.

\section{Loop functions} \label{app:loop}
In this appendix we collect loop functions and other inputs appearing in CLFV processes, $h \rightarrow \gamma \gamma$ and DM direct detection.

The loop functions for the dipole and monopole photon contributions to CLFV processes are
\begin{equation}\label{f-g-loop}
\begin{split}
	f(x)  \; = \; & \frac{1-6x+3x^2+2x^3-6x^2\ln(x)}{6(1-x)^4}\,, \\
	g(x) \; = \; & \frac{2-9x+18x^2-11x^3+6x^3 \ln(x)}{12(1-x)^4} \,. 
\end{split}
\end{equation}
The following ones enter in the box diagrams of trilepton decays
\begin{equation} \label{app:looph}
\begin{split}
	h_1(x) \;=\; & \frac{1-x^2+2x\ln x}{2(x-1)^3}\,,\\
	h_2(x,y) \;=\; & -\frac{ x y}{2(1-x)(1-y)}-\frac{x^2 y \ln
x}{2(1-x)^2(x-y)}-\frac{x y^2\ln y}{2(1-y)^2 (y-x)}\,.
\end{split}
\end{equation}
The numerical values of the overlap integrals $D$ and $V^{(p,n)}$ and the total muon capture rate $\omega_{\mathrm{capt}}$, needed for the computation of $\mu-e$ conversion ratios in nuclei, are shown in Tab.~\ref{tab:overlap} for three different nuclei.  
\begin{table}[tbp]
\centering\setlength{\extrarowheight}{3pt}
\begin{tabular}{|c|ccc|c|}
\hline
 &  $V^{(p)}$ & $V^{(n)}$ & $D$ & $\omega_{\rm capt}(10^6 \text{s}^{-1})$\\
 \hline\hline
 $^{197}_{79}\text{Au}$ & 0.0859 & 0.108  & 0.167 & 13.07 \\
 $^{48}_{22}\text{Ti}$  & 0.0399 & 0.0495 & 0.0870  & 2.59\\
$^{27}_{13}\text{Al}$   & 0.0159 & 0.0169 &  0.0357 & 0.7054 \\
\hline
\end{tabular}
\vspace{1ex}

\begin{minipage}{14cm}
\caption{Overlap integrals in units of $m_{\mu}^{5/2}$ ($V^{(p)}$, $V^{(n)}$ and $D$) and total capture rates ($\omega_{\rm capt}$) for different nuclei \cite{Kitano:2002mt}. 
	The total capture rates are taken from Tab.~8 in Ref.~\cite{Kitano:2002mt}. 
The overlap integrals of $^{197}_{79}\text{Au}$ as well as $^{27}_{13}\text{Al}$ are taken from Tab.~2 and the ones for
$^{48}_{22}\text{Ti}$ are taken from Tab.~4 in Ref.~\cite{Kitano:2002mt}.
 \label{tab:overlap}}
\end{minipage}
\end{table}

The relevant loop function for the DM magnetic dipole moment which gives the dominant contribution to DM direct detection is 
	\begin{align}
		f_{\rm DD}(x ,y, z) &=  1 - \frac{y^2 -z^2}{x^2} \ln \frac{y}{x}
		\nonumber	\\&
\label{eq:fDD}
		+ \frac{y^4 +z^4 -x^2 y^2 -x^2 z^2 -2 z^2 y^2}{x^2 \lambda^{1/2}(z^2,x^2,y^2)}\, \ln \frac{y^2-x^2 +z^2 +  \lambda^{1/2}(z^2,x^2,y^2)}{2y z} 
	\end{align}
with the K\"all\'en-$\lambda$ function $\lambda(x,y,z) = x^2 +y^2 +z^2 -2 xy -2 yz -2 zx$.

In $h \rightarrow \gamma \gamma$ we need the following loop functions for scalars, fermions and gauge bosons $A_i(x)$ $(i=0,1/2,1)$
\begin{align} \label{loopgamma}
A_0(x) &= -x+x^2 \, f \left (\frac{1}{x}\right )\,, \\
A_{1/2}(x) &= 2x+ 2x (1- x) \, f \left (\frac{1}{x}\right )\,, \\
A_{1}(x) &= -2-3x-3x (2-x) \, f \left (\frac{1}{x}\right )\,.
\end{align}

\section{Oblique parameters}
\label{app:ewpt}

 The two inert scalar doublets $\Phi$ and $\Phi^\prime$ contribute to the EWPT at one-loop level. The contribution in our model to the $T$ parameter is given by~\cite{Grimus:2007if, Haber:2010bw}
\begin{align}
	T & = \dfrac{1}{16\pi^2 \alpha_{\rm em} v_H^2} \biggl\{2\,s^2_\theta \mathcal{F}(m_{\eta^+}^2,m_{\eta_0}^2) + 2\,c^2_\theta \mathcal{F}(m_{\eta^+}^2,m_{\eta'_{0}}^2)+ 2\,c^2_\theta \mathcal{F}(m_{\eta'^{+}}^2,m_{\eta_0}^2) + 2\,s^2_\theta \mathcal{F}(m_{\eta'^+}^2,m_{\eta'_{0}}^2)\nonumber
\biggr\} \,,
\end{align}
where the loop function is defined as
\begin{equation}
\mathcal{F}(x^2,\,y^2) = \dfrac{x^2+ y^2}{2} - \dfrac{x^2 y^2}{x^2-y^2}\ln\dfrac{x^2}{y^2}\,.
\end{equation}
The loop function is symmetric in $x$ and $y$. It vanishes in the custodial symmetry limit, $x\to y$, and diverges for $x/y$ going to $0$ or infinity.
Extending the results of Ref.~\cite{Haber:2010bw}, the $S$ parameter reads in our model
\begin{align} \nonumber
S&= \frac{1}{\pi m_Z^2} \biggl\{-\mathcal{B}_{22}(m_Z^2,m^2_{\eta^+},m^2_{\eta^+}) -\mathcal{B}_{22}(m_Z^2,m^2_{\eta^{\prime +}},m^2_{\eta^{\prime +}}) +\mathcal{B}_{22}(m_Z^2,m_{\eta_0}^2,m_{\eta_0}^2) +\mathcal{B}_{22}(m_Z^2,m_{\eta_0^\prime}^2,m_{\eta_0^\prime}^2)
\biggr\} \,.
\end{align}
Similarly, the combination $S+U$ combination results in
\begin{align} \nonumber
S+ U &= \frac{1}{\pi m_W^2} \biggl\{ 2\,s^2_\theta \mathcal{B}_{22}(m_W^2,m^2_{\eta^+},m_{\eta_0}^2) +2\,c^2_\theta \mathcal{B}_{22}(m_W^2,m^2_{\eta^{+}},m_{\eta_0^\prime}^2) \nonumber\\
&+2\,c^2_\theta \mathcal{B}_{22}(m_W^2,m^2_{\eta^{\prime+}},m_{\eta_0}^2) +2\,s^2_\theta \mathcal{B}_{22}(m_W^2,m^2_{\eta^{\prime+}},m_{\eta_0^\prime}^2) \nonumber\\
&- 2 \mathcal{B}_{22}(m_W^2, m^2_{\eta^+},m^2_{\eta^+}) - 2 \mathcal{B}_{22}(m_W^2, m^2_{\eta^{\prime+}},m^2_{\eta^{\prime+}}) \nonumber
\biggr\}\,.
\end{align}
The auxiliary functions $\mathcal{B}_{22}$ and $\mathcal{B}_0$ are defined as
\begin{align} 
\mathcal{B}_{22}(q^2,m_1^2,m_2^2) &\equiv
B_{22}(q^2,m_1^2,m_2^2)-B_{22}(0,m_1^2,m_2^2)\,.\label{b22} 
\end{align}
The Passarino--Veltman function $B_{22}$ 
\cite{Passarino:1978jh} arises from
two-point self-energies. In dimensional regularisation this  function reads~\cite{Haber:2010bw}
\begin{align} 
\label{B}
B_{22}(q^2,m_1^2,m_2^2) &= \tfrac{1}{4}(\Delta+1)(m_1^2+m_2^2-\tfrac{1}{3}q^2)-\frac{1}{2}\int^1_0 X \ln(X-i\epsilon) \, {\rm d}x \,, 
\end{align}
with 
\begin{equation}
X \equiv m_1^2 x + m_2^2(1-x) -q^2x(1-x)\,, \quad
\Delta \equiv \frac{2}{4-d}+\ln 4\pi-\gamma_E
\end{equation}
in $d$ space-time dimensions, where $\gamma_E \simeq 0.577$ is the Euler--Mascheroni constant. Note that $B_{22}$ 
is symmetric in the last two arguments. We use the compact analytic expressions given in App.~B of Ref.~\cite{Herrero-Garcia:2017xdu}.
We have confirmed that the expressions agree with the ones in the inert doublet model~\cite{Barbieri:2006dq}, when taking the appropriate limit.

\bibliography{refs}

\end{document}